\documentclass[aps,prx,twocolumn,superscriptaddress,longbibliography]{revtex4-2}
\usepackage[utf8]{inputenc}
\usepackage{color}
\usepackage{amsmath}
\usepackage{amsfonts}
\usepackage{mathtools}
\usepackage{hyperref}
\usepackage[capitalise]{cleveref}
\usepackage{amsthm}
\usepackage{bm}
\usepackage{float}
\usepackage{accents}
\usepackage[percent]{overpic}
\usepackage{natbib}
\usepackage{braket}
\usepackage{comment}
\usepackage{scalerel}
\usepackage{makecell}
\usepackage{multirow}  
\usepackage[normalem]{ulem}
\usepackage{placeins}

\usepackage[dvipsnames]{xcolor}
\definecolor{mygreen}{HTML}{008000}
\hypersetup{
  colorlinks = true,
  allcolors= black,
}

\usepackage{tikz,pgfplots}
\usepackage{blkarray}
\usetikzlibrary{patterns}
\usetikzlibrary{fadings}

\crefname{appendix}{Appendix}{Appendices}

\makeatletter
\pretocmd{\appendix}{%
  \crefalias{section}{appendix}%
}{}{}
\makeatother

\makeatletter
\def\@bibdataout@aps{%
\immediate\write\@bibdataout{%
@CONTROL{%
apsrev41Control%
\longbibliography@sw{%
    ,author="08",editor="1",pages="1",title="0",year="1"%
    }{%
    ,author="08",editor="1",pages="1",title="",year="1"%
    }%
  }%
}%
\if@filesw \immediate \write \@auxout {\string \citation {apsrev41Control}}\fi 
}

\pgfplotsset{compat=1.18}

\definecolor{orange}{rgb}{1,0.5,0}
\definecolor{dgreen}{rgb}{0,0.68,0.25}
\definecolor{darkgreen}{rgb}{0,0.4,0.1}
\DeclareUnicodeCharacter{300}{à}

\allowdisplaybreaks

\newcommand*{\spindeg}{{\mathfrak{D}}}

\newcommand*{\Nabla}{\ensuremath{\bm{\nabla}}}
\newcommand*\dif{\mathop{}\!\mathrm{d}}

\newcommand*\ssp[1]{{\ensuremath{#1^{\scaleto{\scriptstyle\prime}{0.5ex}}}}} 

\newcommand*{\km}{\ensuremath{\bm{k}m}}
\newcommand*{\kmp}{\ensuremath{\bm{\ssp{k}}\ssp{m}}}
\newcommand*{\kmpp}{\ensuremath{\bm{\ssp{\ssp{k}}}\ssp{\ssp{m}}}}
\newcommand*{\qs}{\ensuremath{\bm{q}s}}
\newcommand*{\qsp}{\ensuremath{\bm{\ssp{q}}\ssp{s}}}
\newcommand*{\KM}{\km}
\newcommand*{\QS}{\qs}
\newcommand*{\KMP}{\ensuremath{\bm{k'}m'}}
\newcommand*{\QSP}{\qsp}

\newcommand*{\qsd}{\qsp}
\newcommand*{\qspp}{\ensuremath{\bm{\ssp{\ssp{q}}}\ssp{\ssp{s}}}}
\newcommand*{\kmqs}{\ensuremath{^{\bm{k}m}_{\bm{q}s}}}
\newcommand*{\kmpqsp}{\ensuremath{^{\bm{\ssp{k}}\ssp{m}}_{\bm{\ssp{q}}\ssp{s}}}}
\newcommand*{\Nk}{\ensuremath{\mathcal{N}_{\bm{k}}}}
\newcommand*{\Nq}{\ensuremath{\mathcal{N}_{\bm{q}}}}

\newcommand*{\V}{\ensuremath{V_{\rm eff}}}
\newcommand*{\Vr}{\ensuremath{V_{\rm eff}(\bm{r})}}

\newcommand*{\ue}[1][]{\ensuremath{\bm{u}_{\rm e}^{#1}}}

\newcommand*{\up}[1][]{\ensuremath{\bm{u}_{\rm p}^{#1}}}

\newcommand*{\Q}{\ensuremath{\bm{Q}(\bm{r})}}
\newcommand*{\J}{\ensuremath{\bm{J}(\bm{r})}}

\newcommand*{\relaxon}[2][\kmqs]{\ensuremath{\ket{\theta^{#2}_{#1}}}}

\newcommand*{\leftrelaxon}[2][\kmqs]{\ensuremath{\bra{\theta^{#2}_{#1}}}}
\newcommand*{\normalket}[3][,\kmqs]{\ensuremath{\ket{\phi^{#2}_{#3#1}}}}
\newcommand*{\normalbra}[3][,\kmqs]{\ensuremath{\bra{\phi^{#2}_{#3#1}}}}

\newcommand*{\elmom}[1]{\ensuremath{A^{\rm e}_{#1}}}
\newcommand*{\phmom}[1]{\ensuremath{A^{\rm p}_{#1}}}

\newcommand*{\spert}[2]{\ensuremath{\ket{z^{#1}_{#2}}}}
\newcommand*{\z}{\spert{}{\kmqs}}

\newcommand*{\zd}{\spert{\delta}{\kmqs}}

\renewcommand{\sout}[1]{}

\begin{document}

\title{Coupled electron-phonon hydrodynamics and viscous thermoelectric equations}

\author{Jennifer Coulter}
\altaffiliation{These authors contributed equally to this work.}
\affiliation{Center for Computational Quantum Physics, Flatiron Institute, New York (USA)}

\author{Bogdan Rajkov}
\altaffiliation{These authors contributed equally to this work.}
\affiliation{Theory of Condensed Matter Group of the Cavendish Laboratory, University of Cambridge (UK)}

\author{Michele Simoncelli}
\email{michele.simoncelli@columbia.edu}
\affiliation{Theory of Condensed Matter Group of the Cavendish Laboratory, University of Cambridge (UK)}
\affiliation{Department of Applied Physics and Applied Mathematics, Columbia University, New York (USA)}

\begin{abstract}
Non-diffusive, fluid-like transport of charge and heat has been observed in several materials, raising the question of whether they can emerge simultaneously and how they are related to bi-component electron-phonon fluids.
Here we introduce a first-principles theory and computational framework to quantitatively describe these phenomena from atomistic to continuum scales in complex device geometries. Starting from the microscopic coupled electron-phonon Boltzmann transport equation, we formalize the emergence of composite “relaxon” electron-phonon excitations, show that they determine the viscosity tensors of the two fluids, and quantify the impact of electron-phonon drag on thermoelectric transport coefficients. 
We then demonstrate that the coupled Boltzmann equation can be coarse-grained into a set of mesoscopic Viscous Thermoelectric Equations, formally unifying Gurzhi's hydrodynamic equation for electrons [Sov. Phys. Usp., 1968] and the recently developed Viscous Heat Equations for phonons [PRX 10, 011019, 2020], while extending them to cover the intermediate regime of mixed electron and phonon fluids. We leverage this framework to elucidate how electron and phonon fluids can coexist or mix, rationalizing pioneering experiments on electron-phonon drag in graphite, and predicting smoking-gun signatures of non-diffusive behavior such as non-harmonic temperature and electric potential fields, and compressible thermoelectric backflow. 
\end{abstract}

\maketitle

\section{Introduction}
\label{sec:intro}

Hydrodynamic transport for heat or charge in solids was theoretically predicted in the sixties \cite{mezhov1966measurement,PhysRevLett.16.789,gurzhi1968} to occur when phonons or electrons undergo crystal momentum-conserving ``normal'' collisions much more frequently than momentum-dissipating ``Umklapp'' collisions, leading to a collective drift velocity analogous to the velocity field in a classical fluid.  
Pioneering experiments in the 1960s found unambiguous hallmarks of hydrodynamic thermal transport in dielectric solids. \citet{mezhov1966measurement} measured a steady-state heat-flow profile akin to the typical parabolic ``Poiseuille'' profile of a viscous fluid that slows down in proximity of the boundaries of a pipe.
Meanwhile, several other works \cite{PhysRevLett.16.789,PhysRevLett.25.26,PhysRevLett.36.480,PhysRevLett.75.2416,PhysRevB.76.075207} investigated heat propagation in the time-dependent domain, observing second sound, a coherent temperature wave in which heat transiently backflows from colder to hotter regions, violating the smoothing property of diffusive transport \cite{RevModPhys.61.41,dragasevic2023}.
Observing hydrodynamic behavior for electrons proved more challenging: the first measurement was made in the 1990s~\cite{PhysRevB.51.13389}, three decades after heat hydrodynamics, and the field did not receive much attention until 2016, when electron-only hydrodynamics was predicted \cite{levitov2016electron} and observed in graphene \cite{bandurin2016,crossno2016observation}, and simultaneously detected in PdCoO$_2$ \cite{moll2016evidence}, renewing interest in the field.

Since then, several experiments have observed viscous behavior for electrons in single-layer graphene \cite{sulpizio2019visualizing}, and for phonons in graphite \cite{Huberman_science_2019,machida2020phonon,jeong_transient_2021,huang2023,huang_graphite_2024}, typically in micrometer-sized devices and at temperatures around 70-120 K. Recently, hydrodynamic behavior has been observed for phonons up to 200 K in isotopically pure graphite \cite{ding2022observation}, and for electrons at 300 K in graphene \cite{palm_observation_2024} as well as around 150 K in bilayer graphene \cite{geursSupersonicFlowHydraulic2025}.

Because the hydrodynamic regime shares analogies with fluid dynamics, early theoretical works attempted to predict geometry- or boundary-condition-induced signatures of hydrodynamic transport using continuum Navier-Stokes-like partial differential equations. In the 1960s, \citet{gurzhi1968} coarse-grained the microscopic electron Boltzmann transport equation (eBTE) into continuum equations for hydrodynamic charge current and potential, which are still used nowadays to describe electron hydrodynamics \cite{tomadin_corbino_2014,Torre2015,bandurin2016,aharon-steinberg_direct_2022,varnavides_charge_2023}. 
Similarly, \citet{guyer1966solution}, \citet{PhysRevB.10.3546}, and others \cite{Sussmann_Thellung_1963,enz1968one,shklovskii1968,shklovskii1969} derived from the phonon Boltzmann transport equation (pBTE) continuum models for heat hydrodynamics. Some early works \cite{prohofsky_collective_1964,prohofsky_stimulated_1964,shklovskii1966} also analyzed the influence of electron-phonon interactions on second sound in semiconductors, focusing on the transient regime and neglecting viscous effects.

The accuracy of approximations used in deriving and empirically parameterizing these continuum models is difficult to control a priori.
Modern computational methods allow one to solve from first principles the eBTE \cite{ponce_epw_2016,phoebe,zhou_ab_2016,giustino2017electron,ponce2020}, the pBTE \cite{chaput2013,fugallo2013,paulatto2013anharmonic,lindsay2016first,allen_temperature_2018,puligheddu_computational_2019,guo_phonon_2021,raya-moreno_bte-barna_2022,zhang_emergence_2022,han_is_2023,phoebe,shang_heat_2020,lucente_vortices_2025}, as well as the coupled electron-phonon Boltzmann transport equation (epBTE) in the steady-state \cite{zhou_ab_2015,protik_electron-phonon_2020,protik2020,protik2022}
or spatially homogeneous and time-dependent \cite{yaoAdvancingSimulationsCoupled2025,mocattiNonequilibriumPhotocarrierPhonon2025} domain, motivating a reexamination of continuum hydrodynamic models. 
For heat transport, this has led to extension of Fourier's law \cite{hua_generalized_2019,hua_space-time_2020} or of the Guyer-Krumhansl equation \cite{sendra_derivation_2021,sendra_hydrodynamic_2022} to describe non-local heat transport in nanostructures, as well as to a set of Viscous Heat Equations (VHE) \cite{simoncelli2020} that encompass previous models for heat hydrodynamics as special limits \cite{dragasevic2023} and, most importantly, describe thermal transport from the hydrodynamic to the diffusive regime with first-principles accuracy.

These developments have spurred the exploration of a physically richer phenomenon --- the possibility of simultaneous electron and phonon hydrodynamics.
Theoretical works \cite{levchenko2020transport,huang_electron-phonon_2021} have used single-band model systems or first-principles simulations\cite{quanCoupledElectronPhononHydrodynamics2025} to investigate the single electron-phonon fluid regime, where continuous momentum exchange between electrons and phonons results in a perfectly mixed fluid, characterized by a single drift velocity.
On the experimental side, recent works focused on probing signatures of electron-phonon hydrodynamics including anomalies in: (i) the temperature dependence of the thermopower \cite{yang2021bifluid} and (ii) the lattice thermal conductivity at ultralow temperature (down to 0.1 K) \cite{jaoui2022bifluid}.
In summary, these pioneering contributions focused on thermoelectric anomalies emerging from a perfectly mixed single electron-phonon fluid, driven by intrinsic materials' properties.

However, in the literature we found old experiments which, combined with recent ones, suggest that non-perfectly-mixed electron and phonon fluids can emerge.
In particular, early experiments in graphite observed \cite{takezawa1969thermoelectric,sugihara_phonon_1970,jay-gerin_phonon_1970,de_combarieu_thermoelectric_1973,ayache_observation_1980,elzinga_thermal_1982,uher_thermopower_1982,sugihara_thermoelectric_1983}, in the same temperature range where phonon-only hydrodynamics has been recently measured \cite{Huberman_science_2019,machida2020phonon,jeong_transient_2021,huang2023,huang_graphite_2024}, anomalous peaks in the thermoelectric responses that have a qualitative behavior and magnitude strongly influenced by charge doping.
These raised several critical questions that cannot be answered with state-of-the-art semi-analytical single-fluid models or first-principles simulation techniques: 
(i) Can semimetallic solids such as graphite, which display phonon hydrodynamics \cite{Huberman_science_2019,machida2020phonon,ding2022observation,jeong_transient_2021,huang2023,huang_graphite_2024},  display the transport physics of multi-component electron-phonon fluids when doped with electrons?  
(ii) Can electron-phonon drag significantly influence phonon hydrodynamics? In particular, can heat hydrodynamics drive charge hydrodynamics, and vice-versa?
(iii) What are the experimental signatures of electron-phonon hydrodynamics, and how do these depend on intrinsic material properties and on extrinsic effects such as device geometry?

Here we introduce the first-principles theoretical and computational framework to quantitatively address these questions. We start by recasting the electron-phonon Boltzmann transport equation (epBTE) in a form amenable to being treated within the relaxon framework \cite{cepellotti2016}. 
This enables us to discuss how microscopic conservation laws govern local equilibrium and transport, showing that standard diffusive thermoelectric coefficients (electrical and thermal conductivity, Seebeck and Peltier coefficients) are not sufficient to fully characterize conduction, and electron and phonon viscosities must be taken into account. 
We formally demonstrate this from a parity argument: electrons and phonons form coupled relaxon excitations that have well-defined (odd or even) parity. Odd components determine standard diffusive heat and charge flows, while the complementary even components determine hydrodynamic viscous flows.
We further show that electron-phonon drag induces a change in the structure of relaxons, mixing otherwise independent electron and phonon relaxon excitations. We rely on these insights to coarse-grain the integro-differential epBTE into mesoscopic viscous thermoelectric equations (VTE) that unify Gurzhi’s \cite{gurzhi1968} equation for electron fluids in metals at constant temperature, and the VHE for phonon fluids in dielectrics \cite{simoncelli2020}, and, most importantly, extend them to cover the coupled thermoelectric hydrodynamic regime that emerges in semimetallic or semiconducting systems.
We rely on the VTE to discuss the conditions under which the temperature and potential fields are no longer harmonic functions, the electronic fluid becomes compressible, and the heat and charge flows violate the similarity property characteristic of diffusive transport.
Most importantly, we quantitatively explore the emergence of non-perfectly-mixed electron and phonon fluids in graphite,
predicting the doping level and device geometry that yield experimentally measurable, unambiguous hydrodynamic anomalies for temperature and voltage profiles, which are inverted compared to those emerging in diffusive transport. We conclude by discussing technological implications for the field of electronics, where viscous violations of the diffusive heat-charge flux similarity property could be exploited to focus the charge flux and the parasitic heat flux in different device regions. 

The manuscript is organized as follows. In \cref{sec:bte} we introduce the epBTE and analyze its symmetries within the relaxon framework. 
In \cref{sec:conductivities_viscosities_and_cross_transport_coefficients} we present the relation between microscopic electron-phonon relaxon excitations, their parity, and the macroscopic transport coefficients. In \cref{sec:viscous_thermoelectric_equations} we rely on microscopic (quasi) conservation laws to coarse-grain the microscopic epBTE into the mesoscopic VTE. 
In \cref{sec:electron_&_phonon_fluids_in_graphite} we show an application of our framework to graphite, discussing how heat hydrodynamics is influenced by electron-phonon couplings as a function of doping.
Finally, in \cref{sec:conclusions} we summarize results and discuss conceptual overlaps of our derivation with other coupled transport phenomena in solids, or with the framework used to describe classical fluids flowing through porous media.

\section{Electron-phonon transport: from diffusion to hydrodynamics}
\label{sec:bte}

\subsection{Electron-phonon Boltzmann transport equation}

We describe the microscopic drift and scattering of electrons and phonons using the semiclassical Boltzmann transport framework. Specifically, we consider the linearized epBTE in the steady state and in the absence of magnetic fields. Furthermore, we rewrite the integro-differential epBTE \cite{landau1981kinetics}
in a physically insightful and numerically amenable matrix form, working in a vector space that is a direct sum of the electron and phonon subspaces \cite{ziman1960}:
\begin{equation}
\begin{aligned}
&\left[\begin{array}{c}
\bm{v}^{\rm el}_{\KM} \cdot \Nabla_{\bm{r}} + \frac{e}{\hbar}[\Nabla_{\bm{r}}\V]\cdot\Nabla_{\bm{k}}\\
\bm{v}^{\rm ph}_{\QS} \cdot \Nabla_{\bm{r}}
\end{array}\right]
\left|\begin{array}{c}
\mathrm{f}_{\KM}\\
\mathrm{n}_{\QS}
\end{array}\right>  \\
&\qquad\qquad = -
\left[\begin{array}{c|c}
\mathrm{S}^{\rm ee}_{\KM,\KMP} & 
\mathrm{D}^{\rm ep}_{\KM,\QSP}\\
\hline
\mathrm{D}^{\rm pe}_{\QS,\KMP} &
\mathrm{S}^{\rm pp}_{\QS,\QSP}\\
\end{array} \right]
\Ket{\begin{array}{c}
\mathrm{f}_{\KMP}\\
\mathrm{n}_{\QSP}
\end{array}},
\end{aligned}
\label{eq:BTE_el_ph}
\end{equation}
where $\mathrm{f}_{\km}$ ($\mathrm{n}_{\qs}$) denotes the deviation of the electron (phonon) population from equilibrium, with $\bm{k}$ and $m$ ($\bm{q}$ and $s$) being the wavevector and band index of the electron (phonon) state. These deviations are defined as $\mathrm{f}_{\km} = \mathrm{F}_{\km} - \bar{\mathrm{F}}_{\km}$ ($\mathrm{n}_{\qs} = \mathrm{N}_{\qs} - \bar{\mathrm{N}}_{\qs}$), where $\mathrm{F}_{\km}$ ($N_{\qs}$) is the total out-of-equilibrium electron (phonon) population and $\bar{\mathrm{F}}_{\km} = (\exp[(\varepsilon_{\km}-\bar{\mu})/(k_B \bar{T})]+1)^{-1}$  ($\bar{\mathrm{N}}_{\qs} = (\exp[\hbar\omega_{\qs}/(k_B \bar{T})]-1)^{-1}$) is the Fermi-Dirac (Bose-Einstein) equilibrium population at temperature $\bar{T}$ for Bloch electrons with energy $\varepsilon_{\km}$ and chemical potential $\bar{\mu}$ (phonons with energy $\hbar\omega_{\qs}$).
$\bm{v}_{\km}$ and $\bm{v}_{\qs}$ are electron and phonon group velocities, respectively, and $\V = V - \mu/e$ denotes the effective electric potential that accounts both for the possible presence of an external electric field and possible variations in chemical potential \cite{ziman1960}. As shown in \cite{landau1981kinetics}, in the linear regime, the force term on electrons due to an electric field $-\Nabla V$ and the drift term due to a chemical potential gradient $\Nabla \mu$ have the same mathematical form, and so the combined effect of these two terms can be described by the gradient of the effective electrical potential $-\Nabla \V$. Here we choose the convention that the zero of the electrical potential $V$ is taken at the charge neutrality point (hence, at equilibrium we have $\V = -\bar{\mu}/e$). 
The right-hand side of Eq.~(\ref{eq:BTE_el_ph}) describes scattering processes involving electrons and/or phonons. We used repeated-index summation convention, as well as a bra-ket notation for electron and phonon distributions, to highlight that the description of scattering in the linear-response regime involves a matrix-vector multiplication:
\begin{multline}
\left[\begin{array}{c|c}
\mathrm{S}^{\rm ee}_{\KM,\KMP} & 
\mathrm{D}^{\rm ep}_{\KM,\QSP}\\
\hline
\mathrm{D}^{\rm pe}_{\QS,\KMP} &
\mathrm{S}^{\rm pp}_{\QS,\QSP}\\
\end{array} \right]
\Ket{\begin{array}{c}
\mathrm{f}_{\KMP}\\
\mathrm{n}_{\QSP}
\end{array}} \\ 
 = 
 \Ket{\begin{array}{c}
{\sum_{\kmp}} \mathrm{S}^{\rm ee}_{\KM,\KMP} \mathrm{f}_{\KMP} {+}{\sum_{\qsp}}\mathrm{D}^{\rm ep}_{\KM,\QSP}\mathrm{n}_{\QSP} \\
\sum_{\kmp}{\rm D}^{\rm pe}_{\QS,\KMP} \mathrm{f}_{\KMP} + \sum_{\qsp}{\rm {S}^{\rm pp}_{\QS,\QSP}\mathrm{n}_{\QSP}}
\end{array}},
\label{eq:matrix_vec}
\end{multline}
where the last equation specifies how we use the repeated-index summation convention.
We note, in passing, that this notation also requires us to define a bra-ket scalar product between electron-phonon distributions (whose usefulness will become apparent later) as follows:
\begin{equation}
    \left<\begin{array}{c}
\mathrm{f}_{\KM}\\
\mathrm{n}_{\QS}
\end{array}    \bigg|\begin{array}{c}
\mathrm{f'}_{\KM}\\
\mathrm{n'}_{\QS}
\end{array}\right> = 
\sum_{\km} \mathrm{f}_{\KM}\mathrm{f'}_{\KM} +
\sum_{\qs} \mathrm{n}_{\QS}\mathrm{n'}_{\QS}.
\label{eq:scalar_prod}
\end{equation}
Within our notation, the term in square brackets on the left-hand side of Eq.~(\ref{eq:BTE_el_ph}) denotes a matrix that is diagonal in both the electron and phonon subspaces.

The notation used for the electron-phonon scattering matrix above highlights the various microscopic mechanisms that we are accounting for. The electronic quadrant $\mathrm{S}^{\rm ee}_{\km,\kmp}$ accounts for: (a) depumping and repumping scattering events involving electrons only, e.g. due to electron-electron interactions \cite{vasko_quantum_2005}; (b) electron depumping due to their interactions with phonons in an equilibrium bath state \cite{landau1981kinetics,ziman1960,mariniEpiqOpensourceSoftware2024,giustino2017electron,ponce2020,caoDominantElectronphononScattering2018,dsouzaElectronphononScatteringThermoelectric2020,margineElectronphononInteractionPairing2016}.
Similarly, the phonon quadrant $\mathrm{S}^{\rm pp}_{\qs,\qsp}$ accounts for: (c) depumping and repumping scattering events involving phonons only, e.g. due to anharmonic phonon-phonon interactions \cite{ziman1960,esfarjaniFundamentalsAdvancesThermal2025,broido2007intrinsic}; (d) phonon depumping due to their interactions with electrons in an equilibrium bath state \cite{landau1981kinetics}.
The off-diagonal quadrants account for electron-phonon drag effects, which are repumping scattering events that balance events (b) and (d). In particular, $\mathrm{D}^{\rm ep}_{\km,\qsp}$ describes phonons' drag on electrons (i.e., out-of-equilibrium phonons that scatter with equilibrium electrons and drive them into an out-of-equilibrium state); analogously, $\mathrm{D}^{\rm pe}_{\qs,\kmp}$, describes electrons' drag on phonons (i.e., out-of-equilibrium electrons that scatter with phonons at equilibrium, driving the latter into an out-of-equilibrium state).
The expressions for scattering rates are derived from Fermi's golden rule, as shown in \cref{app:scattering_rates}.

One can directly verify that when the drag terms in \cref{eq:BTE_el_ph} are neglected the two rows decouple. The first row then reduces to the standard eBTE for out-of-equilibrium electrons interacting with an equilibrium phonon bath \cite{sohier2014phonon,park2014electron,zhou_ab_2016,giustino2017electron,ponce2020}. More generally, the same equation can also include electron-electron scattering and the screened Coulomb collision integral \cite{paulCoulombDragDriven2026} through the electronic collision operator; 
here we neglect these contributions because, in the temperature and doping range considered for graphite, Coulomb-mediated electronic scattering is expected to be subdominant to electron-phonon and phonon-phonon processes, owing to the appreciable screening and the established importance of electron-phonon scattering in graphite \cite{pirauxWeakLocalizationCoulomb1990,akrap_c_2007} \footnote{In several semimetals\cite{narang_effects_2017} sufficiently high temperatures electron-electron interactions are much less frequent compared to electron-phonon interactions, hence the former are negligible compared to the latter.
In particular, for the case of graphite in focus here, experiments \cite{akrap_c_2007} above $\sim$15 K show an electrical conductivity with temperature dependence clearly different from the $T^{-2}$ trend expected for dominant electron-electron scattering \cite{landau1981kinetics}.  
While the works above refer to undoped graphite, these considerations also apply to electron-doped graphite at temperatures comparable to or higher than 70 K, investigated in detail here. In fact, our thermal-conductivity predictions show that at these temperatures phonons are sufficiently populated to yield a phonon-phonon scattering timescale that is comparable to the electron-phonon scattering timescale, already violating the requirement of having negligible phonon populations, which is necessary for entering the regime of dominant electron-electron scattering \cite{peierls1955quantum,landau1981kinetics}.
Finally, we note that the calculation of the full electron-electron collision integral, accounting for both normal and Umklapp electron-electron processes, is still beyond the reach of state-of-the-art methods \cite{abramovitch_respective_2024}, but that our framework would naturally incorporate this contribution as well.}.
The second row in \cref{eq:BTE_el_ph} reduces to the well-known pBTE for out-of-equilibrium phonons interacting among themselves due to anharmonic interactions \cite{fugallo2013,li2014shengbte,carrete2017almabte,chaput2013,phono3py,tadano2014anharmonic} and with an electron bath \cite{liao_significant_2015}. 
Neglecting drag fails to capture the microscopic redistribution of energy and momentum between electrons and phonons, and consequently misses two physical effects: (i) the conservation of the total energy of the electron-phonon system \cite{landau1981kinetics}, which is necessary for rigorously defining temperature \cite{allen_temperature_2018}; (ii) the possibility that the total crystal momentum of the electron and phonon system is quasi-conserved when normal scattering dominates over Umklapp scattering, considering the most general case in which it can be partitioned in a non-symmetric way between the electron and phonon subsystems.
Including these drag terms $\mathrm{D}^{\rm ep}_{\km,\qsp}$, $\mathrm{D}^{\rm pe}_{\qs,\kmp}$ is necessary to overcome these two limitations.
Recent developments \cite{protik2020,protik2022} have thoroughly investigated from first principles the effect (i), through an iterative solution of the epBTE in the diffusive regime that goes beyond the established approximate relaxation-time or partially-decoupled solution methods \cite{fiorentini2016}.
In contrast, effect (ii) remains, to the best of our knowledge, largely unexplored with quantitative first-principles accuracy.
In the following, we will address effect (ii) relying on the relaxon framework for transport \cite{cepellotti2016,simoncelli2020}.

\subsection{From electron and phonon distributions to macroscopic local thermoelectric equilibrium}
\label{sec:local_equilibrium_and_linear_response_rev}

We start by recalling that macroscopic fluid-like behavior for heat and charge originates from microscopic electron and phonon scattering that conserves not only energy and charge, but also quasi-conserves crystal momentum (i.e., from ``normal'' collisions dominating over ``Umklapp'' collisions \cite{ziman1960}). 
Under these conditions, the local equilibrium (LE) ``drifting'' distribution assumed by electrons \cite{tomadin2013,levchenko2020transport,Phan2013} and phonons \cite{gurzhi1968,PhysRevB.10.3546,simoncelli2020} is given by the following expression:
\begin{equation}
    \Ket{\begin{array}{c}
        \mathrm{F}^{\rm LE}_{\km} (\bm{r}) \\[.1cm]
        \mathrm{N}^{\rm LE}_{\qs} (\bm{r})
    \end{array}}
=
    \Ket{\begin{array}{c}
        \left(\exp\big[\tfrac{\varepsilon_{\km}{+}e\V(\bm{r}){-}\hbar\bm{k} {\cdot} \bm{u}_{\rm e}(\bm{r}) }{k_B T(\bm{r})}\big] +1 \right)^{-1} \\
        \left(\exp\big[\frac{\hbar\omega_{\qs} - \hbar\bm{q}{\cdot}\bm{u}_{\rm p}(\bm{r}) }{k_B T(\bm{r})}\big]-1\right)^{-1}
    \end{array}},
\label{eq:local_equilibrium}
\end{equation}
which relates the microscopic energies and momenta of electrons and phonons to macroscopic temperature ($T(\bm{r})$), effective potential ($\V(\bm{r})$), and drift velocities for electrons ($\ue(\bm{r})$) and phonons ($\up(\bm{r})$). 
More specifically, the local temperature $T(\bm{r})$ is the same for electrons and phonons because it represents the Lagrange multiplier associated with total energy conservation \cite{landau1981kinetics}. We note, in passing, that a common temperature for electrons and phonons is also considered in the standard description of thermoelectric transport \cite{protik2020,landau1981kinetics}. 
Similarly, $\V(\bm{r})= -\mu_{\rm eff}(\bm{r})/e$ is associated with charge conservation in all collisions (phonons have zero chemical potential because they are not conserved in number). 
Finally, $\ue(\bm{r})$ and $\up(\bm{r})$ emerge from the quasi-conservation of momentum, and are considered different for generality. 
In fact, allowing different drift velocities for electrons and phonons captures not only the perfectly mixed regime, in which momentum redistribution between electrons and phonons is much quicker than momentum dissipation, implying $\ue(\bm{r})= \up(\bm{r})$ \cite{gurzhi1968,levchenko2020transport}, but also the non-perfectly mixed regime that emerges in semimetals or semiconductors when departures from perfect momentum conservation are significantly asymmetric between the electron and phonon subspaces.
A limiting case of this asymmetric regime is thermoelectric transport in dielectric solids, which feature phonon-phonon scattering in the phonon subspace much stronger than electron-phonon drag, implying $|\!|\up(\bm{r})|\!|\gg |\!|\ue(\bm{r})|\!| \sim 0$.
We will see later that allowing different drift velocities for electrons and phonons is critical to be able to describe the transition from phonon-only hydrodynamics ($|\!|\up(\bm{r})|\!|\gg |\!|\ue(\bm{r})|\!| \sim 0$) to coupled electron-phonon hydrodynamics ($|\!|\up(\bm{r})|\!|\sim |\!|\ue(\bm{r})|\!| > 0$) that emerges in semiconducting systems upon increasing doping.

We note that the first line of \cref{eq:local_equilibrium} represents the local-equilibrium case in which all the fields are space-dependent. In the simpler case of global equilibrium, all the fields in \cref{eq:local_equilibrium} become space independent, $T(\bm{r})=\bar{T}$, {$\V(\bm{r})=-\bar{\mu}/e$,  $\ue(\bm{r})=\bm{0}$, $\up(\bm{r})=\bm{0}$,} and no transport occurs.
For later convenience, we define the local-equilibrium deviational distribution as the difference between the local and global equilibrium distributions:
\begin{equation}
\begin{aligned}
    \Ket{\begin{array}{c}
        \mathrm{f}^{\rm LE}_{\km} (\bm{r}) \\[.1cm]
        \mathrm{n}^{\rm LE}_{\qs} (\bm{r})
    \end{array}}
&=
    \Ket{\begin{array}{c}
        \mathrm{F}^{\rm LE}_{\km} (\bm{r}) \\[.1cm]
        \mathrm{N}^{\rm LE}_{\qs} (\bm{r})
    \end{array}}
-
    \Ket{\begin{array}{c}
        \bar{\mathrm{F}}_{\km} \\[.15cm]
        \bar{\mathrm{N}}_{\qs}
    \end{array}} \\
&\approx
    \Ket{\begin{array}{@{\,}c@{\,}} \mathrm{f}^{T}_{\km} \\[.1cm] \mathrm{n}^T_{\qs} \end{array}}
    + 
    \Ket{\begin{array}{@{\,}c@{\,}} \mathrm{f}^{V}_{\km} \\[.15cm] 0 \end{array}}
    + 
    \Ket{\begin{array}{@{\,}c@{\,}} \mathrm{f}^{\ue}_{\km} \\[.15cm] 0 \end{array}}
    + 
    \Ket{\begin{array}{@{\,}c@{\,}} 0 \\[.1cm] \mathrm{n}^{\up}_{\qs} \end{array}}.
\end{aligned}
\end{equation}
The dependence of the local-equilibrium distribution on macroscopic, space-dependent fields becomes apparent by evaluating the local-equilibrium deviational distributions at linear order: $\vspace{2pt}\mathrm{f}^T_{\km} = [\partial_T \mathrm{F}^{\rm LE}_{\km} |_{\rm eq}] \delta T (\bm{r})$, $\vspace{2pt}\mathrm{f}^{V}_{\km} = [\partial_{V} \mathrm{F}^{\rm LE}_{\km} |_{\rm eq}] \delta\V(\bm{r})$, 
$\vspace{2pt}\mathrm{f}^{\ue}_{\km} = [\partial_{\ue} \mathrm{F}^{\rm LE}_{\km} |_{\rm eq}] \ue(\bm{r})$, and similarly for phonons 
$\mathrm{n}^T_{\qs}=[\partial_T \mathrm{N}^{\rm LE}_{\qs} |_{\rm eq}] \delta T(\bm{r})$ 
and $\mathrm{n}^{\up}_{\qs}=[\partial_{\up} \mathrm{N}^{\rm LE}_{\qs} |_{\rm eq}] \up(\bm{r})$. 
{The subscript ``eq'' indicates that the derivative is evaluated at global equilibrium, while $\delta T(\bm{r}) = T(\bm{r}) - \bar{T}$ and $\delta \V(\bm{r}) = \V(\bm{r}) + \bar{\mu}/e$ are deviations from global equilibrium temperature and effective potential.}%
In the next section we discuss how the (quasi) conservation laws that determine the local-equilibrium distributions~(\ref{eq:local_equilibrium}) are encoded also in the epBTE scattering matrix, and how they affect thermoelectric transport.

\subsection{Electron-phonon relaxons}
\label{sec:coupled_basis}

Within the semiclassical Boltzmann transport equation, the scattering matrix relaxes out-of-equilibrium distributions toward the local-equilibrium distributions defined in the previous section; once local equilibrium is reached, the scattering dynamics leaves it unchanged.
This physical property mathematically emerges as the scattering matrix having a null subspace spanned by the deviational local-equilibrium modes associated with conserved quantities, i.e., eigenvectors with zero eigenvalues. In this section we discuss how this scenario applies to energy and charge, as well as how it can be adapted to describe the hydrodynamic regime in which the Boltzmann dynamics nearly but not perfectly conserves the crystal momentum. 
 
The first step to make these properties explicit is to perform a similarity transformation on the epBTE, after which the coupled electron-phonon scattering matrix becomes symmetric. We do this by generalizing the standard similarity transformation used with the phonon BTE \cite{hardy1970,chaput2013} to our coupled electron-phonon case. In formulas, the action of the similarity transformation $\mathfrak{g}_{\kmqs}$ on the scattering matrix is:
\begin{subequations}
\label{eq:symm_coupled_matrix}
\begin{gather}
\mathfrak{g}_{\kmqs} = 
2 \Big(\!\!\begin{array}{c}
\sqrt{{\spindeg}/({\mathcal{V}\Nk}}) \cosh{[(\varepsilon_{\km}{-}\bar{\mu})/(2k_B\bar{T})]} \\
\sqrt{{1}/{(\mathcal{V}\Nq})} \sinh{[\hbar\omega_{\qs}/(2k_B\bar{T})]}
\end{array} \!\!\Big),
\\
\Omega_{\kmqs,\kmpqsp}=
\mathfrak{g}_{\kmqs}\left[\begin{array}{c|c}
{\mathrm{S}}^{\rm ee}_{\KM,\KMP} & 
{\mathrm{D}}^{\rm ep}_{\KM,\QSP}\\
\hline
{\mathrm{D}}^{\rm pe}_{\QS,\KMP} &
{\mathrm{S}}^{\rm pp}_{\QS,\QSP}\\
\end{array} \right]\mathfrak{g}^{-1}_{\kmpqsp},
\end{gather}
\end{subequations}
where the combined index $\kmqs$ runs over both the electron and phonon states, $\frac{\spindeg}{\mathcal{V}\Nk}$ and $\frac{1}{\mathcal{V}\Nq}$ are normalization factors that account for the electron spin degeneracy $\spindeg$, unit-cell volume $\mathcal{V}$, and number of electron (phonon) wavevectors $\Nk$ ($\Nq$) used to sample the Brillouin zone (BZ).
Correspondingly, after the similarity transformation, the population vector assumes the symmetrised form $\z = \mathfrak{g}_{\kmqs}\Ket{\genfrac{}{}{0pt}{1}{\mathrm{f}_{\km}}{\mathrm{n}_{\qs}}}$. Overall, after the similarity transformation, the epBTE~\ref{eq:BTE_el_ph} can be written as:
\begin{equation}
\bm{v}_{\kmqs} \cdot \Nabla
\z = 
-
\Omega_{\kmqs,\kmpqsp}
\spert{}{\kmpqsp},
\label{eq:BTE_coupled}
\end{equation}
where $\bm{v}_{\kmqs}=\left[\genfrac{}{}{0pt}{1}{\bm{v}_{\km}}{\bm{v}_{\qs}}\right]$ is a vector in the full electron-phonon space containing electron and phonon group velocities.
It is insightful to recast the symmetric full scattering matrix~(\ref{eq:symm_coupled_matrix}) in terms of its eigenvectors $\relaxon{\xi}$ (indexed by $\xi$) and corresponding eigenvalues $1/\tau_\xi$,
\begin{equation}
	\Omega_{\kmqs,\kmpqsp}
=\sum_{\xi}\frac{1}{\tau_{\xi}}\relaxon{\xi}\leftrelaxon[\kmpqsp]{\xi}.
\label{eq:relaxon_spectrum}
\end{equation}
{Since $\Omega$ is symmetric, the eigenvectors $\relaxon[]{\xi}$ form an orthonormal basis, $\braket{\theta^{\gamma} | \theta^{\xi}} = \delta_{\gamma,\xi}$. } It has been shown \cite{cepellotti2016} that the eigenvectors \relaxon[]{\xi} can be interpreted as relaxons, collective excitations that relax towards equilibrium with a lifetime $\tau_\xi$ determined by the inverse of the corresponding eigenvalue.
Importantly, \cref{eq:symm_coupled_matrix} shows that in the presence of electron-phonon drag quadrants, relaxons are hybridized excitations, involving both out-of-equilibrium electrons and out-of-equilibrium phonons; in contrast, when the drag quadrants are neglected, the coupled scattering matrix becomes block-diagonal, and the relaxons become electron- or phonon-only excitations discussed in past work \cite{cepellotti2016,dragasevic2023,lihmNonlinearHallEffect2024}.

The relaxon picture naturally distinguishes collective excitations underlying local equilibrium and out-of-equilibrium states of the system, as they have zero and nonzero eigenvalues, respectively.
In particular, it can be analytically shown \cite{landau1981kinetics} that the deviational local-equilibrium modes appearing in \cref{eq:local_equilibrium} are eigenvectors of the scattering matrix with (quasi)zero eigenvalue.
Specifically, the microscopic conservation of energy implies the emergence of the null (i.e., infinite lifetime) energy relaxon $\Omega \ket{\theta^0} = 0$ \cite{spohn_phonon_2006,landau1981kinetics,hardy1970}, whose entries are related to the electron and phonon energies,
\begin{equation}
\begin{split}
	\relaxon{0} &{=} 
\sqrt{\frac{1}{k_B \bar{T}^2 C_{\rm tot}}}
\left|\!\begin{array}{c}
    \sqrt{\frac{\spindeg}{\mathcal{V}\Nk}} \sqrt{\bar{\mathrm{F}}_{\km}(1{-}\bar{\mathrm{F}}_{\km})} (\varepsilon_{\km}{-}\bar\mu) \\
    \sqrt{\frac{1}{\mathcal{V}\Nq}} \sqrt{\bar{\mathrm{N}}_{\qs}(1{+}\bar{\mathrm{N}}_{\qs})} \hbar\omega_{\qs}
\end{array}\!\!\!\right\rangle,
\label{eq:energy_eigenvector_main}
\end{split}
\end{equation}
and $C_{\rm tot}=C_{\rm el}+C_{\rm ph}$ is the sum of the electron $C_{\rm el} = \frac{\spindeg}{\mathcal{V}\Nk}\sum_{\km}\frac{(\varepsilon_{\km}-\bar\mu)^2}{k_B\bar{T}^2}\bar{\mathrm{F}}_{\km}(1-\bar{\mathrm{F}}_{\km})$ and phonon $C_{\rm ph} = \frac{1}{\mathcal{V}\Nq}\sum_{\qs}\frac{(\hbar\omega_{\qs})^2}{k_B\bar{T}^2}\bar{\mathrm{N}}_{\qs}(1+\bar{\mathrm{N}}_{\qs})$ specific heats at constant volume.
Similarly, the null eigenvector accounting for the conservation of charge is $\Omega \ket{\theta^\star} = 0$, which features nonzero entries only in the electron subspace and is related to the constant electron charge $-e$ \cite{landau1981kinetics},
\begin{equation}
\begin{aligned}
\relaxon{\star}
&= 
\sqrt{\frac{e^2}{k_B\bar{T}\mathfrak{U}}}
\sqrt{\frac{\spindeg}{\mathcal{V}\Nk}}
\left|\begin{array}{c}
    \sqrt{\bar{\mathrm{F}}_{\km}(1-\bar{\mathrm{F}}_{\km})} \\
    0
\end{array}\right\rangle,
\end{aligned}
\label{eq:charge_eigenvector}
\end{equation}
where $\mathfrak{U}=\frac{e^2}{k_B \bar T}\frac{\spindeg}{\mathcal{V}\Nk}\sum_{{\bm k}m} \bar{\mathrm{F}}_{{\bm k}m}(1{-}\bar{\mathrm{F}}_{{\bm k}m})$ is a normalization constant related to the concentration of free carriers --- at zero temperature it is proportional to the electronic density of states at the Fermi level \cite{ponce2020}.

The collective excitations inherent to crystal momentum require more care, since they analytically emerge in the limiting case of vanishing momentum-dissipating Umklapp scattering, but in practice weak Umklapp scattering is present in real systems. We analyze this regime considering four different cases. We start by considering the idealized case (i) in which crystal momentum is perfectly conserved (no Umklapp scattering) both in the electron and phonon subspace, and no redistribution of momentum occurs between these two subspaces.
In this special limit, the scattering matrix will have six momentum eigenvectors with zero eigenvalue, originating from the exact conservation of crystal momentum along the three Cartesian directions for both the electron and phonon systems.
In particular, the phonon momentum eigenvectors satisfy $\Omega \ket{\theta^{\rm p}_i} = 0$, where $i$ is a Cartesian direction and the entries of $\normalket{\rm p}{i}$ are related to the phonon crystal momentum:
\begin{equation}
    \normalket{\rm p}{i}
    {=}
    \sqrt{\frac{1}{k_B \bar{T} \phmom{i}}}
    \sqrt{\frac{1}{\mathcal{V}\Nq}}
    \left|\!\begin{array}{c}
        0 \\
        \sqrt{\bar{\mathrm{N}}_{\qs} (1+\bar{\mathrm{N}}_{\qs})} \hbar q_i
    \end{array}\!\right\rangle,
    \label{eq:phonon_momentum_eigenvector_main}
\end{equation}
where $\phmom{i}=\frac{1}{\mathcal{V}\Nq}\sum_{\qs}\frac{(\hbar  q_i)^2}{k_B\bar{T}}\bar{\mathrm{N}}_{\qs}(1+\bar{\mathrm{N}}_{\qs})$ is the phonon specific momentum \cite{simoncelli2020}.
Similarly, the electron momentum eigenvectors satisfy $\Omega \ket{\theta^{\rm e}_i} = 0$, with $i$ being a Cartesian direction, and have entries related to electron crystal momentum:
\begin{equation}
    \normalket{\rm e}{i}
    {=}
    \sqrt{\frac{1}{k_B \bar{T} \elmom{i}}}
    \sqrt{\frac{\spindeg}{\mathcal{V}\Nk}}
    \left|\!\begin{array}{c}
        \sqrt{\bar{\mathrm{F}}_{\km}(1-\bar{\mathrm{F}}_{\km})} \hbar k_i \\
        0
    \end{array}\!\right\rangle.
    \label{eq:electron_momentum_eigenvector_main}
\end{equation}
Analogously to phonons, the electron specific momentum is defined as $\elmom{i}=\frac{\spindeg}{\mathcal{V}\Nk}\sum_{\km}\frac{(\hbar k_i)^2}{k_B\bar{T}}\bar{\mathrm{F}}_{\km}(1-\bar{\mathrm{F}}_{\km})$. 

A more complex scenario (ii) emerges when the electron and phonon subsystems exchange momentum, so that only the total momentum is perfectly conserved, and the individual electron and phonon components are not. This can be thought of as the particular case of electron and phonon momentum conservation and redistribution that yields $\bm{u}_{\rm e}=\bm{u}_{\rm p}$ in \cref{eq:local_equilibrium}. 
In this case we have only three null eigenvectors, one for each Cartesian direction along which the total momentum is conserved:
\begin{equation}
\begin{aligned}
\normalket{}{i} 
&= \sqrt{\frac{1}{k_B \bar{T} A_i}}
    \left|\!\begin{array}{c}
        \sqrt{\frac{\spindeg}{\mathcal{V}\Nk}} \sqrt{\bar{\mathrm{F}}_{\km}(1-\bar{\mathrm{F}}_{\km})} \hbar k_i \\
        \sqrt{\frac{1}{\mathcal{V}\Nq}} \sqrt{\bar{\mathrm{N}}_{\qs} (1+\bar{\mathrm{N}}_{\qs})} \hbar q_i
    \end{array}\!\right\rangle \\
&= \sqrt{\frac{\phmom{i}}{A_i}} \normalket{\rm p}{i} + \sqrt{\frac{\elmom{i}}{A_i}} \normalket{\rm e}{i};
\end{aligned}
\label{eq:tot_mom_cons}
\end{equation}
the total specific momentum is given by $A_i {=} \phmom{i} {+} \elmom{i}$. We note that these total momentum eigenvectors are linear combinations of the electron momentum and phonon momentum eigenvectors, and as such they lie in the six-dimensional subspace spanned by $\{\normalket[]{\rm c}{i}\}_{^{\rm c=e,p}_{i=1,2,3}} = \{\normalket[]{\rm e}{i},\normalket[]{\rm p}{i}\}_{i=1,2,3}$.
{The quasi-conservation law for the total momentum along the three Cartesian directions implies that this six-dimensional subspace can be split into two three-dimensional subspaces, one spanned by the three total momentum eigenvectors and another spanned by three eigenvectors that describe the redistribution of momentum between electrons and phonons and are determined by linear combinations of $\{\normalket{\rm c}{i}\}_{^{\rm c=e,p}_{i=1,2,3}}$.}

In actual materials crystal momentum is never perfectly conserved due to the presence of momentum-dissipating Umklapp processes at nonzero temperatures. We label as case (iii) the regime in which these Umklapp processes are sufficiently weak to enable the emergence of hydrodynamic behavior. 
The framework to describe this regime with perturbation theory is discussed in the Supplementary Material Sec. I of Ref.~\cite{dragasevic2023} --- it consists in writing the scattering matrix as $\Omega = \Omega^N + \Omega^U$, where $\Omega^N$ accounts only for normal processes and has six or three null eigenvectors in the momentum subspace (these correspond to case (i) and case (ii) above, respectively), while $\Omega^U$ is the perturbation that accounts for Umklapp processes. Umklapp processes lift the degeneracy of the null momentum subspace and induce small, but nonzero, relaxation rates.
The eigenvectors related to momentum (quasi-)conservation and their positive relaxation rates are then found by diagonalizing the projection of $\Omega$ into the momentum subspace: 
\begin{equation}
    \normalbra{\rm c_1}{i} \Omega_{\kmqs,\kmpqsp} \normalket[,\kmpqsp]{\rm c_2}{j}=\frac{{D_{ij}^{\rm c_1 c_2}}}{\sqrt{A_{i}^{\rm c_1}A_{j}^{\rm c_2}}},
	\label{eq:mom_diss}
\end{equation}
{where $\rm c_1,c_2\in\{e,p\}$ are carrier indices. The matrix $D^{\rm c_1c_2}_{ij}$}  determines the dissipation and redistribution of electron and phonon momentum, and we will consider it again when deriving the VTE in \cref{sec:viscous_thermoelectric_equations}.

The final case (iv) occurs in the diffusive regime, where Umklapp momentum dissipation is strong, and the six momentum vectors {$\{\normalket{\rm c}{i}\}_{^{\rm c=e,p}_{i=1,2,3}}$} can no longer be considered eigenvectors of the full scattering matrix. We discuss this regime in \cref{app:diff_limit}, showing that it can be described exactly using the spectrum of the full scattering matrix $\Omega$, which can be computed numerically.

In summary, we have shown how (quasi-)conservation laws in the microscopic BTE dynamics are related to special eigenvectors and macroscopic fields, and later we will demonstrate that these special eigenvectors and the spectrum of the full scattering matrix $\Omega$ allow us to comprehensively describe transport phenomena ranging from the hydrodynamic to the diffusive regime.

\section{Conductivities, viscosities, and cross-transport coefficients}
\label{sec:conductivities_viscosities_and_cross_transport_coefficients}
In this section we discuss how the relaxon formalism allows us to intuitively connect microscopic electron and phonon distributions to macroscopic heat and charge fluxes, as well as to resolve thermoelectric transport coefficients and viscosities in terms of collective microscopic excitations.

\subsection{Heat and charge currents}
\label{ssec:heat_and_charge_currents}

Refs.~\cite{simoncelli2020,dragasevic2023} discussed how, in the hydrodynamic regime of thermal transport, the heat current is made up of a diffusive and a hydrodynamic component, which share formal analogies to the same components used to describe transport in rarefied classical fluids \cite{brenner_navierstokes_2005,sambasivam_numerical_2014,schwarz_openfoam_2023,maurer_second-order_2003,dongari_pressure-driven_2009}. 
To show that these two heat-flux components emerge also in the coupled electron-phonon regime, we express the deviation from global equilibrium of the electron-phonon system in terms of relaxons, 
\begin{equation}
\begin{split}
    \z &= \Ket{z^{\rm LE}_{\kmqs}} + \zd \\
    & {}= \mathfrak{g}_{\kmqs} 
    \Ket{\begin{array}{@{\,}c@{\,}}
        \mathrm{f}^{\rm LE}_{\km} (\bm{r}) \\[.1cm]
        \mathrm{n}^{\rm LE}_{\qs} (\bm{r})
    \end{array}}  + 
    \mathfrak{g}_{\kmqs}
    \Ket{\begin{array}{@{\,}c@{\,}}
        \mathrm{f}^{\delta}_{\km} (\bm{r}) \\[.1cm]
        \mathrm{n}^{\delta}_{\qs} (\bm{r})
    \end{array}} \\
    & \begin{aligned}
        {} = {} &\sqrt{\frac{C_{\rm tot}}{k_B\bar{T}^2}} \delta T (\bm{r}) \relaxon{0}
        - \sqrt{\frac{\mathfrak{U}}{k_B\bar{T}}} \delta \V (\bm{r}) \relaxon{\star} \\
        &+ \sqrt{\frac{\elmom{i}}{k_B\bar{T}}} u_{\rm e}^i (\bm{r}) \normalket{\rm e}{i}
        {} + \sqrt{\frac{\phmom{i}}{k_B\bar{T}}} u_{\rm p}^i(\bm{r}) \normalket{\rm p}{i}
        \\
        & {} + \zd,
    \end{aligned}
\end{split}
\raisetag{40mm}
\label{eq:deviation_expansion}
\end{equation}
where we used the repeated-index summation convention over the Cartesian index $i$.
The terms proportional to $\relaxon[]{0}$, $\relaxon[]{\star}$, $\normalket[]{\rm e}{i}$, $\normalket[]{\rm p}{j}$ relate the local equilibrium distribution to spatial dependence of macroscopic fields, and are just a rewriting of the corresponding terms in \cref{eq:local_equilibrium}. They highlight the relationship between special relaxons and conservation laws that underlie local equilibrium: changing the electron-phonon distribution proportionally to the energy, charge, phonon or electron momentum relaxon induces changes in the corresponding field (in order, {$T$, $\V$, $\bm{u}_{\rm p}$, $\bm{u}_{\rm e}$}).

The special relaxons describing local equilibrium in \cref{eq:deviation_expansion} have well-defined parity --- one can directly verify from \cref{eq:energy_eigenvector_main} that the even parity of the electron and phonon energies in a time-reversal system ($\varepsilon_{\km}=\varepsilon_{-\km}$, $\omega_{\qs}=\omega_{-\qs}$) implies that the energy $\relaxon{0}$ relaxon is even, i.e., $\big|\theta^{0}_{\kmqs}\big>=\big|\theta^{0}_{^{-\km}_{-\qs}}\big>$.
Similarly, the charge relaxon, \cref{eq:charge_eigenvector}, is also even ($\big|\theta^{\star}_{\kmqs}\big>=\big|\theta^{\star}_{^{-\km}_{-\qs}}\big>$).
In contrast, the relaxons describing the momentum of phonons and electrons, \cref{eq:phonon_momentum_eigenvector_main,eq:electron_momentum_eigenvector_main}, are proportional to the crystal momentum and therefore both odd: $\normalket{\rm e}{i}{=}{-}\normalket[,^{-\km}_{-\qs}]{\rm e}{i}$ and $\normalket{\rm p}{i}{=}{-}\normalket[,^{-\km}_{-\qs}]{\rm p}{i}$.
To investigate the parity properties of the out-of-equilibrium distribution $\zd$, we recall that in a system having time-reversal symmetry the scattering matrix is even under wavevector inversion:
\begin{equation}
   \Omega_{\kmqs,\kmpqsp} = \Omega_{^{-\bm{k}m}_{-\bm{q}s}, {^{-\bm{k'}m'}_{-\bm{q'}s'}}},
   \label{eq:inversion_symmetry} 
\end{equation}
resulting in relaxons (i.e., eigenvectors) with odd or even parity: $\ket{\theta^{\xi}_{\kmqs}} {=} {\pm} \ket{\theta_{^{-\bm{k}m}_{-\bm{q}s}}^{\xi}}$ \cite{hardy1970,simoncelli2020}. Therefore, by writing the deviation from equilibrium in the relaxon basis,
\begin{equation}
	\zd= \sum_{\xi \neq \rm LE} c^\xi \relaxon{\xi}=\spert{\delta,E}{\kmqs}{+}\spert{\delta,O}{\kmqs},
	\label{eq:deviation_eq}
\end{equation}
where $c^\xi$ are the coefficients of the expansion, one can decompose the deviation from equilibrium into the sum of even ($\spert{\delta,E}{\kmqs}$) and odd ($\spert{\delta,O}{\kmqs}$) components. In \cref{eq:deviation_eq}, $\xi \neq \rm LE$ means that the sum does not include the local equilibrium subspace spanned by the special eigenvectors.
Refs.~\cite{cepellotti2016,simoncelli2020} discussed a relation between parity of the special relaxons associated to a certain local-equilibrium field, and parity of the out-of-equilibrium distribution obtained by perturbing such a field within linear-response theory.
Specifically, by perturbing local equilibrium associated to even relaxons (here temperature $T(\bm{r})$ related to $\relaxon{0}$, and potential $\V(\bm{r})$ related to $\relaxon{\star}$) one obtains an odd out-of-equilibrium distribution $\spert{\delta,O}{\kmqs}$ proportional to the gradients of the corresponding local-equilibrium fields ($\Nabla T$ and $\Nabla \V$).
Similarly, by perturbing local-equilibrium distributions associated with odd relaxons (here phonon drift velocity $\bm{u}_{\rm p}$ related to $\{\normalket{\rm p}{i}\}_{{i=1,2,3}}$, and electron drift velocity $\bm{u}_{\rm e}$ related to $\{\normalket{\rm e}{i}\}_{{i=1,2,3}}$) one obtains an even out-of-equilibrium distribution $\spert{\delta,E}{\kmqs}$ that is proportional to gradients of local-equilibrium drift-velocity fields (i.e., $\Nabla \bm{u}_{\rm e}$ and $\Nabla \bm{u}_{\rm p}$). 
We conclude this discussion on parity by recalling, for later convenience, the basic property that even functions have odd derivatives, implying that the group velocity 
 $v^i_{\kmqs}=\left[\genfrac{}{}{0pt}{1}{ \hbar^{-1}{\partial_{\bm{k}}} \varepsilon_{\bm{k}m}} {{\partial_{\bm{q}}} \omega_{\bm{q}s}}\right]$
 has odd parity, $v^i_{^{-\km}_{-\qs}}{=}{-}v^i_{^{\km}_{\qs}}$.   
 
\cref{eq:deviation_expansion} can be exploited to relate the microscopic heat-flux expression for an electron-phonon system to special relaxons, and to macroscopic fields: 
\begin{equation}
\begin{split}
    {Q}^i 
    &=\frac{1}{\mathcal{V}\Nq}\sum_{\qs}\hbar\omega_{\qs}{v}^i_{\qs}\mathrm{n}_{\qs}    +\frac{\spindeg}{\mathcal{V}\Nk}\sum_{\km}{(\varepsilon_{\km}-\bar\mu)}{v}^i_{\km}\mathrm{f}_{\km}\\
    &=\sqrt{k_B\bar{T}^2 C_{\rm tot}} \braket{\theta^0_{\kmqs} | v^i_{\kmqs} | z_{\kmqs}} \\
    &= \bar{T} \chi^{i,\rm e}_j u_{\rm e}^j  + \bar{T} \chi^{i,\rm p}_j u_{\rm p}^j - \alpha_D^{ij} \frac{\partial \V}{\partial r^j} - \bar{\kappa}_D^{ij} \frac{\partial T}{\partial r^j}.
\label{eq:energy_flux}
\end{split}
\raisetag{6mm}
\end{equation}
Here, the first line is the usual expression for the phonon \cite{hardy1963energy} and electron \cite{landau1981kinetics} contribution to the heat flux, which can be intuitively resolved in terms of $\mathrm{n}_{\qs}$ phonons (and $\mathrm{f}_{\km}$ electrons) that propagate with velocity ${v}^i_{\qs}$ (${v}^i_{\km}$) and carry the energy $\hbar\omega_{\qs}$ ($\varepsilon_{\km}-\bar\mu$). 
The second line is the total heat flux written in the relaxon formalism; it follows from \cref{eq:scalar_prod,eq:energy_eigenvector_main}, and from the definition of $\z$.
We underscore that only the odd-parity components of $\big|z_{\kmqs}\big>$ contribute to the heat flux. This is because in \cref{eq:energy_flux}, $\relaxon{0}$ has even parity, $\bm{v}_{\kmqs}$ has odd parity, and the scalar product involves summation over a reciprocal-space domain that is symmetric under wavevector inversion, see \cref{eq:scalar_prod}.
In the linear-response regime, these parity arguments determine the macroscopic expression shown in the third line: in particular, the drift-velocity terms emerge from the odd components of $\big|z_{\kmqs}\big>$ that are proportional to $\{\normalket{\rm c}{i}\}_{\mathrm{c},i}$, while the gradients of temperature and potential emerge from the form of $\spert{\delta,O}{\kmqs}$ discussed before.

\Cref{eq:energy_flux} highlights how in general the heat flux has two components: (i) hydrodynamic ``momentum'' components, which are determined by the local-equilibrium drift velocities for electrons and phonons; and (ii) standard, ``diffusion-damped'' components determined by the gradients of the temperature and potential fields.
The coefficients $\chi^{i,\rm e}_j\bar T$ and $\chi^{i,\rm p}_j\bar T$ determine the contributions that electrons and phonons give to the hydrodynamic heat-flux component, respectively; their expressions follow from \cref{eq:deviation_expansion,eq:energy_eigenvector_main}, and are reported in \cref{tab:table_VTE}.
The determination of the diffusion-damped short-circuit thermal conductivity $\bar{\kappa}_D^{ij}$ and Peltier coefficient $\alpha_D^{ij}$, will be discussed later in \cref{sub:conductivities_from_odd_relaxons_and_viscosities_from_even_relaxons}.
We also note that the sign conventions adopted in the last line of \cref{eq:energy_flux} ensure that in the special case of a system with linear-isotropic dispersion, the heat flux becomes entirely determined by the drift velocity that flows in direction opposite to the temperature gradient \cite{dragasevic2023}.

The description of the charge flux is analogous, and starts from the expression that relates the microscopic electronic distribution to the macroscopic charge flux,
\begin{equation}
\begin{split}
    {J}^i &=
    - \tfrac{\spindeg e}{\mathcal{V}\Nk} \sum_{\km}v^i_{\km}\mathrm{f}_{\km}\\
    &=- \sqrt{k_B\bar{T} \mathfrak{U}} \braket{\theta^{\star}_{\kmqs} | {v}^i_{\kmqs} | z_{\kmqs}} \\
    &  =  - \psi^i_j u_{\rm e}^j - \sigma_D^{ij} \frac{\partial \V}{\partial r^j} - [\sigma_DS_D]^{ij} \frac{\partial T}{\partial r^j}.
\end{split}
\label{eq:charge_flux}
\end{equation}
The first line of this equation is the usual expression for the charge flux \cite{landau1981kinetics,fiorentini2016,ponce2020}, which can be intuitively understood in terms of $\mathrm{f}_{\km}$ electrons that propagate with velocity ${v}^i_{\km}$ and carry charge $-e$.
The second line is the total charge flux written in the relaxon formalism, which follows directly from \cref{eq:charge_eigenvector}.
Furthermore, we understand the form of the third line by expanding $\spert{}{\kmqs}$ as in \cref{eq:deviation_expansion}, and noting that even contributions vanish by parity, while the three contributions from $\normalket{\rm p}{i}$ vanish because phonons are charge neutral. Therefore, the total charge current receives contributions from an electronic momentum current (originating from $\normalket{\rm e}{i}$ and proportional to $u_{\rm e}^i$) and a diffusion-damped current (originating from $\spert{\delta,O}{\kmqs}$ and proportional to $\Nabla T$ and $\Nabla \V$). The expression for the coefficient $\psi^i_j$ can be computed by contracting \cref{eq:deviation_expansion,eq:charge_eigenvector}, and is reported in \cref{tab:table_VTE}. The diffusion-damped electrical conductivity $\sigma_D^{ij}$ and Seebeck coefficient $S_D$ will be discussed later in \cref{sub:conductivities_from_odd_relaxons_and_viscosities_from_even_relaxons}.

\cref{eq:energy_flux,eq:charge_flux} describe the heat and charge flux in general, covering both the hydrodynamic and the diffusive regimes, as well as the intermediate cases.
In the hydrodynamic regime, $\bm{u}_{\rm p}$, $\bm{u}_{\rm e}$, $\nabla T$, and $\nabla \V$ are independent. 
In the diffusive regime of strong Umklapp dissipation, or when the second derivatives of the drift velocities are negligible, the electron and phonon drift velocities become proportional to temperature and potential gradients.
In fact, under these conditions, the terms $\bar{T} u_{\rm e}^j \chi^{i,\rm e}_j + \bar{T} u_{\rm p}^j \chi^{i,\rm p}_j$ in \cref{eq:energy_flux} give additive contributions to $\bar{\kappa}_D^{ij}$ and $\alpha_D^{ij}$, and the term $- \psi^i_j u_{\rm e}^j$ in \cref{eq:charge_flux} gives additive contributions to $\sigma_D^{ij}$ and $[\sigma_D S_D]^{ij}$; details are reported in \cref{ssec:dte_limit}.
Therefore, in the diffusive or negligible-viscosity regime, the heat and the charge fluxes \cref{eq:energy_flux,eq:charge_flux} reduce to the standard, gradient-driven diffusive form. 

\subsection{Electron and phonon momentum currents}
We start by recalling that to describe hydrodynamic transport for heat and/or charge in materials ranging from metals to insulators in full generality, it is necessary to have a formalism that allows phonons and electrons to have different drift velocities.
When electron and phonon drift velocities are non-uniform in space, net momentum fluxes emerge \cite{rice_theory_1967,simoncelli2020}; these have an analogous mathematical form for phonons and electrons:
\begin{align}
&\begin{aligned}
    \Pi^{j,\rm p}_i &=\frac{1}{\mathcal{V}\Nq} \sum_{\qs} \hbar q_i v^j_{\qs} \mathrm{n}_{\qs}
    \\
    &=- \sqrt{k_B\bar{T} \phmom{i}} \braket{\phi^{\rm p}_{i,\kmqs} | {v}^j_{\kmqs} | z_{\kmqs}} \\
    &=  \chi^{j,\rm p}_i \delta T (\bm{r})
    - \eta^{jl,\rm pp}_{ik} \frac{\partial u_{\rm p}^k}{\partial r^l}
    - \eta^{jl,\rm pe}_{ik}\frac{\partial u_{\rm e}^k}{\partial r^l}.
\end{aligned} \label{eq:phonon_momentum_flux} \\
&\begin{aligned}
    \Pi^{j,{\rm e}}_i & = \frac{\spindeg}{\mathcal{V}\Nk} \sum_{\km} \hbar k_i{v}^j_{\km} \mathrm{f}_{\km}\\
    &=\sqrt{k_B\bar{T}\elmom{i}} \braket{\phi^{\rm e}_{i,\kmqs} | v^{j}_{\kmqs} | z_{\kmqs}}\\
    & = \chi_{i}^{j,{\rm e}} \delta T (\bm{r}) {-} \psi_i^{j} \delta \V (\bm{r}) {-} \eta^{jl,\rm ee}_{ik} \frac{\partial u_{\rm e}^k}{\partial r^l} {-} \eta^{jl,\rm ep}_{ik} \frac{\partial u_{\rm p}^k}{\partial r^l}.
\end{aligned} \label{eq:electron_momentum_flux} \raisetag{15mm}
\end{align}
The first equality in \cref{eq:phonon_momentum_flux} (\cref{eq:electron_momentum_flux}) shows that the two momentum fluxes are determined by phonons $\mathrm{n}_{\qs}$ (electrons $\mathrm{f}_{\km}$) carrying momentum $\hbar q_i$ ($\hbar k_i$) along direction $i$ and propagating with velocity $v^j_{\qs}$ ($v^j_{\km}$) along direction $j$ \cite{rice_theory_1967,simoncelli2020}. The second equality in each case recasts the momentum fluxes in the relaxon basis, following from \cref{eq:phonon_momentum_eigenvector_main} [\cref{eq:electron_momentum_eigenvector_main}] and definition of $\z$.
The third equality expresses, in both cases, the momentum flux in terms of macroscopic variables.
Analogously to the heat flux, the expressions for the momentum flux are also constrained by parity arguments. Specifically, only the even part of $\z$ contributes to momentum fluxes (since $\bm{k}$, $\bm{q}$, $\bm{v}$ are all odd), so they depend directly on $T$ (with proportionality coefficient 
$\chi^{\rm e}$ and $\chi^{\rm p}$  defined in \cref{eq:energy_flux,eq:charge_flux}) and $\V$ (with proportionality coefficient $\psi$), and on gradients of drift velocities $\bm{u}_e$ and $\bm{u}_p$ through viscosities (more on this later). 

The appearance of the coefficients $\chi^{\rm e}$ and $\chi^{\rm p}$ ($\psi$) in the description of both the heat flux (charge flux) is consistent with a generalized version of the Kelvin-Onsager relation, $\alpha^{ij} = \sigma^{jk} S^{ki} \bar{T}$ \cite{onsager_reciprocal_1931}. Specifically, we see from \cref{eq:energy_flux,eq:charge_flux,eq:phonon_momentum_flux,eq:electron_momentum_flux} that $\chi^{\rm p}$ appears in $\bm{Q}$ and $\Pi^{\rm p}$, $\chi^{\rm e}$ appears in $\bm{Q}$ and $\Pi^{\rm e}$, and $\psi$ appears in $\bm{J}$ and $\Pi^{\rm e}$. In each case, the coefficient is transposed between its two appearances, and the terms appearing in $\bm{Q}$ have an added factor of $\bar{T}$ (exactly as in the Kelvin-Onsager relation).

Finally, in the third line of both \cref{eq:phonon_momentum_flux,eq:electron_momentum_flux} there are four viscosity tensor coefficients, which govern the diffusion of momentum within electron and phonon fluids ($\eta^{\rm pp}$, $\eta^{\rm ee}$) and between the two fluids ($\eta^{\rm pe}$, $\eta^{\rm ep}$). The phonon-phonon viscosity $\eta^{\rm pp}$ and electron-electron viscosity $\eta^{\rm ee}$ have the same meaning as viscosity tensors used in works on phonon- or electron-only hydrodynamics (see \cite{simoncelli2020,sendraDerivationHydrodynamicHeat2021,dragasevic2023,yadavDerivationGeneralizedHeat2025a} for phonons and \cite{gurzhi1968,levitov2016electron,bandurin2016,crossno2016observation,moll2016evidence,aharon-steinberg_direct_2022,sulpizio2019visualizing,palm_observation_2024} for electrons). However, as we discussed in \cref{sec:bte}, momentum can diffuse between electron and phonon subspaces in the presence of electron-phonon drag. We extend these past works by including viscous drag effects. The  viscosity $\eta^{\rm pe}$ captures the diffusion of momentum from the electron fluid into the phonon fluid, and conversely, $\eta^{\rm ep}$ determines the flow of momentum from the phonon fluid into the electron fluid. These two coefficients also obey an Onsager reciprocal relation, namely $\eta^{jl, \rm pe}_{ik} = \eta^{lj, \rm ep}_{ki}$; this relation will be verified from the expressions for the viscosities that we derive in the following subsection.

\subsection{Conductivities from odd relaxons, and viscosities from even relaxons} 
\label{sub:conductivities_from_odd_relaxons_and_viscosities_from_even_relaxons}

Here, we discuss how the relaxon framework allows us to determine---accounting for the full scattering matrix appearing in the epBTE---the linear-response transport coefficients that parameterize the heat, charge, and momentum currents discussed in the previous section. 
We start by presenting the derivation of the transport coefficients in the hydrodynamic regime, where currents receive independent contributions from temperature and potential gradients, and drift velocities. We will show later how the standard diffusive transport coefficients are obtained as a special case of this derivation.

\Cref{eq:deviation_expansion} shows the general relation between the fields describing local hydrodynamic equilibrium $T(\bm{r})$, $\V(\bm{r})$, ${u}^i_{\rm p}(\bm{r})$, ${u}^j_{\rm e}(\bm{r})$ ($i,j=1,2,3$) and the eight special relaxons.
To compute the transport coefficients in the hydrodynamic regime, we employ the standard procedure of inserting the expansion of $\z$ given in \cref{eq:deviation_expansion} into the epBTE, \cref{eq:BTE_coupled}, and consider an out-of-equilibrium, linear-response state $\zd$ that depends only on gradients of local fields \cite{fugallo2013,simoncelli2020,dragasevic2023}.
Then, inverting \cref{eq:BTE_coupled} using the eigenvectors and eigenvalues of the scattering matrix, \cref{eq:relaxon_spectrum}, allows us to determine the out-of-equilibrium distribution $\zd$:
\begin{equation}
\zd = -\sum_{\xi \neq {\rm LE}} \relaxon{\xi} \tau_\xi \braket{\theta^{\xi}_{\kmpqsp} | \bm{v}_{\kmpqsp} \cdot\Nabla | z^{\rm LE}_{\kmpqsp}},
\label{eq:diffusive_response}
\end{equation}
where $\sum_{\xi \neq {\rm LE}} \relaxon{\xi} \tau_\xi \bra{\theta^{\xi}_{\kmpqsp}}$ is the inverse of the scattering matrix $\Omega$ restricted to the subspace orthogonal to the local equilibrium subspace (hence $\xi\neq {\rm LE}$), and $\ket{z^{\rm LE}_{\kmqs}}$ describes the local equilibrium state from \cref{eq:deviation_expansion}, which depends directly on the macroscopic fields.

By combining \cref{eq:deviation_expansion,eq:diffusive_response}, we have completely determined the microscopic electron-phonon population $\z$ as a function of $T,\V,\bm{u}_{\rm e},\bm{u}_{\rm p}$, and their gradients. Therefore, we can now express the fluxes from previous subsections [\cref{eq:energy_flux,eq:charge_flux,eq:phonon_momentum_flux,eq:electron_momentum_flux}], which are all proportional to $\z$, also as a function of local fields. We will show now that this operation yields expressions for transport coefficients in terms of microscopic properties of the electron-phonon system.

\subsubsection{Conductivities and thermoelectric transport coefficients}

To determine the diffusion-damped short-circuit  thermal conductivity $\bar{\kappa}_D$ and Peltier coefficient $\alpha_D$, we {insert the electron-phonon population [\cref{eq:deviation_expansion,eq:diffusive_response}] into the heat flux $\bm{Q}$ (\cref{eq:energy_flux})}, and consider the linear-response regime of gradients constant in space, obtaining:
\begin{align}
    \bar{\kappa}_D^{ij}
&=
    C_{\rm tot} \sum_{\xi\neq {\rm LE}}
    v^i_{0,\xi}v^j_{0,\xi}
    \tau_\xi,
    \label{eq:thermal_K}
\\
    \alpha_D^{ij}
&=
    - \sqrt{{C_{\rm tot}\mathfrak{U}}{\bar T}} \sum_{\xi\neq{\rm LE}}
    v^i_{0,\xi}v^j_{\star,\xi}
    \tau_\xi,
    \label{eq:tildePeltier}
\end{align}
where the parameters appearing in these equations can be determined from first principles (more on this later) and allow us to resolve the transport coefficients in terms of collective electron-phonon relaxon excitations.

It is apparent from \cref{eq:thermal_K} that the diffusion-damped short-circuit thermal conductivity {$\bar{\kappa}_D$} is determined by relaxon excitations that carry heat $C_{\rm tot}=C_{\rm el}+C_{\rm ph}$ with velocity $v^i_{0,\xi}=\braket{\theta^0_{\kmqs} | v^i_{\kmqs} | \theta^{\xi}_{\kmqs}}$, and relax over the characteristic lifetime $\tau_\xi$.
The Peltier coefficient $\alpha_D$ is determined by relaxon excitations that, over the lifetime $\tau_\xi$, carry charge (related to $\mathfrak{U}$) with velocity $v^i_{\star,\xi}=\braket{\theta^{\star}_{\kmqs} | v^i_{\kmqs} | \theta^{\xi}_{\kmqs}}$, and heat (related to $C_{\rm tot}$) with velocity $v^j_{0,\xi}=\braket{\theta^{\xi}_{\kmpqsp} | v^j_{\kmpqsp} | \theta^{0}_{\kmpqsp}}$.
We note, in passing, that because relaxons and velocities of electrons and phonons are all real-valued, the relaxon velocities are symmetric under exchange of relaxon indices, e.g., $v^i_{\star,\xi}=\braket{\theta^{\star}_{\kmqs} | v^i_{\kmqs} | \theta^{\xi}_{\kmqs}}=\braket{\theta^{\xi}_{\kmqs} | v^i_{\kmqs} |\theta^{\star}_{\kmqs}  }=v^i_{\xi,\star}$, and analogous considerations hold when the charge relaxon $\relaxon{\star}$ is replaced with another special relaxon.

The structure of \cref{eq:thermal_K,eq:tildePeltier} shares mathematical analogies to the expressions for thermoelectric transport coefficients obtained from the standard relaxation time approximation (RTA) formulas \cite{fiorentini2016,landau1981kinetics}; most importantly, it rigorously overcomes the two major issues of the RTA:
(i) its inability to describe the exact relaxation dynamics \cite{fugallo_thermal_2014,fiorentini2016,protik2022}; and (ii) its failure to account for the energy and charge conservation in all semiclassical scattering processes---we stress that these conservation laws are necessary to rigorously define the local temperature \cite{allen_temperature_2018} and potential.
A very important consequence of these advancements is that \cref{eq:thermal_K,eq:tildePeltier} describe electron-phonon drag exactly, preserving the exchange of energy and momentum between the electron and phonon systems; we will analyze the relation between structure of electron-phonon relaxons and magnitude of the electron-phonon drag effect later in \cref{sec:elph_relaxons}.

The procedure used to determine $\bar{\kappa}_D^{ij}$ and $\alpha_D^{ij}$ can be repeated for charge transport. Inserting $\zd$ [\cref{eq:diffusive_response}] into the second line of \cref{eq:charge_flux}, and considering the linear-response regime of homogeneous gradients, yields the following expressions for the diffusion-damped electrical conductivity $\sigma_D^{ij}$ and the product $[\sigma_D S_D]^{ij}$ entering the Seebeck coefficient:
\begin{align}
    \sigma_D^{ij} &= \mathfrak{U}\sum_{\xi\neq \rm LE}
    v^i_{\star,\xi}v^j_{\star,\xi} \tau_\xi,
    \label{eq:electrical_sigma}\\
    [\sigma_D S_D]^{ij} &= - \sqrt{\frac{C_{\rm tot} \mathfrak{U}}{\bar T}} \sum_{\xi \neq \rm LE}
    v^i_{\star,\xi}v^j_{0,\xi}
    \tau_\xi.\label{eq:sigma_S_relaxons}
\end{align}
We note how, similarly to the case of the heat-transport coefficients, these expressions allow us to microscopically resolve $\sigma_D^{ij}$ and $[\sigma_D S_D]^{ij}$ in terms of relaxon excitations. In particular, the diffusion-damped electrical conductivity $\sigma_D^{ij}$ is determined by relaxon excitations that carry charge related to the free carrier concentration $\mathfrak{U}$ with velocity $v^i_{\star,\xi}=\braket{\theta^{\star}_{\kmqs} | {v}^i_{\kmqs} | \theta^{\xi}_{\kmqs}}$, and relax over the lifetime $\tau_\xi$.
Focusing on the diffusion-damped Seebeck coefficient $S_D^{ij}$, we underscore how the relaxon formalism makes manifest its Kelvin-Onsager relation with the Peltier coefficient \cref{eq:tildePeltier}, since by comparing these equations one readily verifies that $\alpha_D^{ij} = S_D^{ik}\sigma_D^{kj}\bar{T}$. As such, the microscopic interpretation discussed before for the Peltier coefficient directly translates to the Seebeck one.

Finally, we recall that energy ($\relaxon{0}$) and charge ($\relaxon{\star}$) relaxons have even parity, while the velocity $\bm{v}_{\kmqs}$ has odd parity. Therefore, only odd relaxons $\relaxon{\xi}$ yield nonzero scalar products and contribute to $\bar{\kappa}_D^{ij}$, $\sigma_D^{ij}$, and $\alpha_D^{ij}=S_D^{ik}\sigma_D^{kj}\bar{T}$; even-parity relaxons yield zero scalar product and thus do not contribute to these thermoelectric transport coefficients.

\subsubsection{Viscosities}
The procedure to determine the viscosities is fully analogous to the one employed to derive the thermoelectric transport coefficients, and exposes the mechanisms through which electron and phonon fluids are coupled and exchange momentum.
Inserting $\zd$ [\cref{eq:diffusive_response}] into $\Pi^{j,\rm p}_i$ and $\Pi^{j,\rm e}_i$, \cref{eq:phonon_momentum_flux,eq:electron_momentum_flux}, we find the following expressions for the viscosities:
\begin{align}
    \eta^{jl,\rm pp}_{ik} &= \sqrt{\phmom{i}\phmom{k}} \sum_{\xi \neq \rm LE} v^j_{i(\rm p),\xi}v^l_{k(\rm p),\xi} \tau_\xi, 
    \label{eq:pp_visc_main} \\
    \eta^{jl,\rm ee}_{ik} &= \sqrt{\elmom{i}\elmom{k}} \sum_{\xi \neq \rm LE} v^j_{i(\rm e),\xi}v^l_{k(\rm e),\xi} \tau_\xi,
    \label{eq:ee_visc_main}\\
    \eta^{jl,\rm pe}_{ik}&= \sqrt{\phmom{i}\elmom{k}} \sum_{\xi \neq \rm LE} v^j_{i(\rm p),\xi}v^l_{k(\rm e),\xi} \tau_\xi, \label{eq:pe_visc_main}\\
    \eta^{jl,\rm ep}_{ik} &= \sqrt{\elmom{i}\phmom{k}} \sum_{\xi \neq \rm LE} v^j_{i(\rm e),\xi}v^l_{k(\rm p),\xi} \tau_\xi.\label{eq:ep_visc_main}
\end{align}
where $v^j_{i(\rm c),\xi} = \braket{\phi^{\rm c}_{i,\kmqs}| v^j_{\kmqs} |\theta^{\xi}_{\kmqs}}$ is the velocity of the phonon (electron) momentum relaxon when $\rm c = p$ ($\rm c = e$), see \cref{eq:phonon_momentum_eigenvector_main,eq:electron_momentum_eigenvector_main}. 
Because these momentum relaxons and the velocities $v^j_{\kmqs}$ are 
odd in the wavevectors, only even relaxons $\relaxon{\xi}$ give nonzero contribution to the viscosities.
As we discussed in the previous subsection, 
the four viscosity tensors describe the self and drag viscous stresses that are present within a component or between two components of a two-fluid system. 
One can directly verify from \cref{eq:pp_visc_main,eq:ee_visc_main,eq:pe_visc_main,eq:ep_visc_main} that Onsager-type relations are satisfied by the self viscosities (e.g., $\eta^{jl, \rm ee}_{ik} = \eta^{lj, \rm ee}_{ki}$), while the drag viscosities satisfy $\eta^{jl, \rm ep}_{ik} = \eta^{lj, \rm pe}_{ki}$, as anticipated above.

Since $\bm{v}_{i(\rm p),\xi}$ depends only on the phonon entries of $\relaxon{\xi}$, and $\bm{v}_{i(\rm e),\xi}$ depends only on the electron entries of $\relaxon{\xi}$, it is necessary to have coupled electron-phonon relaxons with nonzero entries in both the electron and phonon sector to have nonzero drag viscosities. In other words, in the absence of drag contributions to the scattering matrix, the drag viscosities vanish. 

The expressions for viscosities have an intuitive microscopic interpretation. For example, when $i = k$ and $j = l$, the viscosity component $\eta^{jl, \rm cc}_{ik}$ can be resolved in terms of relaxons, each of which carries a specific momentum $A^{\rm c}_i$ with velocity $v^j_{i (\rm c), \xi}$ and lifetime $\tau_\xi$.

We note that for an isotropic system the viscosity tensors $\eta^{\rm ee}, \eta^{\rm pp}, \eta^{\rm pe}, \eta^{\rm ep}$ can be rewritten as \cite{el_rot_viscosity,groot_thermodynamics}:
\begin{equation}
  \begin{split}
        \eta^{jl,\rm c c'}_{ik} {=}& (\eta^{\rm c c'}_{\rm vol} {-} \frac{2}{d}\eta_{\rm shr}^{\rm c c'})\delta^{j}_i\delta^{l}_k {+} (\eta^{\rm c c'}_{\rm shr} {+} \eta^{\rm c c'}_{\rm rot})\delta_{ik}\delta^{jl} \\&{+} (\eta^{\rm c c'}_{\rm shr} {-} \eta^{\rm c c'}_{\rm rot})\delta^{l}_i\delta^{j}_k,
  \end{split}
\label{eq:visc_decomposition}
\end{equation}
where $d$ is the effective dimensionality of the system (e.g., for transport taking place in the two-dimensional basal planes of graphite and being constant along the off-plane direction, $d=2$); $\rm c c'\in \{ee,pp,ep,pe\}$ are subsystem indices;
$\eta_{\rm vol}^{\rm c c'}$ and $\eta_{\rm shr}^{\rm c c'}$ are volume and shear viscosities, respectively, which are analogous to those of conventional fluids and describe the stresses arising from compressing and shearing the fluid \cite{landau_fluid}; $\eta_{\rm rot}^{\rm c c'}$ is the rotational viscosity, which arises
when the angular momentum related to the fluid velocity is not conserved, e.g., in electron fluids in solids with non-rotation-invariant Brillouin zone \cite{el_rot_viscosity} and in nematic liquid crystals \cite{groot_thermodynamics}.

\renewcommand{\arraystretch}{1.8} 
\begin{table*}[t]
    \centering
    \begin{tabular}{c|c|c|c|c}
        \hline
        \hline
        Parity & Perturbation & \makecell{Proportional to} & \makecell{Contributes to} & \makecell{Determines \\[2pt] transport coefficient} \\
        \hline    
        \multirow{4}{*}{\centering \rule{0pt}{22mm} odd}    
& \multirow{4}{*}{Odd out-of-equilibrium $\spert{\delta,O}{}$} & \multirow{2}{*}{$\Nabla \V$} & $\bm{J}$ & $\sigma_D^{ij} = \mathfrak{U} \sum_{\xi \neq \rm LE} v^i_{\star,\xi} v^j_{\star,\xi} \tau_\xi$ \\ 
        \cline{4-5} 
        & & & $\bm{Q}$ & $\alpha_D^{ij} = - \sqrt{{\mathfrak{U}C_{\rm tot}\bar{T}}} \sum_{\xi \neq \rm LE} v^i_{0,\xi} v^j_{\star,\xi} \tau_\xi$ \\ 
        \cline{3-5} 
        & & \multirow{2}{*}{$\Nabla T$} & $\bm{J}$ & $\sigma_D^{ik} S_D^{kj} = \alpha^{ij}_D/\bar{T}$ \\ 
        \cline{4-5} 
        & & & $\bm{Q}$ & $\bar{\kappa}_D^{ij} = C_{\rm tot} \sum_{\xi \neq \rm LE} v^i_{0,\xi} v^j_{0,\xi} \tau_\xi$ \\        
        \cline{2-5}
        & \multirow{2}{*}{Electron momentum $\big|\phi^{\rm e}_{i} \big>$, $i$=1,2,3} & \multirow{2}{*}{$\bm{u}_{\rm e}(\bm{r})$} & $\bm{J}$ & $\psi^j_i = \sqrt{\mathfrak{U}\elmom{i}} v^j_{i({\rm e}),\star}$ \\ 
        \cline{4-5} 
        & & & $\bm{Q}$ & $\chi^{j,\rm e}_i\bar{T} = \sqrt{{C_{\rm tot}\elmom{i}}{\bar{T}}} v^j_{i({\rm e}),0}$ \\ 
        \cline{2-5}   
        & {Phonon momentum $\big|\phi^{\rm p}_{i} \big>$, $i$=1,2,3} & {$\bm{u}_{\rm p}(\bm{r})$} & $\bm{Q}$ & $\chi^{j,\rm p}_i\bar{T} = \sqrt{{C_{\rm tot}\phmom{i}}{\bar{T}}} v^j_{i({\rm p}),0}$ \\
        \hline             
        \hline         
\multirow{7}{*}{\centering \rule{0pt}{18mm} even} 
        & \multirow{4}{*}{Even out-of-equilibrium $\spert{\delta,E}{}$} 
                & \multirow{2}{*}{$\Nabla \ue$} 
                    & $\Pi^{\rm e}$ & $\eta^{jl, \rm ee}_{ik} = \sqrt{\elmom{i}\elmom{k}} \sum_\xi v^j_{i({\rm e}),\xi} v^l_{k({\rm e}),\xi} \tau_\xi$ \\
    \cline{4-5} & & & $\Pi^{\rm p}$ & $\eta^{jl, \rm pe}_{ik} = \sqrt{\phmom{i}\elmom{k}} \sum_\xi v^j_{i({\rm p}),\xi} v^l_{k({\rm e}),\xi} \tau_\xi$ \\
   \cline{3-5}& & \multirow{2}{*}{$\Nabla \up$} 
                    & $\Pi^{\rm p}$ & $\eta^{jl, \rm pp}_{ik} = \sqrt{\phmom{i}\phmom{k}} \sum_\xi v^j_{i({\rm p}),\xi} v^l_{k({\rm p}),\xi} \tau_\xi$ \\
    \cline{4-5} & & & $\Pi^{\rm e}$ & $\eta^{jl, \rm ep}_{ik} = \eta^{lj, \rm pe}_{ki}$ \\
        \cline{2-5}
        & \multirow{2}{*}{Temperature $\big|\theta^{0} \big>$} & \multirow{2}{*}{$T(\bm{r})$} & $\Pi^{\rm e}$ & $\chi^{j,\rm e}_i = \sqrt{\frac{C_{\rm tot}\elmom{i}}{\bar{T}}} v^j_{i({\rm e}),0}$ \\ 
        \cline{4-5} 
        & & & $\Pi^{\rm p}$ & $\chi^{j,\rm p}_i = \sqrt{\frac{C_{\rm tot}\phmom{i}}{\bar{T}}} v^j_{i({\rm p}),0}$ \\
        \cline{2-5}    
        & Effective potential $\big|\theta^{\star} \big>$ & $\Vr$ & $\Pi^{\rm e}$ & $\psi^j_i = \sqrt{\mathfrak{U}\elmom{i}} v^j_{i({\rm e}),\star}$ \\ 
        \hline
        \hline        
    \end{tabular}
    \caption{Summary of the coefficients appearing in the VTE, and of their physical meaning.}
    \label{tab:table_VTE}
\end{table*}

The expressions for the transport coefficients discussed in this section are summarized in \cref{tab:table_VTE}. They allow us to parameterize the fluxes of conserved quantities in terms of microscopic properties of the electron-phonon system.
In the next section, we will show that these transport coefficients enter as parameters into partial-differential equations that describe thermoelectric transport in devices.

\section{Viscous Thermoelectric Equations}
\label{sec:viscous_thermoelectric_equations}

In this section we discuss how to exploit the microscopic symmetries and (quasi) conservation laws of the electron-phonon system to coarse-grain the microscopic epBTE~(\ref{eq:BTE_coupled}) into a set of mesoscopic Viscous Thermoelectric Equations (VTE), which allow us to quantitatively predict the behavior of the local fields $T,\V,\ue,\up$ in devices having complex geometries and boundary conditions. We will show that the VTE describe non-diffusive, fluid-like transport phenomena when Umklapp dissipation is weak and in devices having geometry that promotes non-linear temperature and potential profiles; moreover, in the opposite limit of strong Umklapp dissipation,  they reduce to  the standard diffusive thermoelectric equations (DTE).

We start by considering the heat flux, as given in the second line of \cref{eq:energy_flux}; such an expression shows that the evolution of the heat-flux field in space can be determined by contracting both sides of the coupled epBTE \cref{eq:BTE_coupled} with $\bra{\theta^0_{\kmqs}}$ (\cref{eq:energy_eigenvector_main}). Specifically, we show in \cref{app:vte} that the contraction with the drift side of the epBTE yields a term proportional to the divergence of the heat flux $\Nabla \cdot \Q$. The product between $\bra{\theta^0_{\kmqs}}$ and the scattering matrix yields zero, since the energy relaxon~(\ref{eq:energy_eigenvector_main}) is a null eigenvector of the scattering matrix (this stems from the fact that energy is conserved in all microscopic scattering mechanisms). Therefore, we find the continuity equation for energy flow in the steady-state regime $\Nabla \cdot \Q = 0$. Expanding $\Q$ in terms of fields, as in the last line of \cref{eq:energy_flux}, we find energy balance equation \cref{eq:VTE_energy}.

An analogous reasoning can be applied to coarse-grain the epBTE into an equation for the charge flux. Multiplying \cref{eq:BTE_coupled} by $\bra{\theta^{\star}_{\kmqs}}$, and comparing with the second line of \cref{eq:charge_flux}, directly gives $\Nabla \cdot \J = 0$. As with the heat flux, the zero on the right-hand side is obtained because scattering perfectly conserves charge. Expanding $\J$ as in the last line of \cref{eq:charge_flux} gives the second of four equations~(\ref{eq:VTE_charge}).
In contrast to \cref{eq:VTE_energy}, the charge equation does not contain any direct macroscopic contribution from the phonon drift velocity, since phonons are charge neutral. Phonons affect the charge flow only through microscopic interactions with electrons, which influence the magnitude of the transport coefficients.

In the case of electron (phonon) momentum, we can as before multiply \cref{eq:BTE_coupled} with $\normalbra{\rm e}{i}$ ($\normalbra{\rm p}{i}$), obtaining on the left-hand side the divergence of the momentum-flux tensor, $\partial_{r^j}\Pi^{j,\rm e}_i$ ($\partial_{r^j}\Pi^{j,\rm p}_i$). 
In contrast to heat and charge, electron and phonon momentum are dissipated by Umklapp scattering, thus a non-zero value is obtained when $\normalbra{\rm e}{i}$ ($\normalbra{\rm p}{i}$) multiply the scattering side.%
These terms are in general proportional to $\braket{\phi^{\rm c}_{i}|\Omega|\theta^\xi}$, where $\rm c=e,p$ specifies the electron or phonon subsystem and $\relaxon{\xi}$ is any non-special relaxon.
{Recalling} the perturbative regime anticipated in \cref{sec:coupled_basis} --- in which Umklapp causes predominantly momentum dissipation or redistribution within the momentum subspace, and negligible momentum redistribution between the momentum and the other subspaces --- we keep only $\normalket{\rm e}{i}$ and $\normalket{\rm p}{i}$ terms and neglect $\zd$ on the right-hand side. We discuss the validity of this approximation in detail in \cref{app:vte,app:diff_limit}.
Expanding the momentum fluxes as in \cref{eq:electron_momentum_flux,eq:phonon_momentum_flux} yields the macroscopic momentum balance equations for electrons and phonons, \cref{eq:VTE_el_mom,eq:VTE_ph_mom}.

In summary, by contracting the microscopic epBTE with the special eigenvectors describing local equilibrium for energy, charge, electron and phonon momentum, we have obtained the following set of mesoscopic VTE:
\begin{widetext}
\begin{subequations}
\label{eq:VTE}
\begin{gather}
    \bar{T} \chi^{j,\rm e}_i
    \frac{\partial u^i_{\rm e}}{\partial r^j} +
    \bar{T} \chi^{j,\rm p}_i
    \frac{\partial u^i_{\rm p}}{\partial r^j}
    - \alpha_D^{ij} \frac{\partial^2 \V}{\partial r^i\partial r^j}
    - \bar{\kappa}_D^{ij} \frac{\partial^2 T} {\partial r^i \partial r^j}
    = 0,
\label{eq:VTE_energy}\\
    - \psi^j_i   \frac{\partial u^i_{\rm e}}{\partial r^j}
    - \sigma_D^{ij}  \frac{\partial^2 \V}{\partial r^i\partial r^j}
    - [\sigma_D S_D]^{ij}  \frac{\partial^2 T}{\partial r^i\partial r^j}
    = 0,
\label{eq:VTE_charge}\\
    \chi^{j,\rm e}_i \frac{\partial T}{\partial r^j}
    - \psi^j_i \frac{\partial \V}{\partial r^j}
    - \eta^{jl,\rm ee}_{ik} \frac{\partial u_{\rm e}^k}{\partial r^j \partial r^l}
    - \eta^{jl,\rm ep}_{ik} \frac{\partial u_{\rm p}^k}{\partial r^j \partial r^l}
    = - D^{\rm ee}_{ij} u^j_{\rm e} - D^{\rm ep}_{ij} u^j_{\rm p},
\label{eq:VTE_el_mom}\\
    \chi^{j,\rm p}_i \frac{\partial T}{\partial r^j}
    - \eta^{jl,\rm pp}_{ik} \frac{\partial u_{\rm p}^k}{\partial r^j \partial r^l}
    - \eta^{jl,\rm pe}_{ik}\frac{\partial u_{\rm e}^k}{\partial r^j \partial r^l}
    = - D^{\rm pp}_{ij} u^j_{\rm p} - D^{\rm pe}_{ij} u^j_{\rm e}.
\label{eq:VTE_ph_mom}
\end{gather}
\end{subequations}
\end{widetext}
The matrices $D^{\rm ee},\ D^{\rm ep},\ D^{\rm pe},\ D^{\rm pp}$ are momentum dissipation and redistribution rates defined in \cref{eq:mom_diss}, and all the other transport coefficients have been defined in the previous section in terms of microscopic relaxon properties and are reported in \cref{tab:table_VTE}. {We stress that $D^{\rm c_1c_2}$ terms describe local momentum exchange, through direct scattering between different momentum modes; meanwhile, $\eta^{\rm c_1c_2}$ terms describe momentum diffusion, mediated by an even-parity out-of-equilibrium microscopic distribution.} 
We also note that \cref{eq:VTE_energy,eq:VTE_charge} for {heat and charge} are scalar equations, while \cref{eq:VTE_el_mom,eq:VTE_ph_mom} for electron and phonon momentum are vector equations, with one equation for each Cartesian component of the electron or phonon drift velocity. 

In the following we show that the VTE comprehensively describe thermoelectric transport in a coupled electron-phonon system, covering all the regimes ranging from  hydrodynamics to diffusion. In particular, we show that the VTE incorporate as special limits the Viscous Heat Equations, describing hydrodynamic phonon transport of heat \cite{simoncelli2020,lucente_vortices_2025,dragasevic2023}, Gurzhi's equations for hydrodynamic electron transport \cite{gurzhi1968}, as well as the standard Diffusive Thermoelectric Equations (DTE) \cite{landau1981kinetics}. 

\subsection{Special Limits of the VTE}

\subsubsection{Viscous Heat Equations}
\label{ssec:vhe_limit}

The Viscous Heat Equations (VHE) describe hydrodynamic phonon-mediated heat transport in electrical insulators. In this regime, electron transport contributions can be neglected, so that $\mathrm{f}_{\km} = 0$. Comparing with \cref{eq:deviation_expansion}, we find that $\V = -\bar\mu/e$ and $\bm{u}_{\rm e} = 0$. The charge flux $\bm{J}$ and electron momentum flux $\Pi^{\rm e}$ vanish by definition, as they only receive contributions from out-of-equilibrium electrons. Their respective balance equations in the VTE, \cref{eq:VTE_el_mom,eq:VTE_charge}, are trivially satisfied. Under these conditions, the VTE simplify to:
\begin{subequations}
\label{eq:VHE}
\begin{gather}
    \bar{T} \chi^{j,\rm p}_i
    \frac{\partial u^i_{\rm p}}{\partial r^j}
    - \bar{\kappa}_D^{ij} \frac{\partial^2 T} {\partial r^i \partial r^j}
    = 0,
\label{eq:VHE_energy}\\
    \chi^{j,\rm p}_i \frac{\partial T}{\partial r^j}
    - \eta^{jl,\rm pp}_{ik} \frac{\partial u_{\rm p}^k}{\partial r^j \partial r^l}
    = - D^{\rm pp}_{ij} u^j_{\rm p},
\label{eq:VHE_ph_mom}
\end{gather}
\end{subequations}
which are exactly the VHE as given in \cite{dragasevic2023}.

\subsubsection{Gurzhi-type equation for electron-only hydrodynamics}
\label{ssec:gurzhi_limit}

Here we show how the mathematical form of the VTE is adequate to describe also electron-only hydrodynamics driven by an electric field \cite{aharon-steinberg_direct_2022,bandurin2016}.
We start by assuming that phonon transport is negligible, so that $\mathrm{n}_{\qs}=0$; thus $\bm{u}_{\rm p} = \bm{0}$ and \cref{eq:VTE_ph_mom} vanishes. Considering for simplicity an isotropic material, rank-two transport coefficients reduce to scalars ($\sigma^{ij}_D = \sigma_D\delta^{ij}$, etc.), while only the shear component of the viscosity tensor is retained, see \cref{eq:visc_decomposition}. Temperature variations may also be neglected \cite{lucas_hydrodynamics_2018}, so that $T=\bar{T}$. Then, equations for heat and charge, \cref{eq:VTE_charge,eq:VTE_energy}, reduce to $\Nabla \cdot \bm{u} = \nabla^2 \V = 0$, giving the incompressibility condition for the electron drift velocity. \Cref{eq:VTE_el_mom} is rewritten in terms of the experimentally accessible charge flux $\bm{J} = -\psi \bm{u}_{\rm e} - \sigma_D \Nabla \V$, yielding:
\begin{equation}
-\frac{\eta^{\rm ee}_{\rm shr}}{D^{\rm ee}} \nabla^2 \bm{J}+ \bm{J}=-\Big(\frac{\psi^2}{D^{\rm ee}}+\sigma_D\Big) \Nabla \V.
\end{equation}
Here, $\sqrt{\eta^{\rm ee}_{\rm shr}/ {D^{\rm ee}}}=l_G$ is the Gurzhi length \cite{gurzhi1968}; hydrodynamic corrections are only relevant in samples whose size is comparable to this lengthscale \cite{palm_observation_2024,Torre2015}. 
In bulk samples (much larger than the Gurzhi length), the momentum contribution to the charge flux is limited only by Umklapp scattering (rather than viscous effects) and gives a contribution $\psi^2/D^{\rm ee}$ to the electrical conductivity. The $\Big(\frac{\psi^2}{D^{\rm ee}}+\sigma_D\Big)$ term above is then simply the standard diffusive electrical conductivity, consisting of both momentum and non-momentum contributions. We will show this qualitatively later in this section, and quantitatively in \cref{app:diff_limit}.

\subsubsection{Perfectly mixed electron-phonon fluid in Gurzhi's limit}
\label{ssec:perfectly_mixed_gurzhi_limit}

In Ref.~\cite{gurzhi1968}, Gurzhi considered the low-temperature limit of the electrical conductivity, where momentum-conserving interactions between electrons and phonons lead to a `perfectly mixed' electron-phonon fluid, characterized by a single, common drift velocity $\bm{u}_{\rm e} {=} \bm{u}_{\rm p} {=} \bm{u}$. This regime has also been considered, and expanded upon, by \citet{levchenko2020transport}, and here we show how it arises as a special limit of the VTE.

As we discussed in \cref{sec:bte}, we can recover the single-fluid limit in \cref{eq:local_equilibrium} by setting $\bm{u}_{\rm e} = \bm{u}_{\rm p} = \bm{u}$. The conserved quantity corresponding to the common electron-phonon drift velocity is the total momentum flux $\Pi^j_i = \Pi^{j,\rm e}_i + \Pi^{j,\rm p}_i$ \cite{levchenko2020transport}. The balance equation for $\Pi^j_i$ is obtained by summing \cref{eq:VTE_el_mom,eq:VTE_ph_mom} for the electron and phonon fluids ($\bm{u}_{\rm e}$ and $\bm{u}_{\rm p}$ terms are combined):
\begin{equation}
    \chi^j_i \frac{\partial T}{\partial r^j}
    - \psi^j_i \frac{\partial \V}{\partial r^j}
    - \eta^{jl}_{ik} \frac{\partial u^k}{\partial r^j \partial r^l}
    = - D_{ij} u^j.
\label{eq:VTE_tot_mom}
\end{equation}
The coefficients appearing in this equation parameterize the mixed fluid in the same way as the coefficients we defined for the electron and phonon fluids. The expressions for these coefficients also follow from equations for the electron and phonon fluids: $\chi^j_i = \chi^{j, \rm e}_i + \chi^{j, \rm p}_i$, $\eta^{jl}_{ik} = \eta^{jl, \rm ee}_{ik} + \eta^{jl, \rm ep}_{ik} + \eta^{jl, \rm pe}_{ik} + \eta^{jl, \rm pp}_{ik}$, and $D_{ij} = D^{\rm ee}_{ij} + D^{\rm ep}_{ij} + D^{\rm pe}_{ij} + D^{\rm pp}_{ij}$, while $\psi$ remains unchanged. Importantly, one could arrive at the same equation by following the hydrodynamic linear-response derivation outlined in previous sections, replacing momentum eigenvectors $\normalket[]{\rm e}{i}, \normalket[]{\rm p}{i}$ and specific momenta $A^{\rm e}_i,A^{\rm p}_i$ with the total momentum relaxon $\normalket[]{}{i}$ and the total specific momentum $A_i$, defined in \cref{eq:tot_mom_cons}. The corresponding results are:
\begin{subequations}
\label{eq:total_momentum_additive}
\begin{align}
    \chi^j_i &= \sqrt{\tfrac{C_{\rm tot} A_i}{\bar{T}}} v^j_{i,0}, \\
    \psi^j_i &= \sqrt{\mathfrak{U}A_i} v^j_{i,\star}, \\    
    \eta^{jl}_{ik} &= \sqrt{A_i A_k} {} \sum_{\xi} v^j_{i,\xi} v^l_{k,\xi} \tau_\xi, \\
    D_{ij} &= \sqrt{A_i A_j} \braket{\phi_{i,\kmqs} | \Omega_{\kmqs,\kmpqsp} | \phi_{j,\kmpqsp}}. 
\end{align}
\end{subequations}
We defined $v^j_{i, \xi} = \braket{\phi_{i,\kmqs} | v^j_{\kmqs} | \theta^{\xi}_{\kmqs}}$, following the same notation used in the previous sections. Consistency between the two sets of definitions---either as sums of electron and phonon fluid coefficients, or as microscopic sums over relaxons---can be verified straightforwardly by noting that $\sqrt{A_i} \normalket[]{}{i} = \sqrt{\elmom{i}} \normalket[]{\rm e}{i} + \sqrt{\phmom{i}} \normalket[]{\rm p}{i}$, see \cref{eq:tot_mom_cons}.

For completeness, we note that we started from two vector equations [\cref{eq:VTE_el_mom,eq:VTE_ph_mom}] and combined them into a single vector equation. In doing so, we retained the balance equation for total momentum, while the complementary equation describes momentum redistribution between electrons and phonons. We show in \cref{app:diff_limit} that, in a perfectly mixed electron-phonon fluid, momentum redistribution is assumed to occur quickly, and its balance equation can be absorbed into additional contributions to the diffusion-damped thermoelectric coefficients.

Gurzhi's equations follow
from the perfectly mixed electron-phonon fluid limit of the VTE by neglecting temperature variations and  assuming: (i) the system to be isotropic; (ii) the charge flux to be dominated by the drift velocity, $\bm{J} =-\psi \bm{u}$ \cite{gurzhi1968}. 
The first condition allows us to simplify the tensor coefficients to scalars, while the second condition implies that \cref{eq:VTE_charge} reduces to the incompressibility condition, $\Nabla \cdot \bm{u} = 0$, leading to a simplified form of \cref{eq:VTE_tot_mom}:
\begin{equation}
    - \psi \Nabla \V
    - \eta_{\rm shr} \nabla^2 \bm{u}
    = - D \bm{u};
    \label{eq:Gurzhi}
\end{equation}
This equation has the same form as the equations discussed in previous works \cite{gurzhi1968,levchenko2020transport}.

\subsubsection{Diffusive Thermoelectric Equations}
\label{ssec:dte_limit}

Lastly, we show that in the limiting case of momentum dissipation dominating over viscous effects, the VTE reduce to the Diffusive Thermoelectric Equations (DTE).
The DTE are recovered whenever the Gurzhi lengthscales $\sqrt{\eta^{\rm ee}_{\rm shr} / D^{\rm ee}}$ and $\sqrt{\eta^{\rm pp}_{\rm shr}/D^{\rm pp}}$ are much smaller than all characteristic lengthscales of the device, a regime in which the viscous terms become negligible. This condition can be achieved either: (i) in large enough uniform samples, even when momentum is nearly conserved, or (ii) in the high temperature limit, when Umklapp scattering reduces the Gurzhi lengthscales to values much smaller than the device length scale. For the sake of simplicity, here we neglect the effect of drag on momentum dissipation, i.e., the terms $D^{\rm ep}$ and $D^{\rm pe}$; we show in \cref{app:diff_limit} that retaining these terms does not change the mathematical form assumed by the VTE in the diffusive limit.
Under these conditions, \cref{eq:VTE_el_mom} reduces to 
\begin{equation}
   - [D^{\rm ee}]^{-1}_{kl} \chi^{i,\rm e}_l \frac{\partial T}{\partial r^i}
    + [D^{\rm ee}]^{-1}_{kl} \psi^i_l \frac{\partial \V}{\partial r^i}
    =  u^k_{\rm e},
    \label{eq:VTE_el_mom_diff}
\end{equation}
and \cref{eq:VTE_ph_mom} reduces to
\begin{equation}
   -[D^{\rm pp}]^{-1}_{kl} \chi^{i,\rm p}_l \frac{\partial T}{\partial r^i}
    =  u^k_{\rm p}.
\label{eq:VTE_ph_mom_diff}
\end{equation}
Inserting \cref{eq:VTE_ph_mom_diff,eq:VTE_el_mom_diff} into \cref{eq:VTE_energy,eq:VTE_charge} yields:
\begin{subequations}
\label{eq:DTE}
\begin{align}
    - \alpha^{ij} \frac{\partial^2 \V}{\partial r^i\partial r^j}
    - \bar{\kappa}^{ij} \frac{\partial^2 T} {\partial r^i \partial r^j}
    &= 0, 
    \label{eq:DTE_energy}
    \\
    - \sigma^{ij}  \frac{\partial^2 \V}{\partial r^i\partial r^j}
    - [\sigma S]^{ij}  \frac{\partial^2 T}{\partial r^i\partial r^j}
    &= 0,
    \label{eq:DTE_charge}
\end{align}
\end{subequations}
where we find that the standard thermoelectric coefficients $\bar{\kappa},\sigma,\alpha,S$ are given by:
\begin{subequations}
\label{eq:onsager_diffusive_limit}
\begin{align}
	\bar{\kappa}^{ij} &= \bar{\kappa}_D^{ij} {+} \bar{T} \chi^{j,\rm e}_k [D^{\rm ee}]^{-1}_{kl} \chi^{i,\rm e}_l
        {+} \bar{T} \chi^{j,\rm p}_k[D^{\rm pp}]^{-1}_{kl} \chi^{i,\rm p}_l,\label{eq:diff_kappa} \\
    \alpha^{ij} &= \alpha_D^{ij}
        -\bar{T} \chi^{j,\rm e}_k [D^{\rm ee}]^{-1}_{kl} \psi^i_l,\label{eq:diff_alpha}\\ 
	\sigma^{ij} &= \sigma_D^{ij} + \psi^j_k [D^{\rm ee}]^{-1}_{kl} \psi^i_l,\label{eq:diff_sigma} \\
	S^{ij} &= \frac{[\sigma^{-1}]^{ik}\alpha^{kj}}{\bar T}.\label{eq:diff_S}
\end{align}
\end{subequations}
\Cref{eq:DTE_energy,eq:DTE_charge} are the standard diffusive thermoelectric equations. As anticipated in \cref{sub:conductivities_from_odd_relaxons_and_viscosities_from_even_relaxons}, all the transport coefficients appearing in these equations consist of two contributions: (i) the diffusion-damped transport coefficients discussed in \cref{sub:conductivities_from_odd_relaxons_and_viscosities_from_even_relaxons} and 
denoted with subscript ``D''; (ii) additive contributions that originate from drift velocities behaving diffusively (gradient-driven) in the regime where Umklapp dissipation is strong.
In \cref{app:diff_limit} we show that the equations above can be derived considering the local-equilibrium distributions appearing in \cref{eq:deviation_expansion} as being determined only by the scalar potential and temperature fields, and the momentum eigenvectors [\cref{eq:phonon_momentum_eigenvector_main,eq:electron_momentum_eigenvector_main}] as being eigenvectors with finite lifetime determined by $[D^{\rm pp}]^{-1}$ and $[D^{\rm ee}]^{-1}$, respectively. As discussed in detail in Ref.~\cite{dragasevic2023}, this last approximation is justified relying on the perturbative framework of the Schrieffer-Wolff transformation \cite{bravyiSchriefferWolffTransformation2011}.

{In summary, we have derived the VTE by coarse-graining the microscopic epBTE, and relied on linear response to determine the coefficients entering the VTE. Allowing for different electron and phonon drift velocities was crucial to enable the VTE to cover both the electron-only and phonon-only hydrodynamic limits, as well as the perfectly mixed fluid. Finally, we analytically showed that the VTE reduce to the standard DTE in the diffusive limit. So far, the discussion and derivation have been general and independent of material properties. In the next section, we will show that our framework quantitatively describes from first principles experimental signatures of electron-phonon drag in graphite, and also predicts a transition from phonon-only hydrodynamics to a two-fluid regime at high electron doping.}

\section{Case study: electron-phonon bi-component fluid in graphite}
\label{sec:electron_&_phonon_fluids_in_graphite}

\begin{figure*}[htbp!]
    \includegraphics[width=\textwidth]{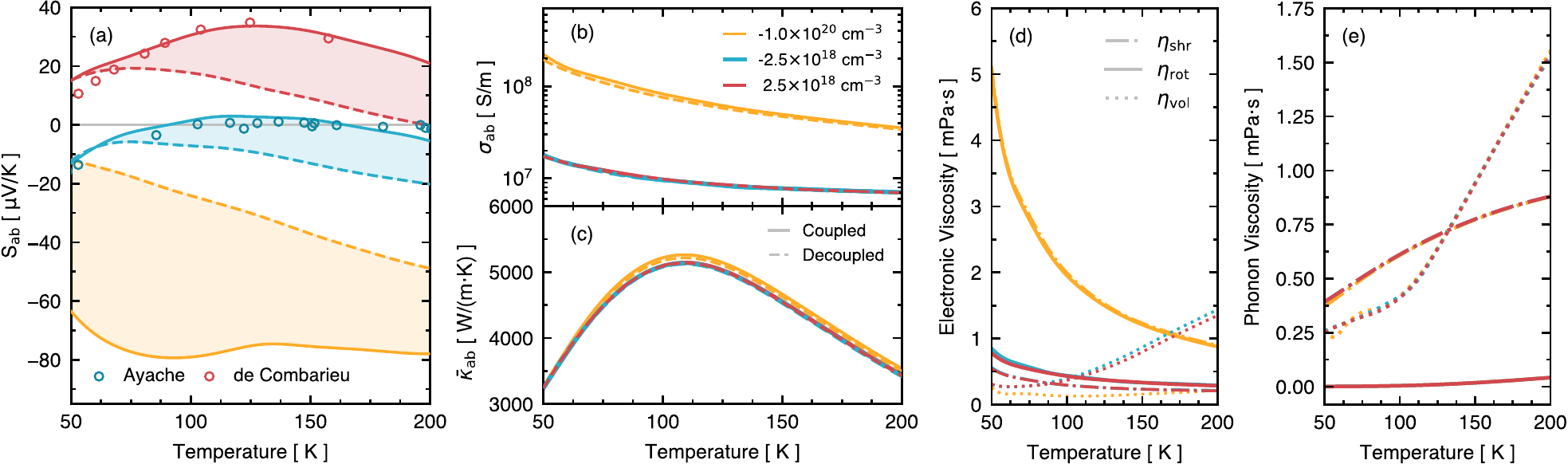}\\[-1mm]
    \caption{\textbf{Influence of electron-phonon drag and doping on bulk transport coefficients in graphite.} We show first-principles predictions for the in-plane Seebeck coefficient (\textbf{a}), electrical conductivity (\textbf{b}), and total thermal conductivity (\textbf{c}), accounting for electron-phonon drag (solid) or not (dashed)---the shaded area quantifies the contribution of electron-phonon drag. (\textbf{d})  and (\textbf{e}) show the independent components of the electron and phonon viscosity tensors ($\eta^{jl,\rm ee}_{ik}$, $\eta^{jl,\rm pp}_{ik}$, respectively) obtained from the coupled epBTE. Colors distinguish doping concentrations: $-10^{20} {\rm cm}^{-3}$ (yellow), $-2.5{\times} 10^{18} {\rm cm}^{-3}$ (blue), and $+2.5{\times} 10^{18} {\rm cm}^{-3}$ (red). 
     Blue circles are experiments from Ref.~\cite{ayache_observation_1980}, and red circles from Ref.~\cite{de_combarieu_thermoelectric_1973} (annealed 520K sample). } 
    \label{fig:bulk_transport}
\end{figure*}

Experimental work from the 1970s suggested that signatures of electron-phonon drag manifest in undoped graphite as an anomalous dip in the temperature dependence of the Seebeck coefficient, $S(T)$, between about 20 and 100 K \cite{takezawa1969thermoelectric,jay-gerin_phonon_1970,sugihara_phonon_1970}. Subsequent measurements found $S(T)$ to be highly sensitive to carrier concentration, revealing its overall increase upon hole doping \cite{de_combarieu_thermoelectric_1973,elzinga_thermal_1982,uher_thermopower_1982,sugihara_thermoelectric_1983} and decrease upon electron doping \cite{akrap_c_2007}; these raised fundamental questions on the mechanisms through which doping and electron-phonon drag influence $S(T)$.

In addition to displaying thermoelectric anomalies, graphite has recently gained attention as a paradigmatic system for the observation of transient and steady-state fluid-like heat transport phenomena. In particular, in the transient domain, experimentalists observed second sound \cite{Huberman_science_2019,ding2022observation} and lattice cooling \cite{jeong_transient_2021,zhang_transient_2021};
in the steady-state regime, recent research measured Poiseuille heat flow \cite{huang2023}, thickness-dependent Poiseuille conductivity peak \cite{machida2020phonon}, and Tesla-valve-like rectification of heat flow \cite{huang_graphite_2024}.
In spite of the combined experimental evidence of heat hydrodynamics and the strong influence of electron-phonon drag on thermoelectric properties, the possible interplay between phonon hydrodynamics and electronic transport in graphite has never been investigated with first-principles accuracy. 
These past findings, and the possibility to tune the electronic-transport properties of graphite via doping \cite{elzinga_thermal_1982,uher_thermopower_1982,sugihara_thermoelectric_1983,akrap_c_2007}, make graphite an ideal system to investigate the possibility of simultaneous heat and charge hydrodynamics. 

\subsection{Transport coefficients from first principles}
\label{ssec:graphite_transport_coefficients}

We employed Density-Functional Theory (DFT) to compute all the microscopic electronic and phononic parameters appearing in the epBTE~(\ref{eq:BTE_coupled}), including the electron and phonon bands, phonon-phonon and electron-phonon couplings that determine the full scattering matrix with drag terms (see \cref{app:computational_details} for details \footnote{In this work, we neglect electron-electron scattering. This choice is justified by starting the analysis from the heat hydrodynamic regime, where phonon-phonon interactions dominate, and then progressively enhancing the role of electrons via doping. This approach ensures that electron-phonon scattering remains stronger than electron-electron scattering.}). The formalism defined above for the epBTE was implemented in the open-source Phoebe code \cite{phoebe}, including the numerical infrastructure to calculate and diagonalize the coupled scattering matrix as well as the ability to evaluate the transport coefficients in Table~\ref{tab:table_VTE}. 

We show in \cref{fig:bulk_transport} the in-plane transport coefficients of graphite as a function of temperature, also evaluating their dependence on charge carrier concentration by comparing three different doping levels: $-10^{20} \rm cm^{-3}$ (strongly electron-doped), $-2.5\times 10^{18} \rm cm^{-3}$ (weakly electron-doped), and $+2.5\times 10^{18} \rm cm^{-3}$ (weakly hole-doped). 
To quantify the impact of electron-phonon drag on the transport coefficients, we compare predictions {obtained using the fully coupled scattering matrix $\Omega$, from \cref{eq:symm_coupled_matrix},} against those obtained neglecting the drag quadrants in the scattering matrix, corresponding to solving decoupled eBTE and pBTE. 

Panel (\textbf{a}) shows a strong impact of electron-phonon drag on the Seebeck coefficient, as defined in \cref{eq:diff_S} for the diffusive regime, for all three charge carrier concentrations analyzed.
First, for the strongly electron-doped case (orange), we see that the influence of electron-phonon drag is maximized at lower temperatures, and decreases steadily with increasing temperature. The decreasing trend  of $S(T)$ upon increasing temperature qualitatively matches the behavior observed in samples with electron doping \cite{akrap_c_2007}.
Moreover, we see that upon decreasing the concentration of electrons to weak electron doping (blue), and further increasing the concentration of holes to weak hole doping (red), the curve $S(T)$ reverses its concavity and progressively shifts towards more positive values. 
Several experiments spanning different doping levels (e.g., via intercalation) are available in the literature \cite{ayache_observation_1980,de_combarieu_thermoelectric_1973, elzinga_thermal_1982,uher_thermopower_1982,sugihara_thermoelectric_1983,akrap_c_2007}. While these works do not directly measure the carrier concentration, they discuss how varying doping (e.g., the intercalation stage) tunes the carrier imbalance from electron-dominated to hole-dominated, which correlates with an unambiguous upward shift of $S(T)$ matching our simulations.

Predictions for the diffusive electrical conductivity, defined in \cref{eq:diff_sigma} and shown in panel (\textbf{b}), also display a significant dependence on doping. Samples with strong electron doping ($-10^{20} \rm cm^{-3}$) show a conductivity that is about one order of magnitude larger than weakly electron-doped and weakly hole-doped samples over the entire temperature range analyzed. 
For all doping concentrations, predictions for the electrical conductivity are negligibly affected by whether electron-phonon drag is included or neglected.

In panel (\textbf{c}), we report the total diffusive thermal conductivity, \cref{eq:diff_kappa}, including electronic and phononic (lattice) contributions. Strong electronic doping induces a small but noticeable ($\lesssim5\%$) increase compared to weakly electron- or hole-doped samples. Similarly to the electronic conductivity, electron-phonon drag has a negligible influence on the thermal conductivity.

The electron and phonon viscosities are shown in (\textbf{d}) and (\textbf{e}), corresponding to \cref{eq:ee_visc_main,eq:pp_visc_main}.
Focusing on the electronic shear viscosities (dash-dotted lines), we see that, for all doping levels, they show a decreasing-with-temperature trend that mirrors that of typical liquids, and their magnitude is sharply reduced upon decreasing doping. In contrast, the phonon shear viscosities are negligibly impacted by doping and increase with temperature, mirroring classical gases.
In strongly electron-doped samples, the electronic volume viscosity (dotted line) is much smaller than the other viscosity-tensor components over the entire temperature range analyzed.
Reducing the doping level to weakly electron-doped or weakly hole-doped yields an electronic volume viscosity with a temperature-activated trend that mirrors gases, and a magnitude that becomes larger than the shear viscosity above $\sim 100$ K. For phonons, the volume viscosity displays, at temperatures higher than 100 K, an increasing-with-temperature trend similar to gases that is independent of doping; below 100 K, doping induces visible but practically unimportant changes in the phonon volume viscosities.
Finally, the rotational viscosity (solid lines) displays very different behavior for electrons and phonons.
For electrons, we see that the rotational viscosity is very similar in magnitude and temperature dependence to the shear viscosity; its nonzero values indicate that the Fermi surface is not rotationally invariant, as discussed in Ref.~\cite{el_rot_viscosity}.
More interestingly, we note that phonons display, for all dopings, a rotational viscosity that is practically negligible (zero) for $T\lesssim 100$ K, and slowly increases at higher temperatures. This can be intuitively understood by noting that, at low temperature, only low-frequency phonons far from the Brillouin-zone (BZ) boundaries are thermally populated; because these are isotropic in the in-plane directions, in-plane phonons at low temperature are excited in a rotationally invariant way, yielding negligible rotational viscosity.
Upon increasing temperature, phonons near the BZ edge are activated and feel the hexagonal (not rotationally invariant) shape of graphite's BZ in the in-plane directions, yielding an $\eta_{\rm rot}$ that starts to deviate from zero.
We conclude by noting that, for viscosity, we report only predictions from coupled calculations with drag, since viscosity is relevant only in the hydrodynamic, low-temperature regime, which requires us to account for the drag contributions to be rigorously described \cite{landau1981kinetics}.

\subsection{Electron-phonon drag \& relaxon structure}
\label{sec:elph_relaxons}

\begin{figure*}[htbp!]
    \centering
    \includegraphics[width=\textwidth]{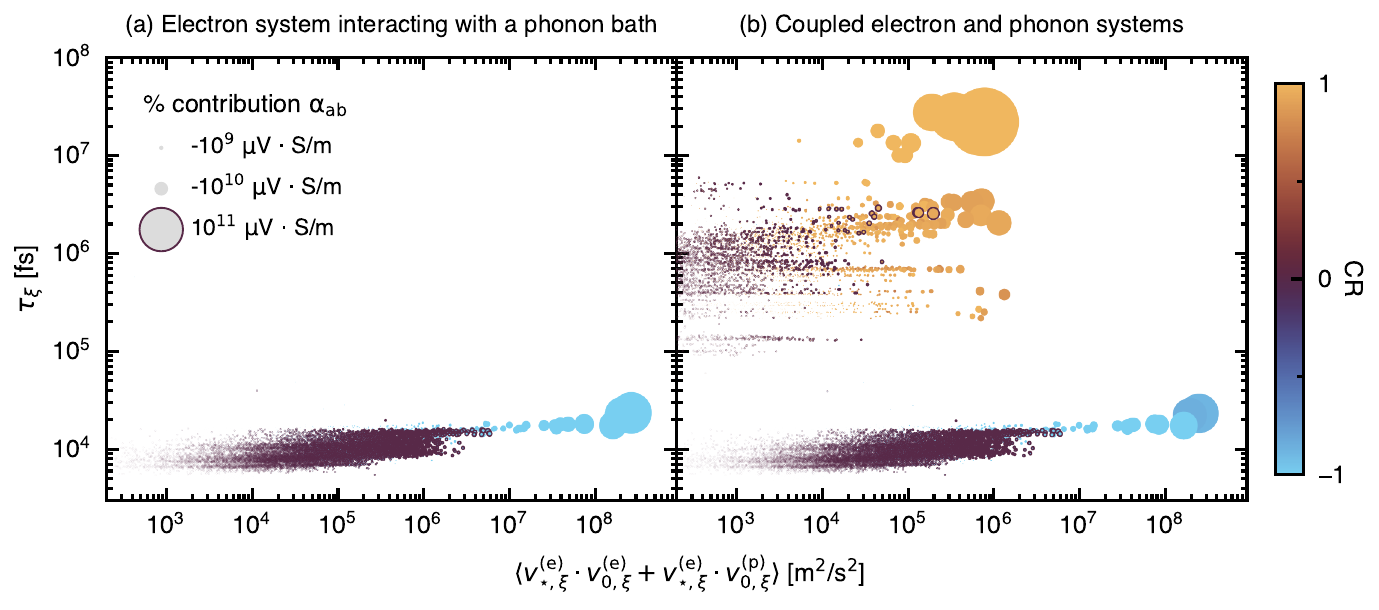}
    \caption{\textbf{Signatures of electron-phonon drag on the structure of relaxon excitations.      
    } 
    Scatter points are contributions to the Peltier coefficient from individual relaxons, shown in (\textbf{a}) without drag and (\textbf{b}) with drag contributions to the scattering matrix, for $n=1\times10^{20} cm^{-3}$ carriers at a temperature of 70 K. The $x$ axis shows the product of velocities for relaxon $\xi$ (\cref{eq:v2eff}), while the $y$ axis shows its lifetime. The area of the scatter points shows the contribution of each relaxon to the Peltier coefficient, with positive (negative) contributions shown as closed (open) circles. The coupling ratio shown by the colorbar quantifies whether a relaxon is electron-dominated (blue), phonon-dominated (orange), or mixed (purple).
    }
    \label{fig:eigenvectors_drag_vs_nodrag}
\end{figure*}
As anticipated in \cref{sec:coupled_basis}, relaxons are collective excitations defined as eigenvectors of the scattering matrix appearing in the Boltzmann transport equation. They were introduced to describe collective phonon excitations in the context of the pBTE \cite{cepellotti2016}, and later adapted to the eBTE to describe electron transport in semiconductors \cite{phoebe} and non-linear Hall effect \cite{lihmNonlinearHallEffect2024}.
One of the main results of this work is to generalize the concept of relaxon excitation to a coupled electron-phonon system, and in this section we analyze analogies and differences between these newly introduced electron-phonon relaxons and the phonon-only \cite{cepellotti2016} or electron-only \cite{phoebe,lihmNonlinearHallEffect2024} relaxons discussed in previous works.

We start by recalling that if electron-phonon drag is neglected, the epBTE scattering matrix in \cref{eq:BTE_el_ph} becomes block-diagonal; consequently, relaxons trivially have non-zero entries either in the electron subspace or in the phonon subspace, and the epBTE decouples into the standard eBTE and pBTE.
In contrast, when electron-phonon drag quadrants are considered in \cref{eq:BTE_el_ph}, relaxons can in principle assume a mixed structure (i.e., featuring non-zero entries simultaneously in the electron and phonon subspaces), and also their corresponding eigenvalue (inverse lifetime) can change.
In this section we investigate quantitatively how this mixing in the relaxon structure and lifetime changes are related to the influence of electron-phonon drag on thermoelectric transport, also analyzing the dependence on doping and temperature.

\begin{figure*}[htbp!]
    \centering
    \includegraphics[width=\textwidth]{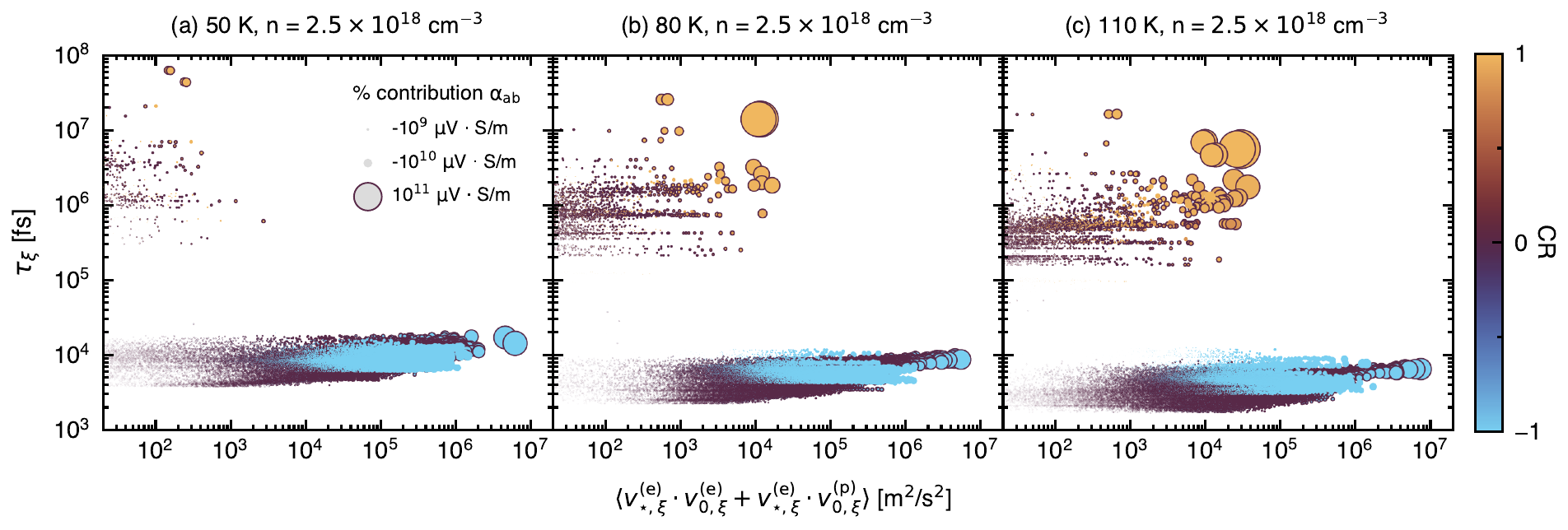}
    \caption{\textbf{Influence of temperature on electron-phonon drag in weakly hole-doped graphite.} Contributions to the Peltier coefficient from individual relaxons as a function of increasing temperature, from (\textbf{a}) 50 K, (\textbf{b}) 80 K, (\textbf{c}) 110 K, for a carrier concentration of $n=2.5\times10^{18} cm^{-3}$. As in \cref{fig:eigenvectors_drag_vs_nodrag}, we plot the relaxation time of each relaxon against a product of relaxon velocities, with scatter point area indicating the contribution of relaxon $\xi$ to the Peltier coefficient. Positive and negative contributions are represented as closed and open circles, respectively. The colorbar shows the coupling ratio (CR), which indicates if a relaxon is primarily electron-like (blue) or phonon-like (orange). As we increase temperature across \textbf{a-c}, the velocity of phonon-like relaxons approaches that of the electron-like ones; this corresponds to a stronger drag effect and a larger phonon contribution to the thermoelectric coefficients.} 
    \label{fig:eigenvectors_contrib_vs_temperature}
\end{figure*}

In the relaxon formalism, transport coefficients (\cref{tab:table_VTE}) are always determined by the product between a lifetime (determined by the inverse eigenvalue) and two velocities (determined by eigenvectors). Therefore, a natural way to resolve how electron-phonon drag influences relaxon structure, lifetime, and transport is to show a scatter plot with relaxon velocities along $x$, relaxon lifetime along $y$, and scatter points with area proportional to how much a single relaxon contributes to transport (product between velocities and lifetimes). 
Because \cref{fig:bulk_transport} shows that electron-phonon drag has the maximum effect on thermoelectric coefficients, we focus on the velocities appearing in the Peltier coefficient, \cref{eq:diff_alpha}. Moreover, since transport is isotropic in the in-plane direction of graphite, we consider the following direction-averaged relaxon velocity products, which have electron-only and electron-phonon-drag components:
\begin{equation}
    \begin{split}
    \big<\bm{v}^{{\rm (e)}}_{\star,\xi}{\cdot} \bm{v}^{{\rm (e)}}_{0,\xi} {+} \bm{v}^{{\rm (e)}}_{\star,\xi}{\cdot} \bm{v}^{{\rm (p)}}_{0,\xi}\big>_{\rm dir} &= 
    \frac{1}{d}\sum_{i=1}^d v^{i}_{0,\xi}v^{i}_{\star,\xi}\\
    &={\frac{1}{d}\sum_{i=1}^d\Big(v^{i ,({\rm e})}_{\star,\xi} v^{i, {\rm (e)}}_{0,\xi} {+}v^{i,({\rm e})}_{\star,\xi} v^{i,({\rm p})}_{0,\xi}\Big)},
    \label{eq:v2eff}
    \end{split}
    \raisetag{20mm}
\end{equation}
where $d$ is the number of Cartesian directions along which the spatial average is performed (here $d=2$ accounts for the two in-plane directions of graphite). In the second line we rewrote the coupled electron-phonon relaxon velocities appearing in \cref{eq:tildePeltier} in an analytically equivalent way that allows us to resolve how much they depend on electrons and phonons: 
$v^{i ,({\rm e})}_{\star,\xi}=
\braket{\theta^{\star}_{i,\kmqs}| v^i_{\km} |\theta^{\xi}_{\kmqs}}$ ($v^{i ,({\rm e})}_{0,\xi}=
\braket{\theta^{0}_{i,\kmqs}| v^i_{\km} |\theta^{\xi}_{\kmqs}}$) is obtained bracketing the electron group velocities between the special charge (energy) relaxon and a non-special relaxon $\xi$; finally,  $v^{i ,({\rm p})}_{0,\xi}=
\braket{\theta^{0}_{i,\kmqs}| v^i_{\qs} |\theta^{\xi}_{\kmqs}}$ is obtained bracketing the phonon group velocities between the special energy relaxon and a generic relaxon $\xi$ \footnote{Note the difference between $v^j_{i(\rm c),\xi}$ defined in \cref{sec:conductivities_viscosities_and_cross_transport_coefficients} and $v^{i ,({\rm c})}_{\star/0,\xi}$, defined here. In the first case, the carrier index $\rm c$ labels the corresponding momentum relaxon, while in the latter it specifies whether we're considering the electron or phonon group velocities.}.
We highlight how in the absence of electron-phonon drag, the unmixed structure of relaxons implies that only the term $v^{i ,({\rm e})}_{\star,\xi} v^{i, {\rm (e)}}_{0,\xi}$ is nonzero in  \cref{eq:v2eff}, while in the presence of drag, the $v^{i,({\rm e})}_{\star,\xi} v^{i,({\rm p})}_{0,\xi}$ term can also be nonzero. 
We also note that these velocities can assume either positive or negative values, associated with a corresponding positive or negative contribution to the Peltier and Seebeck coefficients.

Therefore, this decomposition suggests further resolving how strongly electrons and phonons are coupled and influenced by their mutual drag effect in a relaxon $\xi$ by computing the relative magnitude of these two terms, which we call ``Coupling Ratio'' (CR):
\begin{equation}
    {\rm CR}_\xi = \frac{\sum_{i=1}^d(|v^{i,({\rm e})}_{\star,\xi} v^{i, ({\rm p})}_{0,\xi}|-|v^{i, ({\rm e})}_{\star,\xi} v^{i, ({\rm e})}_{0,\xi}| )}{ 
    \sum_{i=1}^d(|v^{i, ({\rm e})}_{\star,\xi} v^{i, ({\rm e})}_{0,\xi}| + |v^{i,({\rm e})}_{\star,\xi} v^{i, ({\rm p})}_{0,\xi}|)}.
\end{equation}
This descriptor quantifies the reasoning above, providing information on how much a coupled relaxon $\xi$ is determined by electrons and phonons: ${\rm CR}_\xi = -1$ (${\rm CR}_\xi=+1$) implies that the relaxon is dominated by electrons (phonons), while ${\rm CR}_\xi{=}0$ implies a comparable contribution from electrons and phonons.

We show in \cref{fig:eigenvectors_drag_vs_nodrag} these quantities for strongly n-doped graphite ($n=1\times 10^{20}$ cm$^{-3}$) at 70 K. 
Panel (\textbf{a}) shows these quantities when the drag effect is not accounted for, implying that contributions to the in-plane Peltier coefficient arise exclusively from electron-like relaxons (${\rm CR}_{\xi} =-1 \;\forall\; \xi$, in blue). Once drag is taken into account (panel (\textbf{b})), we find contributions to the thermoelectric transport coefficients from a number of phonon-like relaxons, which corresponds to the phonon drag contribution to the strongly n-doped data for $S$ shown in \cref{fig:bulk_transport}(\textbf{a}). 
By comparing the effective velocities in \cref{eq:v2eff} with the square of electron-only velocities and phonon-only velocities, we see that electrons are mostly influenced by electron-phonon drag when the velocities of electron-dominated relaxons are of the same order of magnitude as the velocities of phonon-dominated relaxons. The purple color of small points is due to them having a positive contribution, denoted by a purple border, rather than them having a CR close to 0.

In \cref{fig:eigenvectors_contrib_vs_temperature}, we investigate how relaxon mixing, lifetimes, and contributions to transport depend on doping and temperature, performing the analysis above in weakly hole-doped graphite ($n=2.5 \times 10^{18}$ cm$^{-3}$) at three different temperatures, (\textbf{a}) 50 K, (\textbf{b}) 80 K, and (\textbf{c}) 110 K. 
We highlight how the electron-phonon drag effect transitions from being almost negligible at 50 K to being significant at 110 K, as expected from the macroscopic thermoelectric coefficient in \cref{fig:bulk_transport}{\bfseries{a}}.
Compared to the strongly electron-doped case, weakly hole-doped graphite always shows a smaller electron-phonon drag effect and a smaller electron-phonon mixing in relaxon excitations. 
This corresponds to having a smaller electronic density of states in weakly hole-doped graphite compared to strongly electron-doped graphite around the Fermi level, consistent with weaker electron-phonon interactions and drag. 
The electron-phonon drag effect increases upon increasing temperature, as phase space for electron-phonon scattering increases (see \cref{app:scattering_rates} for details: more phonons with energies comparable to electronic-transition energies become thermally active). 

In summary, we have shown that the electron-phonon drag effect induces a mixing in the electron-phonon relaxons, which acquire entries that are simultaneously non-zero in the electron and phonon subspace. This effect is maximized when the electronic and phononic component of the velocity of electron-phonon relaxons are close to each other, and to maximize the contribution to transport of these relaxons their lifetimes have to be as large as possible.  
Importantly, we have shown that electron-phonon drag can be significant in both strongly electron-doped and weakly hole-doped graphite. In the next subsection, we will show that this is a necessary but not sufficient condition to have coupled electron-phonon hydrodynamics, which is present in strongly electron-doped graphite, but absent in weakly hole-doped graphite.

\subsection{Electron-phonon hydrodynamics in devices}
\label{ssec:tunnel_chamber}

\begin{figure*}[htbp]
    \centering
    \includegraphics[width=0.9\textwidth]{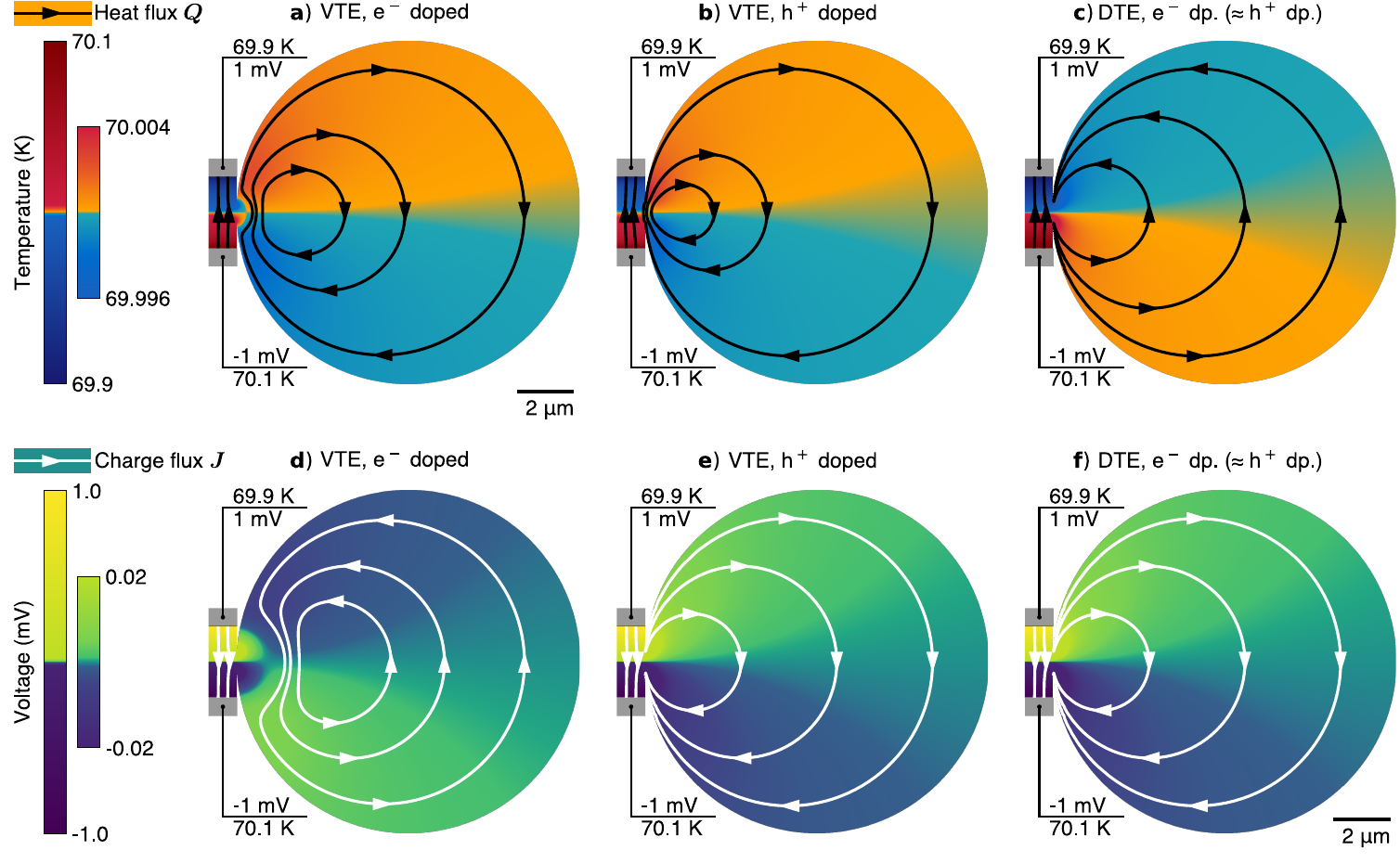}
\caption{\textbf{Doping-dependent heat and charge backflow, and inversion of temperature or voltage} driven by electron-phonon hydrodynamics in a tunnel-chamber device made of graphite. The in-plane device directions coincide with the basal-plane directions of graphite, and the device is very long in the off-plane direction.
Upper row: in-plane heat flux $\bm{Q}$ (streamlines) and temperature field $T$ (colormap) obtained from the VTE solution for high electron doping (carrier concentration $-10^{20}\; {\rm cm}^{-3}$, \textbf{a}) or weak hole doping ($+2.5{\cdot}10^{18}\; {\rm cm}^{-3}$, \textbf{b}), and from the standard DTE (\textbf{c} shows the strongly electron-doped case, $-10^{20}\; {\rm cm}^{-3}$, which is practically indistinguishable from the hole-doped case, see \cref{fig:DTE_doping_effect}).
Bottom row: corresponding in-plane charge current $\bm{J}$ (streamlines) and voltage $V$ (colormap). In all cases, the boundary conditions are: a temperature difference and a potential difference are applied to the tunnel ($T = 70\pm 0.1 \mathrm{\ K}$ and $V = \mp 1\mathrm{\ mV}$ at $y =\mp 1.25 \mathrm{\ \mu m}$); the other boundaries are adiabatic, electrically insulating, and, in the VTE, ``frictionless'' (see text for details).
In strongly electron-doped samples, the VTE predict the emergence of viscous backflow and vortices for both heat (\textbf{a}) and charge (\textbf{d}), accompanied by temperature and voltage gradients that are inverted in the chamber compared to the tunnel.
In weakly hole-doped samples, the VTE predict heat vortices and temperature inversion (\textbf{b}), while no charge backflow and no potential inversion are found (\textbf{e}).
The DTE always predict irrotational flow for heat and charge, and correspondingly no inversion of temperature or potential (\textbf{c} and \textbf{f}).}
\label{fig:vte_vs_dte}
\end{figure*}

In this section we discuss how the VTE allow us to predict macroscopic signatures of non-diffusive thermoelectric transport accounting for both intrinsic material properties (through the transport coefficients entering the VTE) and extrinsic effects such as device geometry and BCs (we also note that when the mean free paths of electrons and phonons become comparable to the device size, extrinsic finite-size effects reduce the transport coefficients, and we account for this effect as detailed in \cref{app:finite_size}).

Recent works have explored how intrinsic material properties and extrinsic device geometries influence the emergence of electron-only \cite{levitov2016electron,Torre2015,aharon-steinberg_direct_2022,bandurin2016,palm_observation_2024} or phonon-only \cite{dragasevic2023,guo_phonon_2021,shang_heat_2020,lucente_vortices_2025} hydrodynamics. Here we extend these studies using the VTE to investigate the simultaneous appearance of electron-phonon hydrodynamics, and discuss how this coupled regime differs both qualitatively and quantitatively from previous electron-only and phonon-only studies.

\subsubsection{Vortices, inversion, and backflow}

To systematically understand how electron and phonon hydrodynamics coexist and mix, it is natural to start from the regime where one of the two dominates---e.g., from phonon-only hydrodynamics in undoped graphite discussed in our previous work \cite{dragasevic2023}---and progressively strengthen the electronic contribution to transport by increasing the electronic density via doping (Fig.~\ref{fig:bulk_transport}). 
We consider extrinsic effects in a tunnel-chamber geometry (Fig.~\ref{fig:vte_vs_dte}), since it was shown in Ref. \cite{dragasevic2023} to drive unambiguous phonon hydrodynamic phenomena in graphite, and electron hydrodynamic phenomena in graphene \cite{palm_observation_2024}. 
The device is very long in the out-of-plane direction (which coincides with the out-of-plane direction of graphite), and its in-plane section consists of a `tunnel' on the left, a rectangular channel with two terminals on either end, and a circular `chamber', connected to the tunnel by a narrow opening.
The sample is at an average (equilibrium) temperature of 70 K; transport of heat and charge is driven by a temperature perturbation of 70$\pm$0.1 K and by a voltage perturbation of $\mp1$ mV applied between the lower and upper terminals of the tunnel. Recalling \cref{sec:conductivities_viscosities_and_cross_transport_coefficients}, this flow comprises: (i)  diffusive components determined by the temperature and voltage gradients; (ii) hydrodynamic components determined by the drift velocities of electrons and phonons. These two components are parallel in the tunnel, implying an analogous behavior for the charge and heat flux. At the chamber's opening the diffusive component of the flow spills into the chamber generating a dipole-like field (clockwise for phonons and counterclockwise for electrons), while the hydrodynamic component is subject to viscous stresses which drive heat and charge flows in directions opposite to the diffusive case (counterclockwise for phonons and clockwise for electrons). Therefore, the chamber serves to set up diffusive and hydrodynamic flow components that behave in opposite ways, and in the regime where the hydrodynamic component has a magnitude stronger than the diffusive one, it yields unambiguous signatures of non-diffusive, hydrodynamic transport phenomena.

Apart from the terminals on the top and the bottom of the tunnel, the other boundaries of the device are thermally and electrically insulating, i.e. no heat or charge flows across them \footnote{The insulating boundary condition can be expressed mathematically as $\bm{Q}\cdot\hat{\bm{n}}=0$ and $\bm{J}\cdot\hat{\bm{n}}=0$, where $\hat{\bm{n}}$ is a unit vector perpendicular to the boundary surface.}.
Boundaries are also assumed to be `frictionless', corresponding to reflective carrier-boundary scattering \cite{ziman1960}. Such boundaries have zero perpendicular drift velocity components and vanishing off-diagonal components of the viscous stress tensor \cite{Torre2015}. 
More details on the geometry, as well as the numerical implementation of frictionless boundaries using the lubrication-layer approach are given in Ref.~\cite{dragasevic2023} and \cref{app:computational_details}.

Considering this setup, in \cref{fig:vte_vs_dte} we
compare the solutions of the VTE, \cref{eq:VTE}, at high electronic doping [panels (\textbf{a}), (\textbf{d}) show carrier concentration -10$^{20}$ cm$^{-3}$] or low hole doping [panels (\textbf{b}), (\textbf{e}) show carrier concentration $+2.5\cdot 10^{18}$ cm$^{-3}$] with the solution of the diffusive thermoelectric equations, \cref{eq:DTE}, [DTE, panels (\textbf{c}) and (\textbf{f})]. We only show the solutions of the DTE at high electronic doping, since the solutions are qualitatively unaffected by doping, as we show in \cref{app:dte_similarity}.
The solutions of the VTE in \cref{fig:vte_vs_dte} display hallmarks of hydrodynamic transport: heat and charge flow vortices, temperature and voltage inversion, and thermoelectric backflow.
Each of these signatures is associated with violations of various properties of diffusive transport, as we discuss in order below.

The heat and charge flows predicted by the VTE form vortices in the chamber, as shown in panels (\textbf{a}) and (\textbf{b}) for the heat flux $\bm{Q}$ and in panel (\textbf{d}) for the charge flux $\bm{J}$. Vortices are an exclusively non-diffusive transport phenomenon associated with non-zero curl ($\Nabla \times \bm{Q}\neq 0$, and $ \Nabla \times \bm{J} \neq \bm{0}$), which is forbidden by the irrotational form of the solution of the DTE (panels (\textbf{c}) and (\textbf{f})). In particular, because in the DTE the flows are gradient-driven, their curl must be zero, implying that streamlines are open and the circulation $\Gamma$ around any closed line $C$ vanishes by Stokes' theorem: $\Gamma(\bm{F}) = \oint_{C} \bm{F}\cdot\mathrm{d}\bm{l} = 0$, whenever $\bm{F}\propto \nabla T +\nabla \V$ is an irrotational field proportional to gradients of scalar fields. 
In contrast, in the VTE case, closed streamlines are in general permitted, and emerge under the conditions shown in \cref{fig:vte_vs_dte}.
Integrating the heat or charge flux along a closed streamline trivially results in a non-zero circulation; therefore, the formation of a vortex with closed streamlines is a sufficient condition to have non-diffusive flow.

Because heat and charge flows are vector fields that are  experimentally more challenging to measure than scalar $T$ and $\V$ fields, we focused on finding signatures of hydrodynamic transport on scalar fields, shown in \cref{fig:vte_vs_dte} as colormaps.
We found that hydrodynamic vortices are related to an inversion of the profile of $T$ (\cref{fig:vte_vs_dte}\textbf{a}) and $\V$ (\cref{fig:vte_vs_dte}\textbf{d}), where we highlight that the gradients of these fields in the chamber assume a direction opposite to that in the tunnel.
We highlight how the inverted temperature and voltage profiles emerging from the VTE are qualitatively opposite to those emerging from the solutions of the DTE, where temperature and voltage change monotonically from one terminal to the other (see \cref{fig:vte_vs_dte}\textbf{c} for temperature and \cref{fig:vte_vs_dte}\textbf{f} for voltage). 
To observe temperature (voltage) inversion, it is sufficient to directly measure the temperature (voltage) in four points of the device. This is in contrast to hydrodynamic signatures related to the heat or charge flux, such as Poiseuille flow, where the flux has to be imaged across the whole sample \cite{geursSupersonicFlowHydraulic2025}. 
As such, these signatures may be particularly suitable for experimental validation. Temperature inversion is also robust with respect to drift velocity BCs \cite{dragasevic2023}, whereas Poiseuille flow is most prominent with no-slip BCs and vanishes with frictionless BCs \cite{Torre2015}. In \cref{app:mixing_geometry}, we discuss thermoelectric Poiseuille flow predicted by the VTE and highlight the distinct length scales governing electron and phonon fluids.

After having discussed electron-doped graphite, where signatures of electron-phonon hydrodynamics are most prominent, we discuss how these depend on the doping level.
To this aim, we compare the solution of the VTE in samples with high electron doping and low hole doping.
The heat vortices and temperature inversion are qualitatively unaffected by electron doping; this indicates that they are mainly driven by phonons, and is consistent with their emergence in undoped samples \cite{dragasevic2023}.
The behavior for charge flow and voltage is completely different --- \cref{fig:vte_vs_dte}\textbf{d},\textbf{e} show that they are strongly influenced by doping.
While a charge vortex and related backflow, as well as voltage inversion, emerge at high electron doping (\textbf{d}), none of these signatures appear at low hole doping (\textbf{e}), which is qualitatively similar to the diffusive flow shown in (\textbf{f}).
This shows that heat and charge hydrodynamics are not necessarily related: doping can drive a transition from heat-only hydrodynamics \cite{dragasevic2023} to coupled electron-phonon hydrodynamics in graphite.

In addition to the signatures of non-diffusive transport that can be detected by imaging inversion in the scalar fields $T,\;\V$ or closed streamlines in the vector fields $\bm{Q},\;\bm{J}$, we note that a hallmark of non-diffusive transport can be detected by investigating the relation between temperature and voltage gradients, and heat and charge currents. In thermal hydrodynamic transport, heat can backflow along the temperature gradient \cite{dragasevic2023,guo_phonon_2021,shang_heat_2020,lucente_vortices_2025}, while in electrical hydrodynamic transport charge can backflow, i.e., flow in direction opposite to that prescribed by the electric potential gradient \cite{levitov2016electron,Torre2015,bandurin2016,aharon-steinberg_direct_2022}. In diffusive heat-only or charge-only transport, both of these effects are forbidden, since fluxes are always driven by temperature and potential gradients, respectively. In the diffusive thermoelectric case, one of the two terms is allowed to be negative as long as their sum is positive\cite{landau1981kinetics}. Hence, to detect non-diffusive thermoelectric backflow, we need to observe heat and charge backflow occurring simultaneously and at the same point in the device. This is exactly what we see at the opening of the chamber in panels (\textbf{a}) and (\textbf{d}). \footnote{We note that this simultaneous backflow of heat and charge does not violate the second law of thermodynamics, describing the entropy produced in this phenomenon requires us to consider also momentum fluxes and their dissipation \cite{guo_phonon_2021}.}

\Cref{fig:vte_vs_dte} thus shows that the VTE can predict established signatures of hydrodynamic transport, while encompassing both the phonon-only and the coupled electron-phonon regime. Each signature provides a different measurement method for establishing the emergence of hydrodynamic flow: vortices are detected by imaging the fluxes, inversion necessitates only a four-point measurement of scalar fields, and backflow is found by imaging both the scalar fields and the fluxes. We now focus specifically on the properties of the mixed electron-phonon regime, and examine how it differs from electron-only hydrodynamics.

\subsubsection{Electron compressibility and non-harmonic potential}
\label{ssec:electron_compressibility}

In the VTE, both the electron and phonon fluids are compressible, $\Nabla \cdot \bm{u}_{e/p} \neq 0$, even at linear response, as can be seen from \cref{eq:VTE}. This is an important difference from the incompressible Gurzhi-type hydrodynamic equations used to describe charge-only hydrodynamics in doped graphene \cite{aharon-steinberg_direct_2022,levitov2016electron,Torre2015,bandurin2016}. In this sense, the electronic transport described by the VTE bears some analogies to the hydrodynamics of nearly neutral graphene ($\mu \ll k_B T$), where the dynamics of charge and energy cannot be decoupled and the electron flow need not be incompressible \cite{gall_electronic_2023}. 

We now investigate experimental signatures of this compressible behavior. We focus on its effect on the effective voltage $\V$,
since it can be imaged in experiments \cite{palm_observation_2024,geursSupersonicFlowHydraulic2025}.
Within the DTE and Gurzhi's equations, the incompressibility of the charge flow implies a harmonic electrical potential profile, i.e., a profile that satisfies $\nabla^2 V {=} 0$ \footnote{For the DTE, this follows directly from \cref{eq:DTE_charge,eq:DTE_energy}, while for Gurzhi's equations one needs to take the divergence of \cref{eq:Gurzhi} and simplify it using the incompressibility condition $\Nabla \cdot \bm{u} = 0$.}. In contrast, the compressibility of electron fluid in the VTE implies that the potential will in general have non-harmonic behavior, $\nabla^2 \V \neq 0$ \footnote{Since the DTE and Gurzhi's equation do not include changes in the chemical potential, the effective and electrical voltage coincide.}. 

\begin{figure}[t]
    \centering
    \includegraphics[width=\columnwidth]{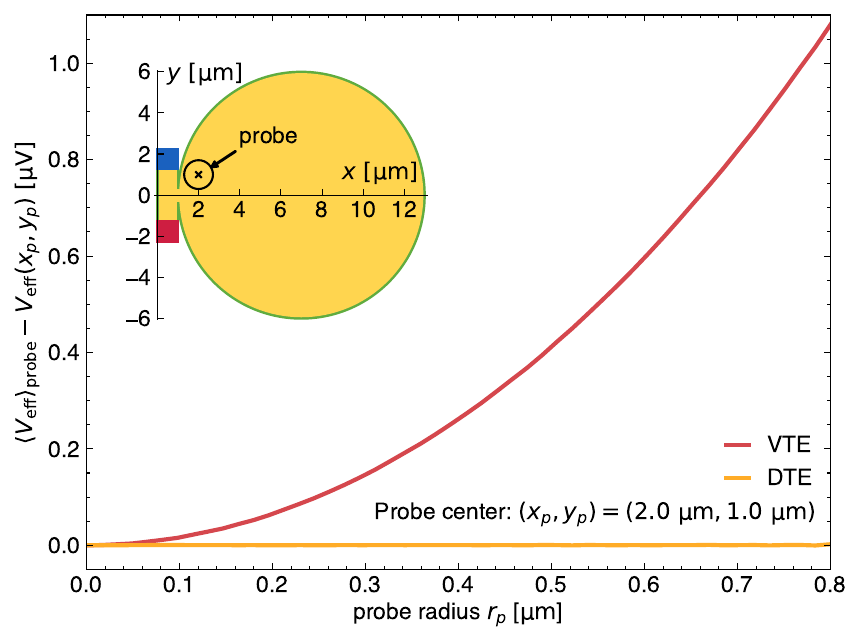}\\[-2mm]
    \caption{\textbf{Non-diffusive transport \& non-harmonic potential.} The difference between the effective potential averaged over a circular probe and the effective potential at the centre of the probe, as a function of the probe radius. Inset: outline of the tunnel-chamber geometry (adapted from \cite{dragasevic2023}) and the position of the probe within the device. By the mean-value property, the average over the probe should be exactly the same as the value at the centre of the probe for a harmonic function, showing that $\V$ is harmonic in the diffusive regime, but non-harmonic in the hydrodynamic regime.} 
    \label{fig:non_harmonic_potential}
\end{figure}
The analysis above shows that it is possible to rely on the different analytical properties between harmonic and non-harmonic scalar-potential functions to detect compressible behavior in thermoelectric hydrodynamic transport.
In particular, as a measure of non-harmonicity, we consider how the effective potential $\V$ deviates from the mean-value property satisfied by a harmonic function.
This property states that, for a harmonic function, the average of the function over a circular area will be the same as the value of the function in the centre of the area. 
In \cref{fig:non_harmonic_potential} we consider a circular probe of radius $r_p$ and compute, for the VTE and DTE, the difference between the
value of the potential in the center of the probe ($\V(x_p, y_p)$) and the
average of the potential over the area of the probe ($\langle \V \rangle_{\rm probe} = \frac{1}{N}\sum_{i=1}^N \V(x_i,y_i)$ where $(x_i,y_i)$ denote points uniformly distributed in the probe circular area). We see that for the DTE potential this difference is always zero for all values of $r_p$, as expected for a harmonic function. In contrast, for the VTE potential, this difference increases as $r_p$ increases---this unambiguously indicates a non-harmonic behavior for $\V$, which in the context of the VTE, implies compressible behavior for the charge flow.

In summary, imaging the effective potential across the device can show differences between incompressible charge flow of electron-only hydrodynamics and compressible charge flow of electron-phonon hydrodynamics.

\section{Conclusions} 
\label{sec:conclusions}

We introduced a theoretical framework to describe thermoelectric transport beyond the diffusive regime with first-principles accuracy and elucidated the conditions under which phonons, electrons, or a mixture of them can display fluid-like behavior. 
Starting from the coupled electron-phonon Boltzmann transport equation (epBTE), we exposed the microscopic symmetries and (quasi-) conservation laws that determine transport of heat in phonon-only systems, of charge in electron-only systems, and thermoelectric phenomena in coupled electron-phonon systems. Specifically, we demonstrated how out-of-equilibrium mixtures of electrons and phonons form relaxon excitations with well-defined parity, with odd relaxons determining the standard diffusive thermoelectric transport coefficients --- electrical conductivity ($\sigma$), thermal conductivity ($\bar{\kappa}$), Peltier and Seebeck coefficients ($\alpha$ and $S$, respectively) --- while the complementary set of even relaxons determine the viscosities of the electron and phonon components of the two-fluid system.

These microscopic insights have allowed us to coarse-grain the epBTE into Viscous Thermoelectric Equations (VTE) that formally unify Gurzhi's \cite{gurzhi1968} equation for electron-only hydrodynamics and the Viscous Heat Equations for phonon-only hydrodynamics \cite{simoncelli2020}, and, most importantly, extend them to cover the intermediate regime of mixed electron and phonon fluids as well as its transition into the diffusive limit in large devices or at high temperature.

As a case study, we applied our quantitatively accurate first-principles framework to several graphite samples with different carrier concentrations, finding that accounting for electron-phonon drag is critical to obtaining quantitative agreement with experimental measurements of the Seebeck coefficient \cite{ayache_observation_1980,de_combarieu_thermoelectric_1973} at different doping levels. 
At the macroscopic, device level, we showed that in strongly electron-doped graphite ($-10^{20}\;{\rm cm}^{-3}$) by engineering device shape and boundary conditions, it is possible to simultaneously induce vortices for heat and charge, which are associated with temperature and potential profiles opposite to those predicted by the standard diffusive thermoelectric transport equations.
We also discussed how these thermoelectric hydrodynamic phenomena depend on the carrier concentration. By reducing the carrier concentration via doping, we found that charge hydrodynamics disappears while heat hydrodynamics persists in the temperature range 70-120 K, consistently with the observation of heat-only hydrodynamics in natural graphite samples at these conditions \cite{Huberman_science_2019,machida2020phonon,huang_graphite_2024}.
Additionally, we found that electron-phonon drag can be significant also at temperatures where charge or heat backflow (and related voltage or temperature inversion) do not emerge.
Therefore, having electron-phonon drag is necessary but not sufficient for the emergence of an electron-phonon bi-component fluid regime.
 
We demonstrated that the electron and phonon fluids can have very different viscosities, and that the electronic viscosity can be controlled through doping. In particular, the electronic shear viscosity decreases with temperature for all doping levels, displaying a trend reminiscent of liquids, while the phonon shear viscosity increases with temperature and is only weakly affected by doping, displaying a gas-like trend. In addition, the electronic volume viscosity changes qualitatively with doping: it increases with temperature in weakly doped graphite; upon increasing electron doping it becomes smaller and nearly temperature independent.

We also discussed how the emergence of compressible electron flow sets apart thermoelectric hydrodynamics from the incompressible electron-only hydrodynamic flow in doped graphene \cite{Torre2015,levitov2016electron}, and showed that the compressibility of electron flow can be tested by experimental measurements of the electrical potential.

Additionally, the theory and methodology developed here for thermoelectric hydrodynamics share several formal analogies with other transport phenomena.
The coexistence of a gradient-driven and a drift-velocity component in both the charge and heat flux is reminiscent of the diffusive and convective contributions to transport in rarefied classical fluids \cite{brenner_navierstokes_2005,sambasivam_numerical_2014,schwarz_openfoam_2023,maurer_second-order_2003,dongari_pressure-driven_2009}; the drift-velocity damping term is reminiscent of the dissipation term appearing in the hydrodynamic equations to describe fluids
flowing through porous media \cite{balasubramanian_darcys_1987,dardis_lattice_1998,bresch_existence_2003,cai_weak_2008,zhang_uniqueness_2011}.
More generally, the framework introduced here could be applied to describe other coupled transport phenomena in solids --- including drag between phonons and photons \cite{glazov_valley_2020}, magnons and electrons \cite{yamaguchi_microscopic_2019}, or phonons and magnons \cite{pan_ab_2023} --- and therefore will inspire further developments and applications. We also release these developments as open-source software to the community in an extension of the Phoebe software suite \cite{phoebe}, enabling future investigations of electron-phonon drag effects and thermoelectric hydrodynamics. 

We conclude by noting that our findings set the stage for engineering thermoelectric transport phenomena that are not simply controlled by temperature and voltage gradients.
As such, they may enable the design of novel electronic devices where viscous degrees of freedom allow unprecedented control and management of mixed heat and charge flows.

\section{Code Availability}
The infrastructure to solve from first principles the microscopic coupled electron-phonon Boltzmann transport equation in the linear-response regime is released as a major update of the open-source Phoebe \cite{phoebe} software suite, available at \url{https://github.com/phoebe-team/}. 

\section{Acknowledgments}
We gratefully acknowledge Bo Peng, Samuel Huberman, Aleksei Sokolov, and Enrico Di Lucente for the useful discussions.
The computational resources were provided by: (i) the Sulis Tier 2 HPC platform (funded by EPSRC Grant EP/T022108/1 and the HPC Midlands+consortium); (ii) the UK National Supercomputing Service ARCHER2, for which access was obtained via the UKCP consortium and funded by EPSRC [EP/X035891/1]; (iii) the Kelvin2 HPC platform at the NI-HPC Centre (funded by EPSRC and jointly managed by Queen’s University Belfast and Ulster University). 
The Flatiron Institute is a division of the Simons Foundation. 
B. R. acknowledges support from Trinity College at University of Cambridge.
M. S. acknowledges support from: (i) the Swiss National Science Foundation (SNSF) project P500PT\_203178; (ii) Gonville and Caius College at University of Cambridge; (iii) Columbia University in the city of New York. 

\raggedbottom
\pagebreak
\appendix

\begin{widetext}

\section{Coupled electron-phonon scattering matrix}
\label{app:scattering_rates}

In this Appendix we derive the scattering rates for electrons and phonons, including drag effects. These determine the scattering matrix on the right-hand side of the epBTE, \cref{eq:BTE_el_ph}. Self quadrants $\mathrm{S}^{\rm ee}_{\km,\kmp},\mathrm{S}^{\rm ph}_{\qs,\qsp}$ of the electron-phonon scattering matrix are reported in \citet{phoebe}; here we verify those expressions and derive the complementary expressions for the drag quadrants $\mathrm{D}^{\rm ep}_{\km,\qsp},\mathrm{D}^{\rm pe}_{\qs,\kmp}$ \cite{levchenko2020transport,fiorentini2016}, and then perform a two-carrier generalization of the symmetrization transformation discussed by \citet{hardy1970}.

\subsection{Electron-phonon scattering rates}

In considering the electron-phonon interaction, we restrict ourselves to the lowest-order collisions, in which a phonon is absorbed or emitted by an electron, followed by an appropriate change of state of the electron. The net scattering rate of the $\km$ state into the $\KMP$ state due to the $\qs$ phonon is given by Fermi's golden rule \cite{ziman1960}. In the linearized epBTE, we consider only the rates at leading linear order in $\mathrm{f}_{\km},\mathrm{n}_{\qs},\textrm{ and }\mathrm{f}_{\kmp}$:
\begin{equation}
\begin{aligned}
    W_{\km,\qs;\kmp}
=
    \frac{2\pi}{\hbar}
    |g_{m,m',s}(\bm{k}, \bm{k'}, \bm{q})|^2
    &\delta_{\bm{k'}{-}\bm{k}{-}\bm{q},\bm{G}}
    \delta(\varepsilon_{\kmp}{-}\varepsilon_{\km}{-}\hbar\omega_{\qs}) \\
    &\times \Big[
        (\bar{\mathrm{N}}_{\qs} + \bar{\mathrm{F}}_{\kmp}) \mathrm{f}_{\km}
        + (\bar{\mathrm{F}}_{\km}-\bar{\mathrm{F}}_{\kmp}) \mathrm{n}_{\qs}
        - (1 - \bar{\mathrm{F}}_{\km} + \bar{\mathrm{N}}_{\qs}) \mathrm{f}_{\kmp}
    \Big].
\label{eq:matrix/scattering_rate}
\end{aligned}
\end{equation}
The Kronecker delta and the Dirac delta function enforce the conservation of crystal momentum and energy, respectively, and $g_{m,m',s}(\bm{k},\bm{k'},\bm{q})$ is the electron-phonon matrix element \cite{giustino2017electron}. We note that the linearization of the scattering rate requires choosing an equilibrium point; in this equilibrium point all scattering rates must vanish. This equilibrium is given by the Fermi-Dirac and Bose-Einstein distributions for electrons and phonons, respectively, at some chosen equilibrium temperature $\bar{T}$ ($\beta=1/k_B\bar{T}$) and equilibrium chemical potential $\bar{\mu}$. We write the net rate as $W_{\km,\qs;\kmp} = \mathcal{P}^{\km}_{\km,\qs;\kmp}\mathrm{f}_{\km} + \mathcal{P}^{\qs}_{\km,\qs;\kmp}\mathrm{n}_{\qs} - \mathcal{P}^{\kmp}_{\km,\qs;\kmp}\mathrm{f}_{\kmp}$, where $\mathcal{P}$ are equilibrium transition rates \cite{ziman1960}. These rates do not depend on the deviation from equilibrium and are always non-negative. We can then express the scattering rates due to the electron-phonon interaction, starting with electrons:
\begin{equation}
\begin{aligned}
    \left.\frac{\dif \mathrm{f}_{\km}}{\dif t}\right|_{\rm coll} 
={}
    &- \frac{1}{\mathcal{V}\Nk} \sum_{\kmp,\qs} \left( \mathcal{P}^{\km}_{\km,\qs;\kmp} + \mathcal{P}^{\km}_{\kmp,\qs;\km} \right) \mathrm{f}_{\km}
    + \frac{1}{\mathcal{V}\Nk} \sum_{\kmp,\qs} \left( \mathcal{P}^{\kmp}_{\km,\qs;\kmp} + \mathcal{P}^{\kmp}_{\kmp,\qs;\km} \right) \mathrm{f}_{\kmp} \\
    &- \frac{1}{\mathcal{V}\Nq} \sum_{\qs,\kmp} \left( \mathcal{P}^{\qs}_{\km,\qs;\kmp} - \mathcal{P}^{\qs}_{\kmp,\qs;\km} \right) \mathrm{n}_{\qs},
\end{aligned}
\label{eq:matrix/el_scattering}
\end{equation}
where the two rates in each bracket account for processes in which an electron absorbs or emits a phonon. We further separate out-scattering terms, which are proportional to $\mathrm{f}_{\km}$ itself, from in-scattering terms that are proportional to the populations of other electron states $\mathrm{f}_{\kmp}$. Comparing \cref{eq:matrix/el_scattering} with the electron part of \cref{eq:BTE_el_ph}, we find the expressions for two upper quadrants of the unsymmetrized scattering matrix:
\begin{gather}
\begin{aligned}
    \mathrm{S}^{\rm ee}_{\km,\kmp}
= 
    \frac{1}{\mathcal{V}\Nk}
    \sum_{\kmpp, \qs} 
    \frac{2\pi}{\hbar}
    &|g_{m,\ssp{\ssp{m}},s}(\bm{k},\bm{k''},\bm{q})|^2
    \delta_{\bm{k''}{-}\bm{k}{-}\bm{q},\bm{G}}
    \Big[ 
    \delta(\varepsilon_{\kmpp}{-}\varepsilon_{\km}{-}\hbar\omega_{\qs}) + 
    \delta(\varepsilon_{\kmpp}{-}\varepsilon_{\km}{+}\hbar\omega_{\qs})
    \Big] \times
    \\
        &\times \frac{1}{2\sinh{[\tfrac{\beta}{2}\hbar\omega_{\qs}]}} \left[
        \delta_{\kmp,\km}
        \frac   {\cosh{[\tfrac{\beta}{2}(\varepsilon_{\km}{-}\bar\mu)]}}
                {\cosh{[\tfrac{\beta}{2}(\varepsilon_{\kmpp}{-}\bar\mu)]}}
        - \delta_{\kmp,\kmpp}
        \frac   {\cosh{[\tfrac{\beta}{2}(\varepsilon_{\kmp}{-}\bar\mu)]}}
                {\cosh{[\tfrac{\beta}{2}(\varepsilon_{\km}{-}\bar\mu)]}}
    \right],
\end{aligned}
\label{eq:matrix/S_el}
\\
\begin{aligned}
    \mathrm{D}^{\rm ep}_{\km,\qs}
= 
    \frac{1}{\mathcal{V}\Nq} \sum_{\kmp}
    \frac{2\pi}{\hbar}
    \Big[&
        |g_{m,m',s}(\bm{k},\bm{k'},\bm{q})|^2
        \delta_{\bm{k'}{-}\bm{k}{-}\bm{q},\bm{G}}
        \delta(\varepsilon_{\kmp}{-}\varepsilon_{\km}{-}\hbar\omega_{\qs}) \\
    &- 
        |g_{m',m,s}(\bm{k'}, \bm{k}, \bm{q})|^2
        \delta_{\bm{k'}{-}\bm{k}{+}\bm{q},\bm{G}}
        \delta(\varepsilon_{\kmp}{-}\varepsilon_{\km}{+}\hbar\omega_{\qs})
    \Big]
    \frac{\sinh{[\tfrac{\beta}{2}\hbar\omega_{\qs}]}}
    {2\cosh{[\tfrac{\beta}{2}(\varepsilon_{\km}{-}\bar\mu)]}
    \cosh{[\tfrac{\beta}{2}(\varepsilon_{\kmp}{-}\bar\mu)]}},
\end{aligned}
\label{eq:matrix/D_el_ph}
\end{gather}
where we expanded the scattering rates using \cref{eq:matrix/scattering_rate} and expressed equilibrium populations in terms of Fermi-Dirac and Bose-Einstein populations. In the first bracket of \cref{eq:matrix/S_el}, we group together absorption of $\qs$ and emission of $-\qs$, while in the second bracket the two terms determine the diagonal and off-diagonal terms of $\mathrm{S}^{\rm ee}_{\km,\kmp}$ \footnote{Note that, due to inversion symmetry of the phonon Brillouin zone, the state $-\bm{q},s$ has the same energy as $\qs$ and we can replace $\bm{q}$ with $-\bm{q}$ under the sum over $\bm{q}$.}. In \cref{eq:matrix/D_el_ph}, the two terms in the bracket account for absorption and emission of the $\qs$ mode, exactly as in \cref{eq:matrix/el_scattering}. 

We proceed in an analogous way to describe the scattering term for phonons:
\begin{equation}
    \left.\frac{\dif \mathrm{n}_{\qs}}{\dif t}\right|_{\rm coll} 
=
    \frac{\spindeg}{\mathcal{V}\Nk} \sum_{\km,\kmp} \left( \mathcal{P}^{\km}_{\kmp,\qs;\km} - \mathcal{P}^ {\km}_{\km,\qs;\kmp} \right) \mathrm{f}_{\km}
    - \frac{\spindeg}{\mathcal{V}\Nk} \sum_{\km,\kmp} \mathcal{P}^{\qs}_{\km,\qs;\kmp}\mathrm{n}_{\qs};
\label{eq:matrix/ph_scattering}
\end{equation}
here a single sum over electrons accounts for both absorption and emission, and we note also that the electron-phonon interaction is diagonal in the phonon subspace. The scattering matrix elements are then: 

\begin{gather}
\begin{aligned}
    \mathrm{D}^{\rm pe}_{\qs, \km}
= 
    \frac{\spindeg}{\mathcal{V}\Nk} \sum_{\bm{k}'m'}
    \frac{2\pi}{\hbar}
    \Big[& 
        |g_{m,m',s}(\bm{k},\bm{k'},\bm{q})|^2
        \delta_{\bm{k'}{-}\bm{k}{-}\bm{q},\bm{G}}
        \delta(\varepsilon_{\kmp}{-}\varepsilon_{\km}{-}\hbar\omega_{\qs}) \\
    &- 
        |g_{m',m,s}(\bm{k'}, \bm{k}, \bm{q})|^2 
        \delta_{\bm{k'}{-}\bm{k}{+}\bm{q},\bm{G}}
        \delta(\varepsilon_{\kmp}{-}\varepsilon_{\km}{+}\hbar\omega_{\qs})
    \Big]
    \frac{\cosh{[\tfrac{\beta}{2}(\varepsilon_{\km}{-}\bar\mu)]}}
    {2\sinh{[\tfrac{\beta}{2}\hbar\omega_{\qs}]}
    \cosh{[\tfrac{\beta}{2}(\varepsilon_{\kmp}{-}\bar\mu)]}},
\end{aligned}
\label{eq:matrix/D_ph_el} \\
    \mathrm{S}^{\rm ph}_{\qs,\qsp} 
= 
    \delta_{\qs,\qsp}
    \frac{\spindeg}{\mathcal{V}\Nk}
    \sum_{\km,\kmp}
    \frac{2\pi}{\hbar}|g_{m',m,s}(\bm{k'}, \bm{k}, \bm{q})|^2{\delta_{\bm{k'}{-}\bm{k}{-}\bm{q},\bm{G}}} 
    \delta(\varepsilon_{\kmp}{-}\varepsilon_{\km}{-}\hbar\omega_{\qs}) 
    \frac{\sinh{[\tfrac{\beta}{2}\hbar\omega_{\qs}]}}
    {2\cosh{[\tfrac{\beta}{2}(\varepsilon_{\km}{-}\bar\mu)]}
    \cosh{[\tfrac{\beta}{2}(\varepsilon_{\kmp}{-}\bar\mu)]}}.
\end{gather}

\subsection{Symmetrization}
\label{sub:symmetrization}

As discussed in \cref{sec:coupled_basis}, the first step of solving the epBTE is recasting the coupled scattering matrix in a symmetric form. We start by symmetrizing the self quadrants $\mathrm{S}^{\rm ee}_{\km, \kmp}, \mathrm{S}^{\rm ph}_{\qs, \qsp}$; these transformations are taken from \citet{phoebe} and \citet{hardy1970,chaput2013}, respectively:
\begin{align}
    \mathrm{S}^{\rm ee}_{\km, \kmp} &= \mathcal{S}^{\rm el}_{\km, \kmp} \frac{\cosh{[\tfrac{\beta}{2}(\varepsilon_{\kmp}{-}\bar\mu)]}}{\cosh{[\tfrac{\beta}{2}(\varepsilon_{\km}{-}\bar\mu)]}}, \\
    \mathrm{S}^{\rm ph}_{\qs, \qsp} &= \mathcal{S}^{\rm ph}_{\qs, \qsp} \frac{\sinh{[\tfrac{\beta}{2}\hbar\omega_{\qsp}]}}{\sinh{[\tfrac{\beta}{2}\hbar\omega_{\qs}]}}.
\end{align}
As can be verified from the expressions above, this transformation produces symmetric self quadrants, so that $\mathcal{S}^{\rm el}_{\km, \kmp} = \mathcal{S}^{\rm el}_{\kmp, \km}, \mathcal{S}^{\rm ph}_{\qs,\qsp}=\mathcal{S}^{\rm ph}_{\qsp,\qs},$ while the symmetrised drag quadrants $\mathcal{D}^{\rm el-ph}_{\km, \qsp}, \mathcal{D}^{\rm ph-el}_{\qs, \kmp}$ need to satisfy $\mathcal{D}^{\rm el-ph}_{\km, \qs} = \mathcal{D}^{\rm ph-el}_{\qs, \km}$. Comparing \cref{eq:matrix/D_el_ph,eq:matrix/D_ph_el}, we find that this condition can be satisfied with the following transformation:
\begin{align}
    \mathrm{D}^{\rm ep}_{\km, \qsp} &= \mathcal{D}^{\rm el-ph}_{\km,\qsp} \sqrt{\frac{\Nk}{\spindeg\Nq}} \frac{\sinh{[\tfrac{\beta}{2}\hbar\omega_{\qsp}]}}{\cosh{[\tfrac{\beta}{2}(\varepsilon_{\km}{-}\bar\mu)]}}, \\
    \mathrm{D}^{\rm pe}_{\qs, \kmp} &= \mathcal{D}^{\rm ph-el}_{\qs,\kmp} \sqrt{\frac{\spindeg\Nq}{\Nk}} \frac{\cosh{[\tfrac{\beta}{2}(\varepsilon_{\kmp}{-}\bar\mu)]}}{\sinh{[\tfrac{\beta}{2}\hbar\omega_{\qs}]}}.
\end{align}
The four relations above motivate the definition of the similarity transformation $\mathfrak{g}_{\kmqs}$ discussed in \cref{eq:symm_coupled_matrix} of the main text:
\begin{equation}
\mathfrak{g}_{\kmqs}{=}
\left(\!\!\!\begin{array}{c}
\sqrt{\frac{\spindeg}{\mathcal{V}\Nk}}\frac{1}{\sqrt{\bar{\mathrm{F}}_{\km}[1{-}\bar{\mathrm{F}}_{\km}]}}\\
\sqrt{\frac{1}{\mathcal{V}\Nq}} \frac{1}{\sqrt{\bar{\mathrm{N}}_{\qs}[\bar{\mathrm{N}}_{\qs}{+}1]}}
\end{array} \!\!\!\right)
{=}
2 \left(\!\!\!\begin{array}{c}
\sqrt{\frac{\spindeg}{\mathcal{V}\Nk}} \cosh{[\tfrac{\beta}{2}(\varepsilon_{\km}{-}\bar\mu)]} \\
\sqrt{\frac{1}{\mathcal{V}\Nq}} \sinh{[\tfrac{\beta}{2}\hbar\omega_{\qs}]}
\end{array} \!\!\!\right).
\label{eq:collision_symmetrisation}
\end{equation}
Note that we included a factor of $\mathcal{V}$ in the definition of $\mathfrak{g}_{\kmqs}$, which has no effect on $\Omega_{\kmqs,\kmpqsp}$, but it will later simplify relations involving symmetrized populations $\z$. Thus, the final form of the symmetrized electron-phonon scattering rates is:
\begin{gather}
\begin{aligned}
    \mathcal{S}^{\rm el}_{\km,\kmp}
= 
    \frac{1}{\mathcal{V}\Nk}
    \sum_{\kmpp, \qs} 
    \frac{2\pi}{\hbar}
    &|g_{m,\ssp{\ssp{m}},s}(\bm{k},\bm{k''},\bm{q})|^2
    \delta_{\bm{k''}{-}\bm{k}{-}\bm{q},\bm{G}}
    \Big[ 
    \delta(\varepsilon_{\kmpp}{-}\varepsilon_{\km}{-}\hbar\omega_{\qs}) + 
    \delta(\varepsilon_{\kmpp}{-}\varepsilon_{\km}{+}\hbar\omega_{\qs})
    \Big] \times
    \\
    &\times \frac{1}{2\sinh{[\tfrac{\beta}{2}\hbar\omega_{\qs}]}} 
    \left[
        \delta_{\kmp,\km}
        \frac   {\cosh{[\tfrac{\beta}{2}(\varepsilon_{\km}{-}\bar\mu)]}}
                {\cosh{[\tfrac{\beta}{2}(\varepsilon_{\kmpp}{-}\bar\mu)]}}
        - \delta_{\kmp,\kmpp}
    \right],
\end{aligned} \\
\begin{aligned}
    \mathcal{D}^{\rm el-ph}_{\km, \qs} =  
    \frac{1}{\mathcal{V}} \sqrt{\frac{\spindeg}{\Nk\Nq}}
    \sum_{\kmp}
    \frac{2\pi}{\hbar}
    \Big[&|g_{m,m',s}(\bm{k},\bm{k'},\bm{q})|^2
    \delta_{\bm{k'}{-}\bm{k}{-}\bm{q},\bm{G}}
    \delta(\varepsilon_{\bm{k}'m'}{-}\varepsilon_{\bm{k}m}{-}\hbar\omega_{\bm{q}s}) \\
    &-|g_{m',m,s}(\bm{k'}, \bm{k}, \bm{q})|^2
    \delta_{\bm{k'}{-}\bm{k}{+}\bm{q},\bm{G}}
    \delta(\varepsilon_{\bm{k}'m'}{-}\varepsilon_{\bm{k}m}{+}\hbar\omega_{\bm{q}s}) \Big]
    \frac{1}{2\cosh{[\tfrac{\beta}{2}(\varepsilon_{\kmp}-\bar\mu)]}},
\label{eq:symm_drag_element}
\end{aligned}
\end{gather}
and $\mathcal{S}^{\rm ph}_{\qs, \qsp} = \mathrm{S}^{\rm ph}_{\qs,\qsp}$ since this quadrant is diagonal.

\end{widetext}

\subsection{Other scattering mechanisms}

The phonon-phonon scattering matrix elements take into account the leading-order anharmonic three-phonon interactions, i.e., coalescence ($\qsd + \qspp \rightarrow \qs$) and decay ($\qs \rightarrow \qsd + \qspp$). The resulting matrix elements are \cite{fugallo2013}:
\begin{gather}
\begin{aligned}
    &\mathcal{S}^{\rm ph-ph}_{\qs, \qsd}\bar{N}_{\qsd}(\bar{N}_{\qsd} + 1) \\
    &= \delta_{\qs, \qsd} \Nq \Gamma^{\rm ph-ph}_{\qs} - \sum_{\qspp} \left(P_{\qs, \qspp}^{\qsd} - P_{\qs, \qsd}^{\qspp} + P_{\qsd, \qspp}^{\qs}\right),
    \raisetag{15mm}
\end{aligned} \\
\Gamma^{\rm ph-ph}_{\qs} = \frac{1}{\Nq}\sum_{\qsd,\qspp}\left( P_{\qsd, \qspp}^{\qsd} + \frac{1}{2}P_{\qspp, \qsd}^{\qs} \right),\hspace*{18mm}
 \raisetag{8mm}
\end{gather}
where $\Gamma^{\rm ph-ph}_{\qs}$ is the linewidth due to phonon-phonon scattering, and the scattering rates for coalescence and decay processes are respectively given by:
\begin{align}
&\begin{aligned}
    P_{\qs, \qsd}^{\qspp} &= \frac{2\pi}{\hbar} |V^{(3)}(\qs, \qsd, -\qspp)|^2 
    \delta_{\bm{q}{+}\bm{q'}{-}\bm{q''},\bm{G}} \\
    &\delta(\hbar\omega_{\qs}{+}\hbar\omega_{\qsd}{-}\hbar\omega_{\qspp})
    \bar{N}_{\qs}\bar{N}_{\qsd}(\bar{N}_{\qspp}{+}1),
\end{aligned} \\
&\begin{aligned}
    P_{\qs}^{\qsd,\qspp} &= \frac{2\pi}{\hbar} |V^{(3)}(\qs, -\qsd, -\qspp)|^2 
    \delta_{\bm{q}{+}\bm{q'}{-}\bm{q''},\bm{G}} \\
    &\delta(\hbar\omega_{\qs}{-}\hbar\omega_{\qsd}{-}\hbar\omega_{\qspp})
    \bar{N}_{\qs}(\bar{N}_{\qsd}+1)(\bar{N}_{\qspp}{+}1).
     \raisetag{10mm}
\end{aligned}
\end{align}
The phonon-phonon matrix element $V^{(3)}$ is the Fourier transform of the third derivative of the total crystal energy per unit cell $\varepsilon^{\rm cell}$ with respect to the atomic displacements \cite{fugallo2013}.

Finally, we include phonon-isotope scattering, whose matrix elements are given by \cite{fugallo2013}:
\begin{equation}
\begin{split}
    \mathcal{S}^{\rm ph-isot}_{\qs,\qsd} &=
    \frac{\pi}{2}\omega_{\qs}\omega_{\qsd}\left[ \bar{N}_{\qs}\bar{N}_{\qsd} + \frac{1}{2}(\bar{N}_{\qs} + \bar{N}_{\qsd}) \right] \hspace*{10mm}\\
    &\times \sum_{m} g_{m}^{\rm isot} 
    \left| \sum_{\alpha} z_{\qsd}^{m\alpha*} \cdot 
    z_{\qs}^{m\alpha}\right|^2
    \delta(\omega_{\qs} - \omega_{\qsd}).
\end{split}
 \raisetag{12mm}
\end{equation}

\section{Deriving the Viscous Thermoelectric Equations}
\label{app:vte}

In \cref{sec:conductivities_viscosities_and_cross_transport_coefficients}, we derived transport coefficients parameterizing the heat, charge, and momentum fluxes. Here, we show the details of the derivation of Viscous Thermoelectric Equations, a set of mesoscopic partial differential equations describing the flow of (quasi-)conserved quantities.

In the hydrodynamic regime, where most of the collisions conserve the crystal momentum, we decompose the full scattering matrix into the normal and Umklapp components, as in \cref{sec:coupled_basis}: $\Omega_{\kmqs,\kmpqsp}=\Omega^N_{\kmqs,\kmpqsp}+\Omega^U_{\kmqs,\kmpqsp}$. Then the electron-phonon epBTE, \cref{eq:BTE_coupled}, becomes:
\begin{equation}
\bm{v}_{\kmqs}\cdot\Nabla_{\bm{r}} \z=
-
\sum_{\kmpqsp}\left(\Omega^N_{\kmqs,\kmpqsp}+\Omega^U_{\kmqs,\kmpqsp}\right)\spert{}{\kmpqsp}.
\label{eq:lbte_coupled_NU}
\end{equation}

We expand $\z$ as in \cref{eq:deviation_expansion}. Then, after noting that all the local-equilibrium terms are proportional to special eigenvectors that are all null eigenvectors of the normal scattering matrix, we have that \cref{eq:lbte_coupled_NU} becomes:
\begin{equation}
\begin{split}
\bm{v}_{\kmqs}{\cdot}\nabla_{\bm{r}} 
\z
=&-
\sum_{\kmpqsp}
\Omega^N_{\kmqs,\kmpqsp}
    \spert{\delta}{\kmpqsp}\\
&-
\sum_{\kmpqsp}
\Omega^U_{\kmqs,\kmpqsp}
    \left( 
    \spert{\ue}{\kmpqsp}
    +\spert{\up}{\kmpqsp}
    +\spert{\delta}{\kmpqsp}
    \right).
\end{split}
\raisetag{22mm}
\label{eq:lbte_coupled_NU_long}
\end{equation}
To derive a set of mesoscopic viscous equations describing transport, we take the scalar product of \cref{eq:lbte_coupled_NU_long} with the 8 special eigenvectors (assuming three spatial dimensions). As a result we obtain four coupled equations describing the evolution of local chemical potential (which is related to the electrical potential), local temperature, and local electron and phonon drift velocities (these last two are equations for vector fields, with three equations each).

\subsection{Projection in the charge subspace} 
\label{sub:projection_in_the_charge_subspace}

Taking the scalar product of \cref{eq:lbte_coupled_NU_long} with 
$\leftrelaxon{\star}$, and recalling the second line of \cref{eq:charge_flux}, we obtain:
\begin{equation}
\begin{aligned}
\braket{\theta^{\star}_{\kmqs} | \bm{v}_{\kmqs} \cdot \Nabla | z_{\kmqs}}
={}& \Nabla \cdot 
\braket{\theta^{\star}_{\kmqs} | \bm{v}_{\kmqs} | z_{\kmqs}}\\
={}& \left(k_B\bar{T}\mathfrak{U}\right)^{-1/2} \Nabla \cdot \bm{J} \\
={}& \braket{\theta^{\star}_{\kmqs} | \Omega_{\kmqs,\kmpqsp} | z_{\kmqs}}\\
={}& 0.
\end{aligned}
\label{eq:continuity_start}
\end{equation}
In the last line, the scalar product vanishes because $\relaxon{\star}$ is a zero eigenvector of $\Omega$. We can now scale all terms by $\sqrt{k_B\bar{T}\mathfrak{U}}$ and insert the form for $\bm{J}$ found in \cref{eq:charge_flux}, obtaining:
\begin{equation}
- \psi^j_i   \frac{\partial u^i_{\rm e}}{\partial r^j}
- \sigma_D^{ij}  \frac{\partial^2 \V}{\partial r^i\partial r^j}
- [\sigma_D S_D]^{ij}  \frac{\partial^2 T}{\partial r^i\partial r^j}
=0.
\label{eq:continuity_charge}
\end{equation}
The physical content of \cref{eq:continuity_charge} is that of a continuity equation for charge. We can rewrite the equation as $\Nabla \cdot \bm{J}=0$, where $\bm{J}$ is defined in \cref{eq:charge_flux}.  

\subsection{Projection in the energy subspace} 
\label{sub:projection_in_the_energy_subspace}

Similarly, we can take the scalar product of \cref{eq:lbte_coupled_NU_long} with 
$\sqrt{C_{\rm tot} k_B \bar T^2}\leftrelaxon{0}$, obtaining:
\begin{equation}
\begin{aligned}
    \sqrt{C_{\rm tot} k_B \bar T^2} &
    \braket{\theta^{0}_{\kmqs} | \bm{v}_{\kmqs} \cdot \Nabla | z_{\kmqs}} \\
={}&
    \Nabla \cdot 
    \left(
        \sqrt{C_{\rm tot} k_B \bar T^2}
        \braket{\theta^{0}_{\kmqs} | \bm{v}_{\kmqs} | z_{\kmqs}}
    \right) \\
={}&
    \Nabla \cdot \bm{Q} \\
={}&
    \sqrt{C_{\rm tot} k_B \bar T^2}
    \braket{\theta^{0}_{\kmqs} | \Omega_{\kmqs,\kmpqsp} | z_{\kmqs}}\\
={}& 0.
\end{aligned}
\label{eq:continuity_energy}
\end{equation}
In going from the second to the third line we used \cref{eq:energy_flux}, and as for charge conservation, the right-hand side vanishes because the energy eigenvector is a zero eigenvector of $\Omega$. 
Now we can insert the form for $\bm{Q}$ given in \cref{eq:energy_flux} to find:
\begin{equation}
\begin{split}
\chi^{j,\rm e}_i
\frac{\partial u^i_{\rm e}}{\partial r^j} +
\chi^{j,\rm p}_i
\frac{\partial u^i_{\rm p}}{\partial r^j}
- \alpha_D^{ij} \frac{\partial^2 \V}{\partial r^i\partial r^j}
- \bar{\kappa}_D^{ij} \frac{\partial^2 T} {\partial r^i \partial r^j}
= 0.
\end{split}
\label{eq:continuity_energy_2}
\end{equation}

The physical meaning of \cref{eq:continuity_energy_2} is that of a continuity equation for energy. Using \cref{eq:energy_flux}, it can be rewritten as $\Nabla \cdot \bm{Q}=0$.

\subsection{Projection in the momentum subspace} 
\label{sub:projection_in_the_momentum_subspace}

\subsubsection{Momentum in the electronic subspace} 
\label{ssub:momentum_in_the_electronic_subspace}

To describe the evolution of the electron momentum, we take the scalar product of \cref{eq:lbte_coupled_NU_long} with 
$\normalbra{\rm e}{i}$ ($i=1,2,3$), finding:
\begin{equation}
\begin{aligned}
& \Braket{\phi^{\rm e}_{i,\kmqs} | v^j_{\kmqs} \cdot \frac{\partial}{\partial r^j} | z_{\kmqs}} \\
=& \frac{1}{\sqrt{k_B \bar{T} \elmom{i}}} \frac{\partial \Pi^{j,\rm e}_i}{\partial r^j}  \\
=&
    -\normalbra{\rm e}{i} \Omega^U_{\kmqs,\kmpqsp}
    \left(
    \spert{\ue}{\kmpqsp}
    + \spert{\up}{\kmpqsp}
    + \spert{\delta,O}{\kmpqsp}
    \right) \\
\approx&
    -\normalbra{\rm e}{i} \Omega^U_{\kmqs,\kmpqsp}
    \left(
    \spert{\ue}{\kmpqsp}
    + \spert{\up}{\kmpqsp}
    \right) \\
=&
    - \sqrt{\frac{\elmom{j}}{k_B\bar{T}}}
    \braket{\phi^{\rm e}_{i,\kmqs} |
    \Omega_{\kmqs,\kmpqsp}
    | \phi^{e}_{j,\kmpqsp}} u_{\rm e}^j (\bm{r}) \\
    &- \sqrt{\frac{\phmom{j}}{k_B\bar{T}}}
    \braket{\phi^{\rm e}_{i,\kmqs} |
    \Omega_{\kmqs,\kmpqsp}
    | \phi^{p}_{j,\kmpqsp}} u_{\rm p}^j (\bm{r}).
\end{aligned}
\end{equation}
As before, we replace the left-hand side with the divergence of the appropriate flux, and scale both sides with $\sqrt{k_B\bar{T}\elmom{i}}$. Here that is the electron momentum flux, as defined in \cref{eq:electron_momentum_flux}. However, on the right-hand side we are left with the $\Omega^U$ term, since $\normalket{\rm e}{i}$ are not eigenvectors of the full scattering matrix. However, assuming that $\Omega^U$ is a small perturbation to $\Omega^N$, we keep only the leading order terms within the momentum subspace, as described in \cref{sec:coupled_basis}, defining the following dissipation tensors:
\begin{align}
    D^{\rm ee}_{ij} &= \sqrt{\elmom{i}\elmom{j}} \braket{\phi^{\rm e}_{i,\kmqs} | \Omega_{\kmqs,\kmpqsp} | \phi^{e}_{j,\kmpqsp}}, \\
    D^{\rm ep}_{ij} &= \sqrt{\elmom{i}\phmom{j}} \braket{\phi^{\rm e}_{i,\kmqs} | \Omega_{\kmqs,\kmpqsp} | \phi^{p}_{j,\kmpqsp}}.
\end{align}
Finally, we can expand the electron momentum flux as in \cref{eq:electron_momentum_flux}, obtaining:
\begin{equation}
\begin{gathered}
    \chi^{j,\rm e}_i \frac{\partial T}{\partial r^j}
    - \psi^j_i \frac{\partial \V}{\partial r^j}
    - \eta^{jl,\rm ee}_{ik} \frac{\partial u_{\rm e}^k}{\partial r^j \partial r^l}
    - \eta^{jl,\rm ep}_{ik} \frac{\partial u_{\rm p}^k}{\partial r^j \partial r^l} \\
    = - D^{\rm ee}_{ij} u^j_{\rm e} - D^{\rm ep}_{ij}u_{\rm p}^j.
\end{gathered}
\end{equation}

\subsubsection{Momentum in the phonon subspace} 
\label{ssub:momentum_in_the_ph_subspace}

Finally, we project \cref{eq:lbte_coupled_NU_long} onto the phonon momentum eigenvectors and follow the same steps as for electron momentum. Specifically, using \cref{eq:phonon_momentum_flux}, we find:
\begin{equation}
\begin{aligned}
& \Braket{\phi^{\rm p}_{i,\kmqs} | v^j_{\kmqs} \cdot \frac{\partial}{\partial r^j} | z_{\kmqs}} \\
=& \frac{1}{\sqrt{k_B \bar{T} \phmom{i}}} \frac{\partial \Pi^{j,\rm p}_i}{\partial r^j}  \\
=&
    -\normalbra{\rm p}{i} \Omega^U_{\kmqs,\kmpqsp}
    \left(
    \spert{\ue}{\kmpqsp}
    + \spert{\up}{\kmpqsp}
    + \spert{\delta,O}{\kmpqsp}
    \right) \\
\approx&
    -\normalbra{\rm p}{i} \Omega^U_{\kmqs,\kmpqsp}
    \left(
    \spert{\ue}{\kmpqsp}
    + \spert{\up}{\kmpqsp}
    \right) \\
=&
    - \sqrt{\frac{\phmom{j}}{k_B\bar{T}}}
    \braket{\phi^{\rm p}_{i,\kmqs} |
    \Omega_{\kmqs,\kmpqsp}
    | \phi^{p}_{j,\kmpqsp}} u_{\rm p}^j (\bm{r}) \\
    &- \sqrt{\frac{\elmom{j}}{k_B\bar{T}}}
    \braket{\phi^{\rm p}_{i,\kmqs} |
    \Omega_{\kmqs,\kmpqsp}
    | \phi^{e}_{j,\kmpqsp}} u_{\rm e}^j (\bm{r}).
\end{aligned}
\end{equation}
As for the electron momentum flux, we define phonon-phonon and phonon-electron momentum dissipation rate tensors:
\begin{align}
 D^{\rm pp}_{ij} &= \sqrt{\phmom{i}\phmom{j}} \braket{\phi^{\rm p}_{i,\kmqs} | \Omega_{\kmqs,\kmpqsp} | \phi^{p}_{j,\kmpqsp}}, \\
 D^{\rm pe}_{ij} &= \sqrt{\phmom{i}\elmom{j}} \braket{\phi^{\rm p}_{i,\kmqs} | \Omega_{\kmqs,\kmpqsp} | \phi^{e}_{j,\kmpqsp}}.
\end{align} 
Using \cref{eq:phonon_momentum_flux}, we express the phonon momentum equation in terms of conjugate variables, obtaining the continuity equation for momentum in the phonon subspace:
\begin{equation}
    \chi^{j,\rm p}_i \frac{\partial T}{\partial r^j}
    - \eta^{jl,\rm pp}_{ik} \frac{\partial u_{\rm p}^k}{\partial r^j \partial r^l}
    - \eta^{jl,\rm pe}_{ik}\frac{\partial u_{\rm e}^k}{\partial r^j \partial r^l}
    = - D^{\rm pp}_{ij} u^j_{\rm p} - D^{\rm pe}_{ij} u^j_{\rm e}.
\end{equation}

\section{The limit of diffusive thermoelectric transport}
\label{app:diff_limit}

We start by recalling that in the main text we discussed how momentum eigenvectors, $\normalket[ ]{e}{}$ and $\normalket[ ]{p}{}$, have long lifetimes in the hydrodynamic regime, and thus lead to deviations from diffusion. We derived the VTE relying on the hypothesis that these six modes have long lifetimes and using the Schrieffer-Wolff formalism\cite{bravyiSchriefferWolffTransformation2011,dragasevic2023}. Here, we show that, by defining the coefficients entering the VTE and separating the momentum and diffusion-damped contributions as in \cite{dragasevic2023}, the VTE remain valid also in the diffusive regime. 

\subsection{Simplification of the VTE in bulk materials}

In \cref{ssec:dte_limit} we outlined how the VTE have the same form as the DTE in the bulk limit where viscous effects are negligible.
Here we complete the proof by considering the effect of $D^{\rm pe}$ and $D^{\rm ep}$ and by showing that the thermoelectric fluxes predicted by the VTE are the same as those predicted by the DTE with transport coefficients determined by the standard linear-response solution of the BTE \cite{phoebe}.
In the VTE, \cref{eq:VTE}, hydrodynamic corrections are prominent at lengthscales comparable to Gurzhi's length $l_G^2 = \eta/D$, which is the lengthscale over which the drift velocities reach their bulk value. When the sample is uniform and much larger than this lengthscale, the viscous terms in \cref{eq:VTE_ph_mom,eq:VTE_el_mom} become negligible and those two vector equations reduce to:
\begin{equation}
    - \left[\begin{array}{cc}
        D^{\rm ee}_{ij} & D^{\rm ep}_{ij} \\ 
        D^{\rm pe}_{ij} & D^{\rm pp}_{ij}
    \end{array} \right]
    \left[\begin{array}{c}
        u^j_{\rm e} \\
        u^j_{\rm p}
    \end{array} \right] = 
    \left[\begin{array}{c}
        \chi^{j,\rm e}_i \frac{\partial T}{\partial r^j}
        - \psi^j_i \frac{\partial \V}{\partial r^j} \\
        \chi^{j,\rm p}_i \frac{\partial T}{\partial r^j}
    \end{array} \right].
    \label{eq:diff_limit_drift_velocities}
\end{equation}
By rewriting these equations in terms of the microscopic perturbation from equilibrium $\ket{z}$, we can show that they correspond exactly to the solution of the BTE for momentum modes in the diffusive regime. Recalling \cref{eq:deviation_expansion}, we see that the drift velocities are exactly the coefficients that determine the deviation from equilibrium in the momentum subspace:
\begin{equation}
\begin{aligned}
    P^M_{\kmqs,\kmpqsp} \ket{z_{\kmpqsp}} = \sum_{i=1}^3
        &\sqrt{\frac{\elmom{i}}{k_B\bar{T}}} \normalket{\rm e}{i} u_{\rm e}^i \\
        &+ \sqrt{\frac{\phmom{i}}{k_B\bar{T}}} \normalket{\rm p}{i} u_{\rm p}^i,
\end{aligned}
\end{equation}
where we defined a projector operator for the momentum subspace:
\begin{gather}
    P^M_{\kmqs,\kmpqsp} = \sum_{i=1}^3 \normalket{\rm e}{i}\normalbra[,\kmpqsp]{\rm e}{i} + \normalket{\rm p}{i}\normalbra[,\kmpqsp]{\rm p}{i}.
\end{gather}
Similarly, we can recall the definitions of $D^{\rm c_1c_2}_{ij}$ [\cref{eq:mom_diss}], $\chi^{j,\rm c}_i$ and $\psi^j_i$ [\cref{tab:table_VTE}], and $\ket{z^T}$ and $\ket{z^V}$ [\cref{eq:deviation_expansion}]. Inserting these into \cref{eq:diff_limit_drift_velocities} above, we find:
\begin{equation}
\begin{split}
    \normalbra[,\kmqs]{\rm c}{i}& \Omega_{\kmqs,\kmpqsp} P^M_{\kmpqsp,^{\kmpp}_{\qspp}} \big|z_{^{\kmpp}_{\qspp}}\big>\\
        &= - \normalbra[,\kmqs]{\rm c}{i} v^j_{\kmqs}| \frac{\partial}{\partial r^j} \left(\big|z^T_{\kmqs} \big>+ \big|z^V_{\kmqs}\big>\right) \\
        &= - \normalbra[,\kmqs]{\rm c}{i} v^j_{\kmqs} |\frac{\partial \big| z^{\rm LE}_{\kmqs}\big>}{\partial r^j},
\end{split}
\end{equation}
where the six scalar equations are now found by iterating over $i=1,2,3$ and $\rm c=e,p$, and the second line follows from the odd parity of momentum modes. This can be further rewritten in compact matrix notation as:
\begin{equation}
    P^M \Omega P^M \ket{z} = - P^M \bm{v} \cdot \Nabla \ket{z^{\rm LE}}.
\end{equation}
We solve this equation the same way as the BTE, by writing down the spectrum of $P^M \Omega P^M$:
\begin{equation}
    P^M \Omega P^M = \sum_{m=1}^6 \frac{1}{\tau_m} \Ket{\widetilde{\phi}^m_{\kmqs}}\Bra{\widetilde{\phi}^m_{\kmpqsp}},
\end{equation}
This allows the solution in the momentum subspace to be written as:
\begin{equation}
    P^M_{\kmqs,\kmpqsp} \ket{z_{\kmpqsp}} = -\sum_{m = 1}^6 \Ket{\widetilde{\phi}^m_{\kmqs}} \tau_m \Braket{\widetilde{\phi}^m_{\kmpqsp} | \bm{v}_{\kmpqsp} \cdot \Nabla | z^{\rm LE}_{\kmpqsp}}.
    \label{eq:momentum_diffusive_response}
\end{equation}
This equation is the analogue of \cref{eq:diffusive_response} for the momentum subspace. Following the same steps that we performed to find $\bar{\kappa}_D, \alpha_D, \sigma_D,$ and $\sigma_DS_D$, we can find the bulk transport contributions from the momentum subspace:
\begin{subequations} \label{eq:momentum_contribution_bulk}
\begin{align}
    \bar{\kappa}^{ij}_M &=
        C_{\rm tot} \sum_{m} v^i_{0,m}v^j_{0,m} \tau_m, \\
    \alpha^{ij}_M &=
        - \sqrt{{C_{\rm tot}\mathfrak{U}}{\bar T}} \sum_{m} v^i_{0,m}v^j_{\star,m} \tau_m, \\
    \sigma^{ij}_M &=
        \mathfrak{U}\sum_{m} v^i_{\star,m}v^j_{\star,m} \tau_m, \\
    [\sigma_M S_M]^{ij} &= - \sqrt{\frac{C_{\rm tot} \mathfrak{U}}{\bar T}} \sum_{m} v^i_{\star,m}v^j_{0,m} \tau_m.
\end{align}
\end{subequations}
The notation $v^i_{0/\star,m} = \Braket{\theta^{0/\star}_{\kmqs} | v^i_{\kmqs} | \widetilde{\phi}^m_{\kmqs}}$ is a straightforward extension from the main text.
Note that these expressions have been derived directly from the VTE, without reference to the properties of the BTE.

We can now verify that the VTE are consistent with the DTE in the bulk. As explained in the main text, in the bulk, the heat and charge fluxes are composed of additive contributions from the momentum and the diffusion-damped subspaces. The total bulk transport coefficients are then found by adding together \cref{eq:thermal_K,eq:tildePeltier,eq:electrical_sigma,eq:sigma_S_relaxons} and \cref{eq:momentum_contribution_bulk}:
\begin{subequations} \label{eq:mom_and_diff_contributions}
\begin{align}
    \bar{\kappa}^{ij} &= \bar{\kappa}^{ij}_M + \bar{\kappa}^{ij}_D, \\
    \alpha^{ij} &= \alpha^{ij}_M + \alpha^{ij}_D, \\
    \sigma^{ij} &= \sigma^{ij}_M + \sigma^{ij}_D, \\
    [\sigma S]^{ij} &= [\sigma_M S_M]^{ij} + [\sigma_D S_D]^{ij}.
\end{align}
\end{subequations}
\Cref{eq:mom_and_diff_contributions} generalizes \cref{eq:onsager_diffusive_limit} by including the contributions from $D^{\rm pe}$ and $D^{\rm ep}$.
When the Schrieffer-Wolff framework can be applied, i.e., the fast normal-process timescale can be decoupled from the slow Umklapp momentum-dissipation timescale, the momentum modes $\Ket{\widetilde{\phi}^m_{\kmqs}}$ can be considered  eigeivectors of $\Omega$ \cite{dragasevic2023}. Therefore, the expressions for total transport coefficients simplify to sums over all relaxons:
\begin{subequations}
\begin{align}
    \bar{\kappa}^{ij} &=
        C_{\rm tot} \sum_{\xi} v^i_{0,\xi}v^j_{0,\xi} \tau_\xi, \label{eq:bulk_kappa} \\
    \alpha^{ij} &=
        - \sqrt{{C_{\rm tot}\mathfrak{U}}{\bar T}} \sum_{\xi} v^i_{0,\xi}v^j_{\star,\xi} \tau_\xi, \\
    \sigma^{ij} &=
        \mathfrak{U}\sum_{\xi} v^i_{\star,\xi}v^j_{\star,\xi} \tau_\xi, \\
    [\sigma S]^{ij} &= - \sqrt{\frac{C_{\rm tot} \mathfrak{U}}{\bar T}} \sum_{\xi} v^i_{\star,\xi}v^j_{0,\xi} \tau_\xi. \label{eq:bulk_sigmaS}
\end{align}
\end{subequations}
These are exactly the expressions found when one solves the BTE in the diffusive regime \cite{phoebe}. As expected, the VTE and the DTE agree exactly when quasi-degenerate perturbation theory (QDPT) within the Schrieffer-Wolff formalism is applicable.

\subsection{Bulk VTE in the diffusive regime}
However, we can improve the VTE if we use \cref{eq:mom_and_diff_contributions} as the \textit{definition} of diffusion-damped coefficients:
\begin{subequations}
\label{eq:diffusive_components}
\begin{align}
    \bar{\kappa}^{ij}_D &\equiv \bar{\kappa}^{ij} - \bar{\kappa}^{ij}_M, \\
    \alpha^{ij}_D &\equiv \alpha^{ij} - \alpha^{ij}_M, \\
    \sigma^{ij}_D &\equiv \sigma^{ij} - \sigma^{ij}_M, \\
    [\sigma_D S_D]^{ij} &\equiv [\sigma S]^{ij} - [\sigma_M S_M]^{ij}.
\end{align}
\end{subequations}
From the discussion above, it should be clear that, when QDPT is valid, this definition coincides with the definitions given in the main text, \cref{eq:thermal_K,eq:tildePeltier,eq:sigma_S_relaxons,eq:electrical_sigma}. With this definition, the bulk solution of the VTE will always quantitatively agree with the DTE, regardless of whether we are in the hydrodynamic regime (where QDPT is valid) or in the diffusive regime (where it is not).

\subsection{Single fluid models as limits of the VTE}
Finally, we consider what happens when there are fewer than six momentum eigenvectors. In the main text, we assumed that the normal scattering matrix $\Omega^N$ has six null momentum eigenvectors, corresponding to separate electron momentum and phonon momentum conservation laws. However, there are three important cases in which this assumption is not valid. The first two cases correspond to only electron momentum or only phonon momentum being conserved, and the third is when only total momentum is conserved. In each of these cases, the normal scattering matrix will only have three null eigenvectors. We will now argue that the VTE are equally valid in these cases as when there are six null eigenvectors of $\Omega^N$.

If an eigenvector $\ket{\theta^m}$ of $\Omega$ lies in the momentum subspace, then $P^M \ket{\theta^m} = \ket{\theta^m}$ and $\ket{\theta^m}$ will also be an eigenvector of $P^M \Omega P^M$. For these eigenvectors, the derivation in the main text applies and the VTE correctly predict their relaxation time and viscosity. For eigenvectors $\Ket{\widetilde{\phi}^m}$ of $P^M \Omega P^M$ that are not eigenvectors of $\Omega$, we can assume that they are not related to null eigenvectors of $\Omega^N$; otherwise, they would be well described by QDPT and also be eigenvectors of $\Omega$. In the absence of any (quasi-)conservation laws, we assume that the relaxation times of these eigenvectors $\Ket{\widetilde{\phi}^m}$ are comparable to scattering times of individual electrons and phonons. In this case, the length scale associated with $\Ket{\widetilde{\phi}^m}$ will also be comparable to mean free paths of electrons and phonons and certainly much smaller than the scale of the device, hence causing no hydrodynamic behaviour. These modes will be well described by the bulk limit we discussed above, and their contribution to \cref{eq:VTE_ph_mom,eq:VTE_el_mom} will, by definition, cancel with corresponding terms in \cref{eq:diffusive_components}.

Therefore, having a momentum subspace with a dimension larger than the actual number of hydrodynamic modes introduces no error into the prediction of the heat and charge fluxes. It does introduce spurious momentum balance equations, whose validity is dependent on QDPT being applicable to the corresponding momentum modes. However, the breakdown of QDPT occurs exactly when momentum is no longer well conserved, in which case it is not possible to define a momentum balance equation \cite{dragasevic2023}.

We have thus shown that the VTE will remain valid as long as all hydrodynamic effects can be related to relaxons in the momentum subspace. By carefully defining the coefficients parameterizing the VTE, we have ensured their validity from the hydrodynamic to the diffusive regime, and the choice of modes that form the momentum subspace ensures validity in single-fluid limiting cases.

\section{Finite size effects on transport coefficients}
\label{app:finite_size}

In devices featuring a smallest characteristic size $L_S$ much larger than the mean free path of electrons and phonons, the transport coefficients are given by the bulk expressions derived in \cref{sec:conductivities_viscosities_and_cross_transport_coefficients}. 
Conversely, in the opposite limit---when $L_S$ is much smaller than the carrier mean free path---the transport coefficients are determined by extrinsic boundary scattering, often described using the expression proposed by Casimir \cite{ziman1960}. When the two lengthscales become comparable, one has to account for the interplay between these two scattering mechanisms. Here, we discuss an approximate approach to account for finite-size (ballistic) corrections to the transport coefficients. The approach is discussed in Ref.~\cite{simoncelli2020} and is formally analogous to the Bosanquet-type regularization used to describe corrections to fluids' transport coefficients in the transition regime \cite{michalis_rarefaction_2010} (corresponding to 0.1 $\lesssim {\rm Knudsen\ number}\lesssim 10$, i.e. to mean free path of the fluid molecule of the order of the characteristic length of the device).

\subsection{Rank-two thermoelectric coefficients}

In the ballistic limit, the scattering matrix is dominated by boundary scattering, for which the scattering matrix elements are given by $\Omega^{\rm boundary}_{\kmqs,\kmpqsp} = |\bm{v}_{\kmqs}|/L_S \delta_{\kmqs,\kmpqsp}$. Accounting only for boundary scattering, we find that the thermoelectric transport coefficients are:
\begin{subequations}
\label{eq:ballistic_limit}
\begin{align}
    \sigma^{ij}_{\rm ball} &= \frac{\spindeg}{\Nk\mathcal{V}} \sum_{\km} \frac{e^2}{k_B\bar{T}} \bar{F}_{\km}(1-\bar{F}_{\km}) \frac{v^i_{\km}v^j_{\km}}{|\bm{v}_{\km}|} L_S, \\
    [\sigma S]^{ij}_{\rm ball} &= - \frac{\spindeg}{\Nk\mathcal{V}} \sum_{\km} \frac{e(\varepsilon_{\km}{-}\bar\mu)}{k_B \bar{T}^2} \bar{F}_{\km}(1-\bar{F}_{\km}) \frac{v^i_{\km}v^j_{\km}}{|\bm{v}_{\km}|} L_S, \\
    \bar{\kappa}_{\rm ball}^{ij, \rm ee} &= \frac{\spindeg}{\Nk \mathcal{V}} \sum_{\km} \frac{(\varepsilon_{\km}{-}\bar\mu)^2}{k_B\bar T^2} \bar{F}_{\km}(1{-}\bar{F}_{\km}) \frac{v^i_{\km}v^j_{\km}}{|\bm{v}_{\km}|} L_S, \label{eq:k_el_FS} \\
    \bar{\kappa}_{\rm ball}^{ij, \rm pp} &= \frac{1}{\Nq \mathcal{V}} \sum_{\qs} \frac{(\hbar\omega_{\qs})^2}{k_B\bar T^2} \bar{N}_{\qs}(1+\bar{N}_{\qs})\frac{v^i_{\qs}v^j_{\qs}}{|\bm{v}_{\qs}|} L_S. \label{eq:k_ph_FS}
\end{align}
\end{subequations}
To simplify the discussion, we assume that transport is isotropic, which holds for the basal plane of graphite and is thus sufficient for the scope of the present work. 
Isotropic rank-two thermoelectric transport coefficients are proportional to the identity matrix, e.g. $\sigma^{ij} = \sigma\delta^{ij}$. Starting from the electrical conductivity, we compute the effective conductivity using the common Bosanquet-type regularization \cite{michalis_rarefaction_2010,dragasevic2023}:
\begin{equation}
    \frac{1}{\sigma_{\rm eff}} = \frac{1}{\sigma} + \frac{1}{\sigma_{\rm ball}}.
\end{equation}
This expression smoothly interpolates between the two limits, giving an effective conductivity that is always lower than the bulk limit due to boundary scattering. It also tends to the bulk limit for large sample sizes, while for small sample sizes we recover the ballistic conductivity. The effective electrical conductivity also remains isotropic, as desired.
We note that the expression above has been reported for the standard diffusive coefficient $\sigma$. To correct the diffusion-damped coefficient $\sigma_D$ (as well as the other transport coefficients, $\bar{\kappa}_D$, $S_D$, and $\alpha_D$) appearing in the VTE, which is related to the standard $\sigma$ via \cref{eq:diff_sigma} (see \cref{eq:diff_kappa,eq:diff_S,eq:diff_alpha} for the other coefficients), we assume that the ratio between the ballistic and bulk-limit values is the same for $\sigma_D$ and $\sigma$.
In formulas, we compute $\sigma_{D,\rm ball}=\sigma_{\rm ball}\frac{\sigma_D }{\sigma}$ and then we combine it with $\sigma_D$ using the same Bosanquet-type combination:
\begin{equation}
    \frac{1}{\sigma_{D,\rm eff}} = \frac{1}{\sigma_D} + \frac{1}{\sigma_{D,\rm ball}}.
\end{equation}
The electrical conductivity captures the response of electrons to the electric field. In contrast, for thermal conductivity, a temperature change perturbs both electrons and phonons, and both kinds of carriers constitute the heat flux. This distinction is relevant for the finite-size correction in graphite because the bulk thermal conductivity is dominated by phonons, while the ballistic thermal conductivity is dominated by electrons, $\bar{\kappa}^{\rm ee}_{\rm ball} \gg \bar{\kappa}^{\rm pp}_{\rm ball}$. It follows that finite-size corrections become relevant for phonons at much larger length scales than for electrons. We account for this by dividing the thermal conductivity tensor between electron and phonon parts, starting from \cref{eq:bulk_kappa}:
\begin{equation}
\begin{aligned}
    \bar\kappa^{ij}
        &= C_{\rm tot} \sum_{\xi} v^i_{0,\xi}v^j_{0,\xi} \tau_\xi \\
        &= C_{\rm tot} \sum_{\xi} (v^{i,\rm(e)}_{0,\xi} + v^{i,\rm(p)}_{0,\xi})(v^{j,\rm(e)}_{0,\xi} + v^{j,\rm(p)}_{0,\xi}) \tau_\xi \\
        &\begin{aligned}
            {}={} &C_{\rm tot} \sum_{\xi} v^{i,\rm(e)}_{0,\xi}v^{j,\rm(e)}_{0,\xi} \tau_\xi
                + C_{\rm tot} \sum_{\xi} v^{i,\rm(e)}_{0,\xi}v^{j,\rm(p)}_{0,\xi} \tau_\xi \\
                &+ C_{\rm tot} \sum_{\xi} v^{i,\rm(p)}_{0,\xi}v^{j,\rm(e)}_{0,\xi} \tau_\xi
                + C_{\rm tot} \sum_{\xi} v^{i,\rm(p)}_{0,\xi}v^{j,\rm(p)}_{0,\xi} \tau_\xi
        \end{aligned} \\
        &= \bar\kappa^{ij,\rm ee} + \bar\kappa^{ij,\rm ep} + \bar\kappa^{ij,\rm pe} + \bar\kappa^{ij,\rm pp},
\end{aligned}
\end{equation}
where we recalled the definitions of $\bm{v}^{\rm(e)}_{0,\xi}$ and $\bm{v}^{\rm(p)}_{0,\xi}$ from \cref{sec:elph_relaxons}.
An analogous decomposition can be performed on $\bar\kappa_D$, providing definitions for $\bar\kappa^{\rm ee}_D, \bar\kappa^{\rm ep}_D, \bar\kappa^{\rm pe}_D$, and $\bar\kappa^{\rm pp}_D$. The two ``cross'' components $\bar\kappa^{\rm ep}_D$ and $\bar\kappa^{\rm pe}_D$ are much smaller than both $\bar\kappa^{\rm ee,pp}_D$ and $\bar\kappa^{\rm ee,pp}_{\rm ball}$ (for length scales investigated in this work), and so we neglect them. 
The decomposition of $\bar\kappa$ into electron and phonon parts allows us to apply finite-size corrections to each component, following the same approach as for the electrical conductivity:
\begin{align}
    \frac{1}{\bar{\kappa}^{\rm ee}_{D,\rm eff}} &= \frac{1}{\bar{\kappa}^{\rm ee}_{D}} + \frac{1}{\bar{\kappa}^{\rm ee}_{D,\rm ball}} = \frac{1}{\bar{\kappa}^{\rm ee}_{D}} + \frac{1}{\bar{\kappa}^{\rm ee}_{\rm ball} \frac{\bar{\kappa}^{\rm ee}_D}{\bar{\kappa}^{\rm ee}}}, \\
    \frac{1}{\bar{\kappa}^{\rm pp}_{D,\rm eff}} &= \frac{1}{\bar{\kappa}^{\rm pp}_{D}} + \frac{1}{\bar{\kappa}^{\rm pp}_{D,\rm ball}} = \frac{1}{\bar{\kappa}^{\rm pp}_{D}} + \frac{1}{\bar{\kappa}^{\rm pp}_{\rm ball} \frac{\bar{\kappa}^{\rm pp}_D}{\bar{\kappa}^{\rm pp}}}, \\
    \bar{\kappa}_{D,\rm eff} &= \bar{\kappa}^{\rm ee}_{D,\rm eff} + \bar{\kappa}^{\rm pp}_{D,\rm eff}.
\end{align}
Defining $\bar{\kappa}_{D,\rm eff}$ in this way retains the correct limits, namely $\bar{\kappa}_{D, \rm ball} = \bar{\kappa}^{\rm ee}_{D, \rm ball} + \bar{\kappa}^{\rm pp}_{D, \rm ball}$ in the small sample limit and $\bar\kappa_D \approx \bar\kappa^{\rm ee}_D + \bar\kappa^{\rm pp}_D$ in the bulk limit.

Next, we consider the thermoelectric coefficient $[\sigma S]^{ij}$. Since this coefficient gives a contribution to the charge current, it explicitly depends only on the electron response. However, the electron response will be determined, through scattering, by both thermally driven electrons and thermally driven phonons. In this sense, the thermoelectric coefficient can also be decomposed into an electron and a phonon component:
\begin{equation}
\begin{aligned}
    [\sigma S]^{ij}
        &= - \sqrt{\frac{C_{\rm tot} \mathfrak{U}}{\bar T}} \sum_{\xi} v^i_{\star,\xi}v^j_{0,\xi} \tau_\xi \\
        &= - \sqrt{\frac{C_{\rm tot} \mathfrak{U}}{\bar T}} \sum_{\xi} v^i_{\star,\xi} (v^{j, \rm (e)}_{0,\xi} + v^{j,(\rm p)}_{0,\xi}) \tau_\xi \\
        &= - \sqrt{\frac{C_{\rm tot} \mathfrak{U}}{\bar T}} \sum_{\xi} v^i_{\star,\xi} v^{j, \rm (e)}_{0,\xi} \tau_\xi - \sqrt{\frac{C_{\rm tot} \mathfrak{U}}{\bar T}} \sum_{\xi} v^i_{\star,\xi} v^{j,(\rm p)}_{0,\xi} \tau_\xi \\
        &= [\sigma S]^{ij,\rm e} + [\sigma S]^{ij,\rm p},
\end{aligned} \raisetag{20mm}
\end{equation}
and as for $\bar{\kappa}$, we can analogously define the corresponding electron and phonon components of $[\sigma_D S_D]^{ij}$. Under the isotropy assumption adopted above, these reduce to the scalar quantities $\sigma_D S^{\rm e}_D$ and $\sigma_D S^{\rm p}_D$.
Recalling the discussion in \cref{sec:elph_relaxons}, the phonon component is determined by electron-phonon drag. In the ballistic case, there is no drag and $[\sigma S]_{\rm ball}$ is determined by electrons only; therefore, $[\sigma S]_{\rm ball}$ cannot be used to estimate the effect of ballistic scattering on the phonon component of $\sigma S$. To overcome this issue, we take inspiration from the partially decoupled solution to the epBTE \cite{fiorentini2016}. In this approximation, the dependence of the phonon response on thermally driven electrons is neglected. This approximation is supported by phonon thermal conductivity being essentially unaffected by electrons, i.e. $\bar{\kappa}^{\rm pe} \ll \bar{\kappa}^{\rm pp}$. Then, both $\sigma S^{\rm p}$ and $\bar{\kappa}^{\rm pp}$ are fully determined by thermally driven phonons and are affected in approximately the same way by ballistic scattering; we can approximate the effective phonon component as $[\sigma_D S^{\rm p}_{D}]_{\rm eff} = \sigma_D S^{\rm p}_D \frac{\bar{\kappa}^{\rm pp}_{D,\rm eff}}{\bar{\kappa}^{\rm pp}_{D}}$. The total effective thermoelectric coefficient is written again as a sum between electron and phonon parts, where the electron part can be corrected the same way as $\sigma$:
\begin{align}
    \frac{1}{[\sigma_D S^{\rm e}_{D}]_{\rm eff}} &= \frac{1}{\sigma_D S^{\rm e}_{D}} + \frac{1}{[\sigma S]_{\rm ball}} \\
    [\sigma_D S_D]_{\rm eff} &= [\sigma_D S^{\rm p}_{D}]_{\rm eff} + [\sigma_D S^{\rm e}_{D}]_{\rm eff}
\end{align}
As before, $[\sigma_D S_D]_{\rm eff}$ becomes linear in $L_S$ at low $L_S$ (for $[\sigma_D S^{\rm p}_{D}]_{\rm eff}$, this dependence is because of $\bar{\kappa}^{\rm pp}_{D,\rm eff}$), and for large $L_S$ both the electron and phonon parts recover their respective bulk limits. Finally, for the Peltier coefficient we simply enforce the Kelvin-Onsager condition: $\alpha_{D,\rm eff} = [\sigma_D S_D]_{\rm eff} \bar T$.

\subsection{Rank-four viscosities}

In the small $L_S$ limit, where $\Omega^{\rm boundary}$ dominates, the electron and phonon viscosities are given by:
\begin{align}
    \eta^{jl, \rm ee}_{ik, \rm ball} &= \frac{\spindeg}{\Nk\mathcal{V}} \sum_{\km} \frac{\hbar^2 k_ik_k}{k_B\bar{T}} \bar{F}_{\km}(1-\bar{F}_{\km})
    \frac{v^j_{\km}v^l_{\km}}{|\bm{v}_{\km}|}L_S, \\
    \eta^{jl, \rm pp}_{ik, \rm ball} &= \frac{1}{\Nq\mathcal{V}} \sum_{\qs} \frac{\hbar^2 q_iq_k}{k_B\bar{T}} \bar{N}_{\qs}(1+\bar{N}_{\qs})
    \frac{v^j_{\qs}v^l_{\qs}}{|\bm{v}_{\qs}|} L_S.
\end{align}
When the viscosities assume an isotropic form, their tensors can have only four different non-zero components; for $\eta^{\rm ee}$, these are $\eta^{ii, \rm ee}_{ii},\ \eta^{ij, \rm ee}_{ij},\ \eta^{ji, \rm ee}_{ij}$, and $\eta^{jj, \rm ee}_{ii}$. Furthermore, these components satisfy the following relation $\eta^{ii, \rm ee}_{ii} = \eta^{ji, \rm ee}_{ij} + \eta^{jj, \rm ee}_{ii} + \eta^{ij, \rm ee}_{ij}$. We interpolate the three independent components in the same Matthiessen-like way used for thermoelectric coefficients:
\begin{align}
    \frac{1}{\eta^{ij, \rm ee}_{ij, \rm eff}} &= \frac{1}{\eta^{ij, \rm ee}_{ij}} + \frac{1}{\eta^{ij, \rm ee}_{ij, \rm ball}},
\end{align}
and analogously for $\eta^{ji, \rm ee}_{ij}$ and $\eta^{jj, \rm ee}_{ii}$, while for $\eta^{ii, \rm ee}_{ii}$ we enforce the isotropy condition: $\eta^{ii, \rm ee}_{ii, \rm eff} = \eta^{ij, \rm ee}_{ij, \rm eff} + \eta^{ji, \rm ee}_{ij, \rm eff} + \eta^{jj, \rm ee}_{ii, \rm eff}$. The properties discussed for the rank-two case hold also in the rank-four case; notably, the effective viscosity remains isotropic and tends to the ballistic and bulk values in the respective limits. The same procedure is repeated for $\eta^{\rm pp}_{\rm eff}$. Finally, we note that for drag viscosities the ballistic correction is zero, since $\Omega^{\rm boundary}$ is diagonal in the electron-phonon basis, while $\normalket[]{e}{i}$ and $\normalket[]{p}{j}$ lie in the electron and phonon subspace, respectively. The bulk drag viscosities are also, in graphite, much smaller than $\eta^{\rm ee}_{\rm eff}$ and $\eta^{\rm pp}_{\rm eff}$ for all $L_S$ used in this work, so they have negligible effect on the solutions of the VTE and we do not correct them for finite-size effects.

\section{Self-similarity of diffusive heat and charge transport}
\label{app:dte_similarity}

\begin{figure}[htbp]
    \centering
    \includegraphics[width=\columnwidth]{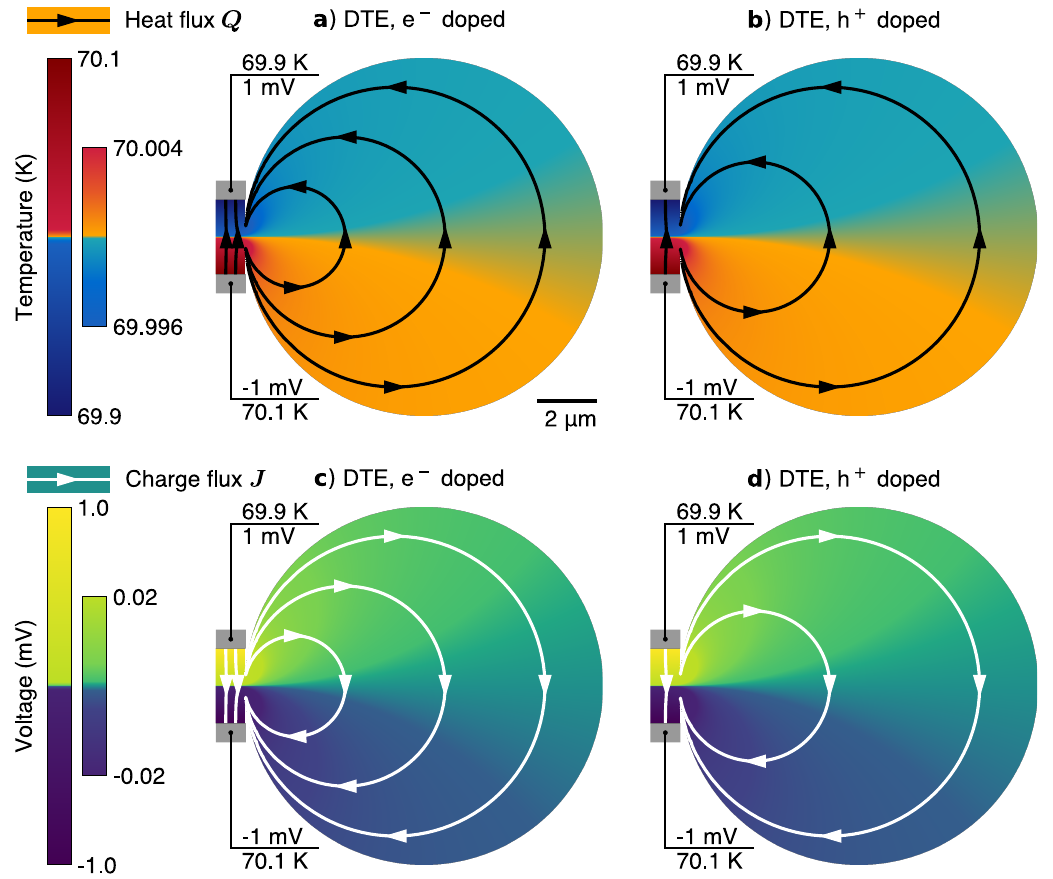}
    \caption{Solutions of the DTE in a tunnel-chamber geometry made of electron-doped graphite (left column, carrier concentration $-10^{20} \rm cm^{-3}$) and hole-doped graphite (right column, carrier concentration $+2.5\cdot10^{18} \rm cm^{-3}$).
    Upper row [panels (\textbf{a}), (\textbf{b})]: heat flux $\bm{Q}$ (streamlines) and temperature field $T$ (colormap) for electron and hole doping, respectively. Bottom row [panels (\textbf{c}), (\textbf{d})]: charge current $\bm{J}$ (streamlines) and voltage field $V$ (colormap) for electron and hole doping. The boundary conditions are identical in all four panels: a temperature gradient and a potential difference are applied to the tunnel ($T = 70\pm 0.1 \mathrm{\ K}$ and $V = \mp 1\mathrm{\ mV}$ at $y =\mp 1.25 \mathrm{\ \mu m}$), while the other boundaries are adiabatic and electrically insulating.
    The solutions for scalar fields $T$ and $V$ are identical, since harmonic functions are determined fully by the boundary conditions.
    Accordingly, the fluxes $\bm{Q}$ and $\bm{J}$ are equal in both cases, up to a uniform scaling determined by doping-dependent transport coefficients. This change is qualitatively shown by the reduced streamline density for the hole-doped case.
    The graphite sample is oriented so that the $x$ and $y$ axes of the device lie in the basal plane of graphite.}
    \label{fig:DTE_doping_effect}
\end{figure}

In \cref{fig:vte_vs_dte}, we highlighted the dependence of hydrodynamic transport on doping. Diffusive transport, governed by the DTE described in \cref{ssec:dte_limit}, is instead qualitatively insensitive to doping, as shown in \cref{fig:DTE_doping_effect} for the same doping levels used in \cref{fig:vte_vs_dte}. The mathematical similarity of the heat and charge flows emerging from the DTE solution follows directly from the form of the DTE.
Recalling \cref{eq:DTE}, it is straightforward to see that, for the homogeneous in-plane isotropic case considered here, the two equations imply $\nabla^2 T = \nabla^2 \V = 0$.
Hence, the equations for $T$ and $\V$ are independent of the transport coefficients. Since a change in doping influences the transport coefficients, it produces no change in the equations for $T$ and $\V$, and no change in their solutions either.
Note that the boundary conditions are set in terms of $T$ and $\V$ and are thus also doping-independent. Furthermore, since both variables satisfy Laplace's equation and their boundary conditions are related by a dimensional constant, the solutions for $T$ and $\V$ are proportional to each other. This similarity extends to the heat flux, $Q^i = - \bar{\kappa}^{ij} \partial_{r^j} T - \alpha^{ij} \partial_{r^j} \V$, and the charge flux, $J^i = - \sigma^{ij} \partial_{r^j} \V - [\sigma S]^{ij} \partial_{r^j} T$. The proportionality between $T$ and $\V$ implies that the streamlines for heat and charge flow are identical and independent of doping. The only dependence of the fluxes on doping is through the transport coefficients, which scale the fluxes by a position-independent factor.

\begin{figure*}[t!]
    \centering
    \includegraphics[width=\textwidth]{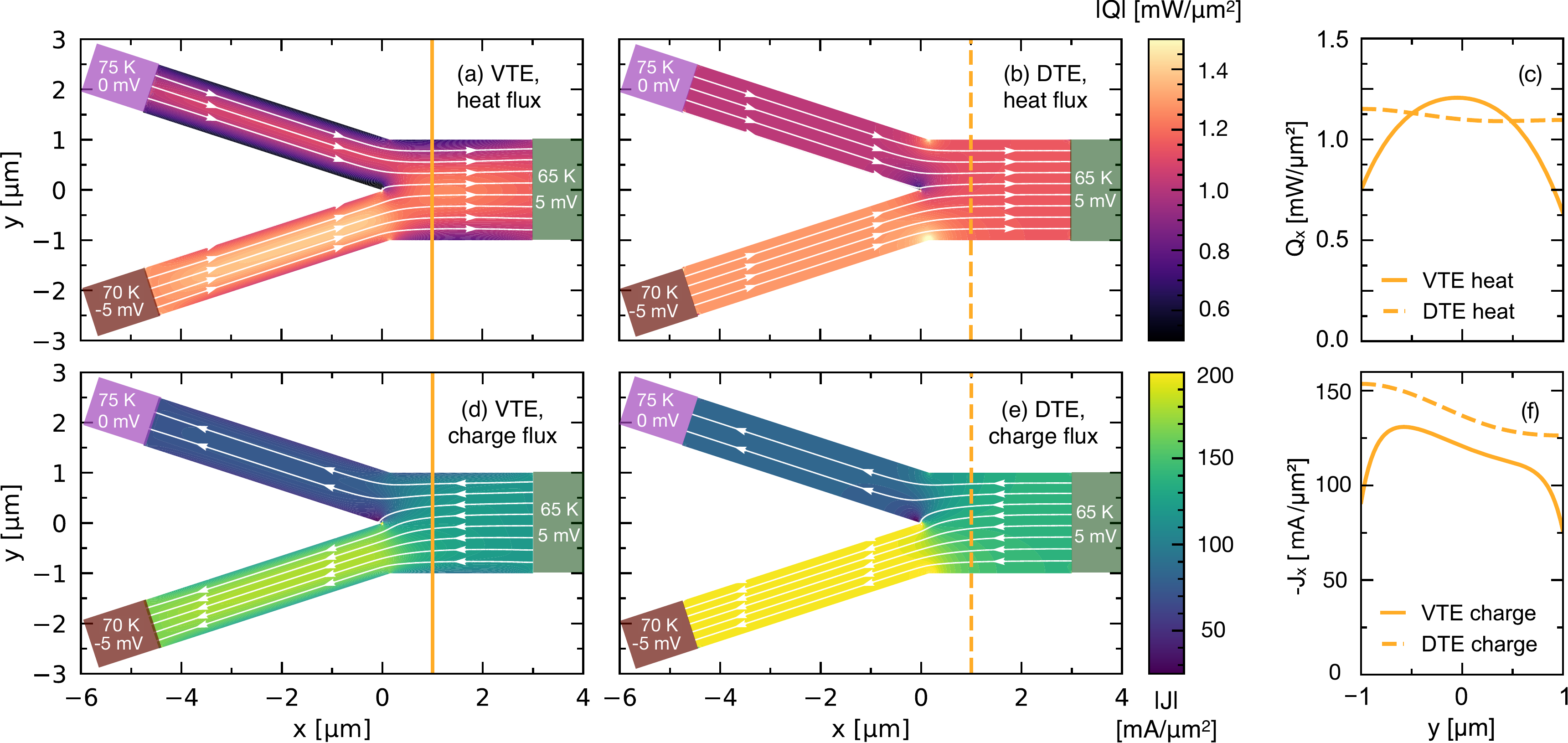}
    \caption{\textbf{Macroscopic signatures of viscous electron-phonon hydrodynamics} 
    in a fluid-mixing device made of highly electron-doped graphite ($-10^{20}$ cm$^{-3}$). Color bars indicate the heat [panels (\textbf{a}),(\textbf{b})] or charge [panels (\textbf{d}),(\textbf{e})] flux magnitude, for in-plane ($x,y$) components of the fluxes for a three dimensional device treated as bulk (infinitely long) in the $z$ direction. 
    Top row, heat flux predicted from (\textbf{a}) the VTE, or (\textbf{b}) DTE and (\textbf{c}) their comparison along the section $x = 1\ \mu m$. Bottom row, charge flux predicted from (\textbf{d}) the VTE, (\textbf{e}) DTE, and (\textbf{f}) their comparison along the section $x = 1\ \mu m$.
    Boundary conditions were applied with temperature (potential) set at 75 K (0 mV) on the upper thin terminal, 70 K ($-$5 mV) on the lower thin terminal, and at 65 K (5 mV) at the wide terminal. The other boundaries are assumed to be electrically insulating and adiabatic, and in the VTE case, electron and phonon drift velocities are zero on all boundaries (no-slip BCs).  
    }
    \label{fig:fluxes}
\end{figure*}

These self-similarity properties of the DTE are missing from the VTE, allowing for qualitatively different (i.e., not mathematically similar) heat and charge flow profiles. As discussed in \cref{ssec:tunnel_chamber}, in the hydrodynamic regime the streamline pattern, and in particular the presence or absence of vorticity, depends on how strong the drift-velocity component of the flow is relative to the diffusive component. This ratio is determined by the device geometry and the coefficients entering the VTE, and controls the appearance, or absence, of hydrodynamic transport signatures. In \cref{app:mixing_geometry}, we compare the flow profiles predicted by the DTE and the VTE in a mixing geometry and discuss the asymmetry between heat and charge flow in the hydrodynamic regime.

\section{Heat-charge mixing geometry}
\label{app:mixing_geometry}

In \cref{ssec:tunnel_chamber}, we showed that signatures of thermoelectric hydrodynamics can be established by comparing the solutions of the VTE and the DTE. Here, we extend our discussion to a geometry which illustrates coexistence, interaction, and mixing between electron and phonon fluids. We show that both fluids exhibit Poiseuille flow, and that they are governed by comparable but distinct length scales as they mix.

As previous works have studied devices containing phonon-only fluids \cite{simoncelli2020,raya-moreno_hydrodynamic_2022,dragasevic2023,lucente_vortices_2025}, or electron-only fluids \cite{bandurin2016,aharon-steinberg_direct_2022,levitov2016electron,palm_observation_2024}, the geometry should allow verification of both the known phonon-only and electron-only fluid limits while also illustrating what happens when the fluids are flowing together.
The two-fluid-mixing device made of highly doped graphite shown in \cref{fig:fluxes} is an ideal setup to investigate all these regimes. 
In this device we employ boundary conditions (BCs), described in the caption of \cref{fig:fluxes}, that drive a phonon-dominated hydrodynamic heat flow in the upper channel and a hydrodynamic electron flow in the lower channel, as well as the merging of the two flows at the intersection of the channels. This enables us to investigate how bi-component electron-phonon fluids emerge from mixing electronic and phononic fluids with non-negligible viscosity.

\begin{figure*}
    \centering
    \includegraphics[width=\textwidth]{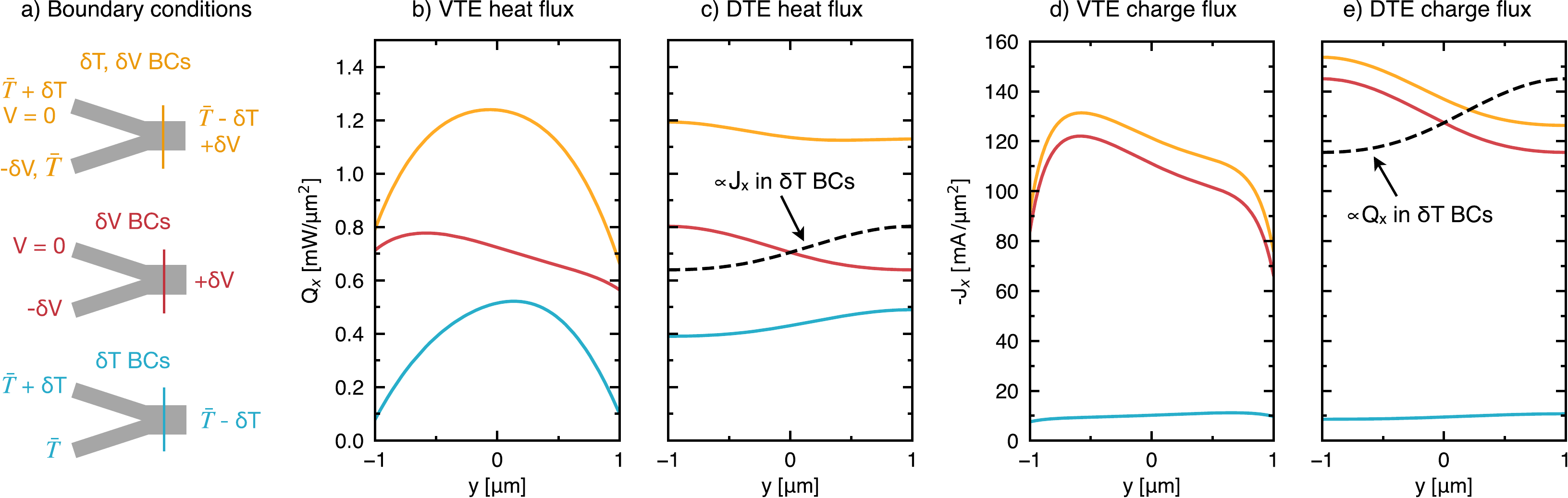}
    \caption{\textbf{Violation of the similarity between heat and charge flow in the thermoelectric hydrodynamic regime.}
    We show the horizontal heat [$Q^x(y)$] and charge [$J^x(y)$] flux along the section $x=1 \ \mathrm{\mu m}$ [vertical slice of the device, as shown in (\textbf{a})] predicted by the VTE [(\textbf{b}) and (\textbf{d})] or DTE [(\textbf{c}) and (\textbf{e})] in the presence of different boundary conditions (BCs), which are detailed in the schematic shown in (\textbf{a}). The flux for temperature difference-only BCs is blue [middle inset in (\textbf{a})], for potential difference-only BCs is red [bottom inset in (\textbf{a})], and for combined temperature and potential difference BCs is yellow [top inset in (\textbf{a})], as in \cref{fig:fluxes}. 
    In the VTE, the boundary conditions influence the shape of the non-diffusive flux profile. In contrast, in the DTE, the type of boundary conditions affects the magnitude but not the shape of the flux profile. 
    This is demonstrated by the dashed line in panels (\textbf{b}) and (\textbf{d}), which show that, after rescaling and mirroring with respect to $y$, the heat and charge fluxes obtained in the presence of temperature-only boundary conditions are mathematically similar to those obtained from the potential-only boundary conditions.}
    \label{fig:viscous_asymmetry}
\end{figure*}

The results of \cref{fig:fluxes}(\textbf{a}) and (\textbf{d}) show heat and charge flow in the chosen mixing device simulated using the VTE, with respective color maps demonstrating heat- and charge-flow profiles. Panel (\textbf{a}) demonstrates phonon-dominated hydrodynamic heat flow in the upper channel and (\textbf{d}) shows hydrodynamic charge flow in the lower channel, each displaying the characteristic Poiseuille profile, where the flow is stronger at the center of the channel and weaker at the boundaries. The hydrodynamic features observed in the VTE flows are absent in the DTE flows, shown in \cref{fig:fluxes}(\textbf{b}) and (\textbf{e}). Specifically, the heat and charge fluxes emerging from the DTE have a near-constant magnitude in the narrow channels.
In addition, we find that, in the upper narrow channel, the charge current is much smaller than in the lower one, as expected from the smaller potential difference across the upper channel. However, the heat flux in the lower channel is larger than in the upper one, despite the stronger temperature difference applied at the boundaries of the upper channel. This large heat flux in the lower channel is driven by the electric field through the Peltier coefficient $\alpha_D$ in \cref{eq:VHE_energy}.

Next, we consider the wide channel in which the electron and phonon flows mix. \cref{fig:fluxes}(\textbf{c}) and (\textbf{f}) show the heat and charge fluxes predicted by the VTE and the DTE in the wide channel along the section $x=1\ \mu m$. The sections of the heat and charge flows are plotted using a solid (VTE) or dashed (DTE) yellow line in (\textbf{c}) and (\textbf{f}), respectively, corresponding to the device section marked by the same lines in panels (\textbf{a}), (\textbf{d}) and (\textbf{b}), (\textbf{e}). In both (\textbf{c}) and (\textbf{f}), the viscous heat and charge currents are minimized at the boundaries, while the diffusive fluxes approach the boundaries with zero derivative \footnote{This formally follows from the DTE constraint that $T(\bm{r})$ and $\V(\bm{r})$ have to be harmonic functions, and also from the requirement that the boundaries have to be electrically insulating and adiabatic.}. However, in contrast to the symmetric behavior observed in the narrow channels, we see that in the wide channel the heat and charge current profiles are visibly different, with a more pronounced asymmetry in the charge flux than in the heat flux.
We note that in \cref{fig:fluxes}, finite-size (ballistic) corrections to the transport coefficients are taken into account following the approach discussed in \cref{app:finite_size}.
In the following subsection, we discuss how the electron and phonon fluids interact in the wide channel and the signatures they produce.

\subsection{Hydrodynamic violations of the similarity between heat and charge flows}

We now investigate how the heat and charge currents in \cref{fig:fluxes} depend on boundary conditions (BCs). In addition to the BCs shown in \cref{fig:fluxes} [top inset in \cref{fig:viscous_asymmetry}(\textbf{a})], we consider transport driven by temperature-difference BCs only [middle inset], or by effective-potential BCs only [bottom inset]. In all three cases we take $\bar T = 70 \rm K$, $\delta \V = 5 \rm mV$, $\delta T = 5 \rm K$, as in \cref{fig:fluxes}.

In \cref{fig:viscous_asymmetry} we show that the behavior predicted by the VTE violates the mathematical similarity property that relates the heat and charge fluxes emerging from the DTE. As discussed in \cref{app:dte_similarity}, the DTE imply that both temperature and effective potential are harmonic functions satisfying $\nabla^2 \V=0$ and $\nabla^2 T=0$. In particular, with $\delta \V$ BCs only, we have $T = \bar T$ everywhere and the effective potential fully determines the profile for both the heat flux ($\bm{Q} = {-}\alpha \nabla \V$), shown in red in panel (\textbf{c}), and the charge flux ($\bm{J} ={-}\sigma \nabla \V$), shown in red in (\textbf{e}). The resulting flux profiles are then related by a constant scaling, as can be verified from the figure. Analogously, with $\delta T$ BCs only, within the DTE the temperature fully determines the profile of both the heat flux ($\bm{Q}=-\bar{\kappa} \nabla T$), shown in blue in (\textbf{c}), and the charge flux ($\bm{J}=-\sigma S\nabla T$), shown in blue in (\textbf{e}). These fluxes will therefore also have profiles related by a constant scaling.

In addition, because the BCs used in the $\delta \V$- and $\delta T$-only cases are the same up to a mirroring of the $y$ coordinate [see \cref{fig:viscous_asymmetry}(\textbf{a})], the fluxes in these two cases are also related by a mirroring and a rescaling. We verify this by rescaling solutions of the DTE in the $\delta T$-only case and comparing them with DTE fluxes driven by the electric field. As a dashed black line in panel (\textbf{c}), we plot the charge flux driven by temperature gradient [blue line in (\textbf{e})] rescaled by $\alpha \delta \V / \sigma S \delta T$; we see that it is equivalent to the electric-field-driven heat flux up to a mirroring. Similarly, as a black dashed line in (\textbf{e}) we plot the temperature-driven heat flux scaled by $\sigma \delta \V / \bar \kappa \delta T$, seeing again that it matches the electric-field-driven charge flux up to a mirroring. We have thus verified that diffusive transport is qualitatively the same regardless of whether it is driven by a temperature difference or an electric field, as discussed in \cref{app:dte_similarity}.

In contrast, the heat and charge fluxes predicted by the VTE are qualitatively different in both $\delta \V$-only [red lines in panels (\textbf{b}),(\textbf{d})] and $\delta T$-only [blue lines in panels (\textbf{b}),(\textbf{d})] setups. In both cases, zero drift-velocity BCs are applied to all boundaries. This difference in the fluxes comes from the different length scales entering the VTE. The electron and phonon fluids each have an inherent Gurzhi length $l^{\rm e/p}_G = \sqrt{\eta^{\rm ee/pp}/D^{\rm ee/pp}}$ that determines the shape of the heat and charge flux profiles. The qualitative difference between heat-flux shapes in \cref{fig:viscous_asymmetry}(\textbf{b}) shows that the viscous flow is described by two distinct length scales, i.e. that the electron and phonon fluids flow independently and do not form a well mixed electron-phonon fluid. Imaging temperature- and electric-field-driven viscous transport can thus serve as a probe of different regimes of electron-phonon hydrodynamics.

In summary, along with Poiseuille flow---an established signature of hydrodynamic transport---the VTE also demonstrate the possible coexistence of electron and phonon fluids in strongly electron-doped graphite, whose heat and charge flow profiles violate the mathematical similarity that characterizes standard diffusive transport.

\section{Computational details}
\label{app:computational_details}

\subsection{DFT and Phoebe calculations}

To obtain the electronic and phononic band structures, as well as the electron-phonon and phonon-phonon couplings used in the construction of the scattering matrix, we used a 32$\times$32$\times$8 k-point grid, an 8$\times$8$\times$2 supercell, a pseudopotential generated with the Rappe-Rabe-Kaxiras-Joannopoulos (RRKJ) \cite{rappe1990optimized} method within the local density approximation \cite{dal2014pseudopotentials}, and a 70 Ry plane-wave energy cutoff. To compute electron-phonon matrix elements, we utilized the open-source density-functional theory (DFT) JDFTx code \cite{sundararaman2017jdftx}. For the generation of third-order phonon-phonon coupling matrix elements, we used the d3q density-functional perturbation theory (DFPT) module of Quantum ESPRESSO \cite{qe,D3Q_package}, as detailed in Refs.~\cite{fugallo_thermal_2014,simoncelli2020}.

To calculate the transport coefficients from the linear-response solution of the electron-phonon BTE, we computed the full scattering matrix~\cref{eq:symm_coupled_matrix} using temperature-dependent discrete grids of $k$ and $q$ points, as at low temperatures a much finer mesh of electron states is required for convergence. 
For phonons, we consistently used a 27$\times$27$\times$11 $q$-point mesh, and for electrons, we chose commensurate $k$-point meshes that resulted in transport properties converged to within 5\% of the total value, ranging from 567$\times$567$\times$231 at 30 K to 297$\times$297$\times$121 at 210 K. We note that to obtain an accurate relaxon solution, these meshes must be odd to respect the wavevector-inversion symmetry. To broaden the Dirac delta functions appearing in the scattering rate expressions, we applied a Gaussian smearing approximation, with a broadening of 5 meV for all scattering involving electrons (el-ph, ph-el, and drag calculations) and 2 meV for phonons (phonon-phonon, phonon-isotope). 

These inputs were then read in by the Phoebe code, where the calculation of the coupled Boltzmann transport equation was performed. The calculation produces JSON files containing standard thermoelectric coefficients: $\bar{\kappa}$, $\sigma$, $\alpha$, and $S$, as well as the additional coefficients that parameterize the VTE: $\eta^{\rm c_1 c_2}$, $D_{ij}^{\rm c_1 c_2}$, $\chi^j_{i,\mathrm{c}}$, and $\psi^j_i$.

\subsection{Numerical properties of the coupled electron-phonon scattering matrix}
To have a physically sound description of transport, it is necessary for the scattering matrix to be positive semi-definite. We can verify this by writing the scattering matrix in the form used in the appendix of Ref.~\cite{fugallo2013}: 
\begin{equation}
    A = \sum_{s,s'} P^{s''}_{s,s'} \mathbf{C}^{s''}_{s,s'},
\end{equation}
where $s,s',s''$ are a permutation of ${\km,\kmp,\qs}$, and $\mathbf{C}$ can be determined using the electron-phonon processes in this work, including $\mathrm{S}^{\rm ee}_{\km,\kmp}$, $\mathrm{D}^{\rm ep}_{\km,\qs}$, $\mathrm{D}^{\rm pe}_{\qs}$, and $\mathrm{S}^{\rm ph}_{\qs, \qsp}$, 
\begin{equation}
\mathbf{C}_{\km, \kmp}^{\qs}=
\begin{blockarray}{c*{3}{c}}
& \km & \kmp & \qs \\
\begin{block}{c!{\,}(ccc)}
\km\ \ & \frac{\mathfrak{D}}{N_k} & -\frac{\mathfrak{D}}{N_k} & \sqrt{\frac{\mathfrak{D}}{N_k N_q}} \\
\kmp\ \ & -\frac{\mathfrak{D}}{N_k} & \frac{\mathfrak{D}}{N_k} & -\sqrt{\frac{\mathfrak{D}}{N_k N_q}} \\
\qs\ \  & \sqrt{\frac{\mathfrak{D}}{N_k N_q}} & -\sqrt{\frac{\mathfrak{D}}{N_k N_q}} & \frac{\mathfrak{D}}{N_k} \\
\end{block}
\label{eq:psd-matrix}
\end{blockarray} 
\end{equation}%
If $N_k = N_q$, the eigenvalues of the above matrix are 3 and 0 with two-fold degeneracy, guaranteeing our coupled scattering matrix is positive semi-definite. Using different k- and q-meshes removes this guarantee and means it is in some cases possible to find negative eigenvalues. 

However, because of the different energy scales of electrons and phonons, achieving numerical convergence by fixing the $k$ and $q$ grids to the same value is computationally impossible. 
It is standard to use different meshes for electron and phonon transport calculations~\cite{ponce2020}. For electrons, particularly at low temperatures, one needs a grid on the order of $100^3$ k-points, while for phonons, grids on the order of $10^3$ are enough. For a calculation of graphite at 210 K, our smallest electronic mesh, we used a $k$-grid of $297 \times 297 \times 121$. In this case, the phonon part of the matrix would be of dimension $(297 \times 297 \times 121 \times 12)^2$ (where 12 is the number of phonon modes), and would require $\approx 130$ petabytes of memory. Additionally, the cost to calculate this many drag quadrant scattering matrix elements using electron-phonon Wannier interpolation would be prohibitively large. 

The similarity transformation we introduce in \cref{sec:coupled_basis} symmetrizes the drag terms, as reflected in the form of $\mathbf{C}_{\km, \kmp}^{\qs}$ in \cref{eq:psd-matrix}. 
It does not change the eigenvalues of the scattering matrix and does not repair the positive semi-definite nature of the matrix. 
We therefore must choose between two practical options: (1) maintain the positive semi-definite nature of the matrix using very coarse but equal $k$- and $q$-meshes, which would severely compromise the quality of electron scattering contributions; or (2) construct a scattering matrix where $N_k \gg N_q$, as in \cite{protik2022} and many past works, which breaks the positive semi-definite character of the matrix and may yield negative eigenvalues.

We compromise between points (1) and (2) above by starting from the equal-mesh case (1) at high temperature, where it is possible to converge electron-transport calculations with coarse meshes.
As a paradigmatic test case, we consider electron-doped silicon (carrier concentration $-$2.7e+19 cm$^{-3}$) at room temperature, where a reasonably converged calculation with $N_k = N_q$ is possible.
Starting from this calculation and increasing the electronic mesh so that $N_k \neq N_q$, while applying the similarity transformation, we verified that the final transport coefficients are preserved to reasonable numerical accuracy, as shown in Table~\ref{tab:Nkq}. 
\begin{table}[H]
\centering
\begin{tabular}{|c|c|c|}
\hline
$N_k$  & $N_q$ & $S$ [$\mu$V/K]  \\ \hline
30$^3$          & 30$^3$          & -277.00    \\ \hline
30$^3$          & 15$^3$          & -289.62    \\ \hline
60$^3$          & 15$^3$          & -299.65    \\ \hline
120$^3$         & 15$^3$          & -308.45    \\ \hline
\end{tabular}
\caption{Convergence of the Seebeck coefficient of electron-doped silicon ($-$2.7e+19 cm$^{-3}$) at 300K, as a function of the number of sampling points of the electronic ($N_k$) and phononic ($N_q$) Brillouin zones.}
\label{tab:Nkq}
\end{table}
It is worth mentioning that we also found that in graphite, negative eigenvalues can emerge from scattering processes involving phonon states with energies smaller than the Gaussian smearing value used in the calculation. 

Using Gaussian smearing to represent Dirac delta functions is an approximation to physical collisional broadening. Recent work \cite{castellano2025fluctuation,DiLucente2025,lihmSelfconsistentElectronLifetimes2024} has shown that, in the non-overdamped regime of transport, it is unphysical to use a smearing larger than the broadening linewidth and therefore larger than the energy of a quasiparticle. The simultaneous implementation of collisional broadening for electron-phonon and phonon-phonon scattering is an active area of research beyond the scope of thiswork, for the small number of phonon states with $\hbar\omega_{\rm ph} < \sigma_{\rm Gaussian}$ (as graphite has unusually low-energy out-of-plane modes along the $\Gamma$-$A$ direction), we set the normalization to $\frac{\mathfrak{D}}{N_{k}}$, corresponding to the value of the same matrix elements in the ideal limit $N_k=N_q$, which ensures a positive-definite matrix. In practice, these corrections might lead to an underestimation of the drag effect, as they reduce a few unphysically large matrix elements in the drag quadrant, so the signatures that we predicted might be considered as a lower bound on those that could be observed in very clean, pristine crystals. 

\subsection{Solving the VTE}

The real-space solutions of the VTE and the DTE were calculated using the same approach as in \cite{dragasevic2023}, including a lubrication layer used to simulate slipping boundary conditions. The width of the lubrication layer was $0.005 \rm \mu m$; within this layer, the viscosity tensors were reduced by a factor of 1000.


%

\end{document}